\documentclass{sig-alternate-2013}
\usepackage{algorithm,algpseudocode,cite,amsmath,amssymb,amsfonts,dsfont,multirow,multicol,enumitem,epsfig,url,array,makecell,balance,tabularx}
\usepackage{times}

\newtheorem{definition} {Definition}
\newtheorem{example} {Example}
\newtheorem{theorem} {Theorem}
\newtheorem{lemma} {Lemma}
\newtheorem{corollary} {Corollary}

\newcommand{\pushleft}[1]{\ifmeasuring@#1\else\omit$\displaystyle#1$\hfill\fi\ignorespaces}
\newcommand{\poly}{{\rm poly}}

\newcommand{\G}{{\mathcal G}}
\newcommand{\R}{{\mathcal R}}
\newcommand{\K}{{\mathcal K}}
\newcommand{\T}{{\mathcal T}}
\newcommand{\E}{{\mathbb E}}
\newcommand{\V}{{\mathcal V}^*}
\DeclareMathOperator*{\argmax}{arg\,max}

\algtext*{EndFor}
\algtext*{EndWhile}
\algtext*{EndIf}

\def\done{\hspace*{\fill} {$\square$}}
\def\header{\vspace{2mm} \noindent}
\def\pheader{\vspace{2.5mm} \noindent}

\def\figcapup{\vspace{-2mm}}
\def\figcapdown{\vspace{-3mm}}

\def\tbldown{\vspace{-2mm}}

\newfont{\mycrnotice}{ptmr8t at 7pt}
\newfont{\myconfname}{ptmri8t at 7pt}
\let\crnotice\mycrnotice%
\let\confname\myconfname%

\permission{Permission to make digital or hard copies of all or part of this work for personal or classroom use is granted without fee provided that copies are not made or distributed for profit or commercial advantage and that copies bear this notice and the full citation on the first page. Copyrights for components of this work owned by others than the author(s) must be honored. Abstracting with credit is permitted. To copy otherwise, or republish, to post on servers or to redistribute to lists, requires prior specific permission and/or a fee. Request permissions from permissions@acm.org.}
\conferenceinfo{SIGMOD'14,}{June 22--27, 2014, Snowbird, UT, USA. \\
{\mycrnotice{Copyright is held by the owner/author(s). Publication rights licensed to ACM.}}}
\copyrightetc{ACM \the\acmcopyr}
\crdata{978-1-4503-2376-5/14/06\ ...\$15.00.\\
http://dx.doi.org/10.1145/2588555.2593670}

\begin{document}
\begin{sloppy}

\allowdisplaybreaks

\title{Influence Maximization: Near-Optimal Time Complexity Meets Practical Efficiency}

\numberofauthors{1}

\author{
\alignauthor
Youze Tang $\qquad \qquad$ Xiaokui Xiao $\qquad \qquad$ Yanchen Shi \\
\affaddr{School of Computer Engineering} \\
\affaddr{Nanyang Technological University}\\
\affaddr{Singapore}\\
\email{tangyouze@gmail.com $\quad$ xkxiao@ntu.edu.sg $\quad$ shiy0017@e.ntu.edu.sg}
}

\maketitle

\begin{abstract}
Given a social network $G$ and a constant $k$, the {\em influence maximization} problem asks for $k$ nodes in $G$ that (directly and indirectly) influence the largest number of nodes under a pre-defined diffusion model. This problem finds important applications in viral marketing, and has been extensively studied in the literature. Existing algorithms for influence maximization, however, either trade approximation guarantees for practical efficiency, or vice versa. In particular, among the algorithms that achieve constant factor approximations under the prominent {\em independent cascade (IC)} model or {\em linear threshold (LT)} model, none can handle a million-node graph without incurring prohibitive overheads.

This paper presents {\em TIM}, an algorithm that aims to bridge the theory and practice in influence maximization. On the theory side, we show that {\em TIM} runs in $O((k+\ell) (n+m) \log n / \varepsilon^2)$ expected time and returns a $(1-1/e-\varepsilon)$-approximate solution with at least $1 - n^{-\ell}$ probability. The time complexity of {\em TIM} is near-optimal under the IC model, as it is only a $\log n$ factor larger than the $\Omega(m + n)$ lower-bound established in previous work (for fixed $k$, $\ell$, and $\varepsilon$). Moreover, {\em TIM} supports the {\em triggering model}, which is a general diffusion model that includes both IC and LT as special cases. On the practice side, {\em TIM} incorporates novel heuristics that significantly improve its empirical efficiency without compromising its asymptotic performance.
We experimentally evaluate {\em TIM} with the largest datasets ever tested in the literature, and show that it outperforms the state-of-the-art solutions (with approximation guarantees) by up to four orders of magnitude in terms of running time. In particular, when $k = 50$, $\varepsilon = 0.2$, and $\ell = 1$, {\em TIM} requires less than one hour on a commodity machine to process a network with $41.6$ million nodes and $1.4$ billion edges. This demonstrates that influence maximization algorithms can be made practical while still offering strong theoretical guarantees.
\end{abstract}


\vspace{-1mm}

%\vspace{1mm}
\section{Introduction} \label{sec:intro}

Let $G$ be a social network, and $M$ be a probabilistic model that captures how the nodes in $G$ may {\em influence} each other's behavior. Given $G$, $M$, and a small constant $k$, the {\em influence maximization} problem asks for the $k$ nodes in $G$ that can (directly and indirectly) influence the largest number of nodes. This problem finds important applications in {\em viral marketing} \cite{DomingosR01,RichardsonD02}, where a company selects a few {\em influential} individuals in a social network and provides them with incentives (e.g., free samples) to adopt a new product, hoping that the product will be recursively recommended by each individual to his/her friends to create a large cascade of further adoptions.

%Following Kempe et al.'s work, there has been

Kempe et al.\ \cite{KempeKT03} are the first to formulate influence maximization as a combinatorial optimization problem. They consider several probabilistic cascade models from the sociology and marketing literature \cite{GoldenbergLM01a,GoldenbergLM01b,Granovetter78,Schelling06}, and present a general greedy approach that yields $(1-1/e-\varepsilon)$-approximate solutions for all models considered, where $\varepsilon$ is a constant. This seminal work has motivated a large body of research on influence maximization in the past decade \cite{KimKY13,WangCW12,ChenYZ10,GoyalBL11,JungHC12,ChenWY09,ChenWW10,Wang2010community,leskovec2007cost,ChenLZ12,BharathiKS07,SeemanS13,KempeKT03,KempeKT05,Borgs14}.

Kempe et al.'s greedy approach is well accepted for its simplicity and effectiveness, but it is known to be computationally expensive. In particular, it has an $\Omega\left(k m n\cdot \poly\left(\varepsilon^{-1}\right)\right)$ time complexity \cite{Borgs14} where $n$ and $m$ are the numbers of nodes and edges in the social network, respectively. Empirically, it runs in days even when $n$ and $m$ are merely a few thousands \cite{ChenWY09}. Such inefficiency of Kempe et al.'s method has led to a plethora of algorithms \cite{KimKY13,WangCW12,ChenYZ10,GoyalBL11,JungHC12,ChenWY09,ChenWW10,Wang2010community,leskovec2007cost} that aim to reduce the computation overhead of influence maximization. Those algorithms, however, either trade performance guarantees for practical efficiency, or vice versa. In particular, most algorithms rely on heuristics to efficiently identify nodes with large influence, but they fail to achieve any approximation ratio under Kempe et al.'s cascade models; there are a few exceptions \cite{leskovec2007cost,GoyalLL11Celf,ChenWY09} that retain the $(1-1/e-\varepsilon)$-approximation guarantee, but they have the same time complexity with Kempe et al.'s method and still cannot handle large networks.

%have the same time complexity with Kempe et al.'s method and remain inapplicable for large networks. In other words, they either trade asymptotic guarantees for practical efficiency, or vice versa.

%The inefficiency of Kempe et al.'s method has motivated a large body of research \cite{} on reducing the computation overhead of influence maximization. The solutions proposed, however, either fail to guarantee any approximation ratio under Kempe et al.'s cascade models, or have the same time complexity with Kempe et al.'s method and remain inapplicable for large networks. In other words, they either trade asymptotic guarantees for practical efficiency, or vice versa.

%Furthermore, for those algorithms \cite{} that offer $(1-1/e-\varepsilon)$-approximate results (including Kempe et al.'s method), there is no formal analysis on how we can set the parameters of the algorithms for a given $\varepsilon$. Instead, it is only shown that once we fix the other parameters, there exists a certain (unknown) $\varepsilon$ for which the $1 - 1/e-\varepsilon$ approximation ratio holds \cite{KempeKT03,KempeKT05}.
%Say something about why this is bad

Very recently, Borgs et al.\ \cite{Borgs14} make a theoretical breakthrough and present an $O(k \ell^2 (m + n) \log^2 n /\varepsilon^3)$ time algorithm\footnotemark for influence maximization under the {\em independent cascade (IC)} model, i.e., one of the prominent models from Kempe et al.\ \cite{KempeKT03}. Borgs et al.\ show that their algorithm returns a $(1-1/e-\varepsilon)$-approximate solution with at least $1 - n^{-\ell}$ probability, and prove that it is {\em near-optimal} since any other algorithm that provides the same approximation guarantee and succeeds with at least a constant probability must run in $\Omega(m + n)$ time \cite{Borgs14}. Although Borgs et al.'s algorithm significantly improves upon previous methods in terms of asymptotic performance, its practical efficiency is rather unsatisfactory, due to a large hidden constant factor in its time complexity. In short, no existing influence maximization algorithm can scale to million-node graphs while still providing non-trivial approximation guarantees (under Kempe et al.'s models \cite{KempeKT03}). Therefore, any practitioner who conducts influence maximization on sizable social networks can only resort to heuristics, even though the results thus obtained could be arbitrarily worse than the optimal ones.

\footnotetext{The time complexity of Borgs et al.'s algorithm is established as $O(\ell^2 (m + n) \log^2 n /\varepsilon^3)$ in \cite{Borgs14}, but our correspondence with Borg et al.\ shows that it should be revised as $O(k \ell^2 (m + n) \log^2 n /\varepsilon^3)$, due to a gap in the proof of Lemma 3.6 in \cite{Borgs14}.}
%\header
%{\bf Motivation.}

%Furthermore, as we discuss in Section~\ref{sec:compare-RIS}, there is a gap in Borgs et al.'s proofs, due to which their algorithm cannot ensure the claimed success probability of $1 - n^{-\ell}$. It remains unclear whether the gap can be fixed without considerable increasing the algorithm's time complexity and further degrading its empirical performance.

%(would further degrade its empirical performance
%We show that those gaps can be fixed, but not without considerably increasing the algorithm's time complexity, which would further degrade its empirical performance.

%We show that those gaps can be fixed by revising Borg et al.'s algorithm, but the resulting time complexity is increased to $O(kn^\ell(n+m) \log n / \varepsilon^3)$, where $k$ is the number of influential nodes selected. Therefore, the revised algorithm incurs tremendous computation cost even when $\ell = 1$.

%This is a severe limitation in practice since online social networks in practice (e.g., Facebook, Twitter, Google+) can easily have millions or even billions of users.

\header
{\bf Our Contributions.} This paper presents {\em Two-phase Influence Maximization (TIM)}, an algorithm that aims to bridge the theory and practice in influence maximization. On the theory side, we show that {\em TIM} returns a $(1-1/e-\varepsilon)$-approximate solution with at least $1 - n^{-\ell}$
probability, and it runs in $O((k+\ell)(m+n) \log n / \varepsilon^2)$ expected time.
%In addition, its expected space consumption is $O(m + k\ell n \log n / \varepsilon^2)$.
The time complexity of {\em TIM} is near-optimal under the IC model, as it is only a $\log n$ factor larger than the $\Omega(m + n)$ lower-bound established by Borgs et al.\ \cite{Borgs14} (for fixed $k$, $\ell$, and $\varepsilon$). Moreover, {\em TIM} supports the {\em triggering model} \cite{KempeKT03}, which is a general cascade model that includes the IC model as a special case.
%Say something more

On the practice side, {\em TIM} incorporates novel heuristics that result in up to $100$-fold improvements of its computation efficiency, without any compromise of theoretical assurances. We experimentally evaluate {\em TIM} with a variety of social networks, and show that it outperforms the state-of-the-art solutions (with approximation guarantees) by up to four orders of magnitude in terms of running time.
%empirical efficiency is unmatched by any existing solution that provides non-trivial approximation guarantees.
In particular, when $k = 50$, $\varepsilon \ge 0.2$, and $\ell = 1$, {\em TIM} requires less than one hour to process a network with $41.6$ million nodes and $1.4$ billion edges. To our knowledge, this is the first result in the literature that demonstrates efficient influence maximization on a billion-edge graph.

In summary, our contributions are as follows:
\begin{enumerate}[topsep=2mm, partopsep=0pt, itemsep=1mm]
\item We propose an influence maximization algorithm that runs in near-linear expected time and returns $(1 - 1/e - \varepsilon)$-approximate solutions (with a high probability) under the triggering model.

\item We devise several optimization techniques that improve the empirical performance of our algorithm by up to $100$-fold.

\item We provide theoretical analysis on the state-of-the-art solutions with approximation guarantees, and establish the superiority of our algorithm in terms of asymptotic performance.

%    point out a crucial gap in the proofs of an important recent work.

\item We experiment with the largest datasets ever used in the literature, and show that our algorithm can efficiently handle graphs with more than a billion edges. This demonstrates that influence maximization algorithms can be made practical while still offering strong theoretical guarantees.
\end{enumerate}

\section{Preliminaries} \label{sec:prelim}

In this section, we formally define the influence maximization problem, and present an overview of Kempe et al.\ and Borgs et al.'s solutions \cite{KempeKT03,Borgs14}. For ease of exposition, we focus on the {\em independent cascade (IC)} model \cite{KempeKT03} considered by Borgs et al. \cite{Borgs14}. In Section~\ref{sec:ext-general}, we discuss how our solution can be extended to the more general {\em triggering model}.

\subsection{Problem Definition} \label{sec:prelim-def}

Let $G$ be a social network with a node set $V$ and a directed edge set $E$, with $|V| = n$ and $|E| = m$. Assume that each directed edge $e$ in $G$ is associated with a {\em propagation probability} $p(e) \in [0, 1]$. Given $G$, the independent cascade (IC) model considers a time-stamped influence propagation process as follows:
%The IC model assumes that each directed edge $\langle u, v\langle$ in $G$ is associated with a {\em propagation probability} $p_{u, v} \in [0, 1]$, and it considers an time-stamped influence propagation process at follows:
\begin{enumerate}[topsep=2.5mm, partopsep=0pt, itemsep=0mm]
\item At timestamp $1$, we {\em activate} a selected set $S$ of nodes in $G$, while setting all other nodes {\em inactive}.

\item If a node $u$ is first activated at timestamp $i$, then for each directed edge $e$ that points from $u$ to an inactive node $v$ (i.e., $v$ is an inactive outgoing neighbor of $u$), $u$ has $p(e)$ probability to activate $v$ at timestamp $i + 1$. After timestamp $i+1$, $u$ cannot activate any node.

\item Once a node becomes activated, it remains activated in all subsequent timestamps.
\end{enumerate}
Let $I(S)$ be the number of nodes that are activated when the above process converges, i.e., when no more nodes can be activated. We refer to $S$ as the {\em seed set}, and $I(S)$ as the {\em spread} of $S$. Intuitively, the influence propagation process under the IC model mimics the spread of an infectious disease: the seed set $S$ is conceptually similar to an initial set of infected individuals, while the activation of a node by its neighbors is analogous to the transmission of the disease from one individual to another.

For example, consider a propagation process on the social network $G$ in Figure~\ref{fig:def-G}, with $S = \{v_2\}$ as the seed set. (The number on each edge indicates the propagation probability of the edge.) At timestamp $1$, we activate $v_2$, since it is only node in $S$. Then, at timestamp $2$, both $v_1$ and $v_4$ have $0.01$ probability to be activated by $v_2$, as (i) they are both $v_2$'s outgoing neighbors and (ii) the edges from $v_2$ to $v_1$ and $v_4$ have a propagation probability of $0.01$. Suppose that $v_2$ activates $v_4$ but not $v_1$. After that, at timestamp $3$, $v_4$ will activate $v_1$ since the edge from $v_4$ to $v_1$ has a propagation probability of $1$. After that, the influence propagation process terminates, since no other node can be activated. The total number of nodes activated during the process is $3$, and hence, $I(S) = 3$.

Given $G$ and a constant $k$, the influence maximization problem under the IC model asks for a size-$k$ seed set $S$ with the maximum expected spread $\E\left[I(S)\right]$. %over the randomness in the propagation process.
In other words, we seek a seed set that can (directly and indirectly) activate the largest number of nodes in expectation.

%Consider adding an example here. Also consider adding the intuitive explanation of viral transmission here.

\begin{figure}[t]
\centering
\begin{small}
\begin{minipage}[h]{0.49\linewidth}
\centering
\includegraphics[height=11.8mm]{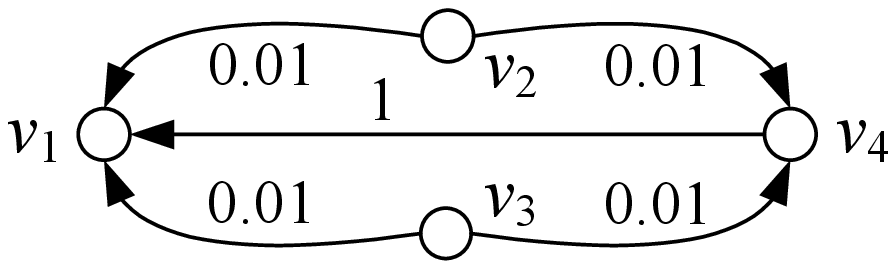}
\figcapup
%\vspace{-4mm}
\caption{Social network $G$.}
\label{fig:def-G}
\figcapdown
\end{minipage}
\hspace{0.03\linewidth}
\begin{minipage}[h]{0.46\linewidth}
\centering
%\hspace{5mm}
\includegraphics[height=11.8mm]{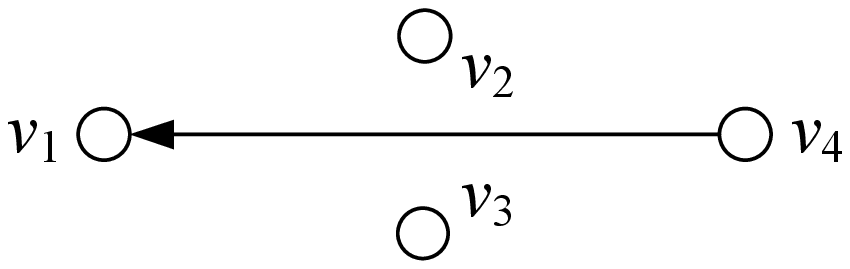}
%\vspace{-4mm}
\figcapup
\caption{Random graph $g_1$.}
\figcapdown
\label{fig:def-random}
\end{minipage}
\end{small}
\end{figure}

\subsection{Kempe et al.'s Greedy Approach} \label{sec:prelim-greedy}

In a nutshell, Kempe et al.'s approach \cite{KempeKT03} (referred to as {\em Greedy} in the following) starts from an empty seed set $S = \emptyset$, and then iteratively adds into $S$ the node $u$ that leads to the largest increase in $\E\left[I(S)\right]$, until $|S| = k$. That is,
\begin{eqnarray}
u &=& \argmax_{v \in V} \Big(\E\big[I\big(S \cup \{v\}\big)\big] - \E\big[I(S)\big]\Big). \nonumber
% \nonumber \\
%&=& \argmax_{v \in V} \E\big[I\big(S \cup \{v\}\big) - I(S)\big]. \nonumber
\end{eqnarray}
%For convenience, we define $\delta_S(v) = \E\big[I\big(S \cup \{v\}\big)\big] - \E\big[I(S)\big]$.
%For convenience, we define $I(v|S) = I(S \cup \{v\}) - I(S)$, and refer to $I(v|S)$ as the {\em marginal spread} of $v$ given $S$.

{\em Greedy} is conceptually simple, but it is non-trivial to implement since the computation of $\E[I(S)]$ is $\#$P-hard \cite{ChenWW10}. To address this issue, Kempe et al.\ propose to estimate $\E[I(S)]$ to a reasonable accuracy using a Monte Carlo method. To explain, suppose that we flip a coin for each edge $e$ in $G$, and remove the edge with $1 - p(e)$ probability. Let $g$ be the resulting graph, and $R(S)$ be the set of nodes in $g$ that are {\em reachable} from $S$. (We say that a node $v$ in $g$ is reachable from $S$, if there exists a directed path in $g$ that starts from a node in $S$ and ends at $v$.) Kempe et al.\ prove that the expected size of $R(S)$ equals $\E[I(S)]$. Therefore, to estimate $\E[I(S)]$, we can first generate multiple instances of $g$, then measure $R(S)$ on each instance, and finally take the average measurement as an estimation of $\E[I(S)]$.

Assume that we take a large number $r$ of measurements in the estimation of each $\E[I(S)]$. Then, with a high probability, {\em Greedy} yields a $(1 - 1/e - \varepsilon)$-approximate solution under the IC model \cite{KempeKT03}, where $\varepsilon$ is a constant that depends on both $G$ and $r$ \cite{KempeKT05,Borgs14}. In general, {\em Greedy} achieves the same approximation ratio under any cascade model where $\E[I(S)]$ is a {\em submodular} function of $S$ \cite{KempeKT05}. To our knowledge, however, there is no formal analysis in the literature on how $r$ should be set to achieve a given $\varepsilon$ on $G$. Instead, Kempe et al.\ suggest setting $r = 10000$, and most follow-up work adopts similar choices of $r$. In Section~\ref{sec:compare}, we provide a formal result on the relationship between $\varepsilon$ and $r$.

Although {\em Greedy} is general and effective, it incurs significant computation overheads due to its $O(kmnr)$ time complexity. Specifically, it runs in $k$ iterations, each of which requires estimating the expected spread of $O(n)$ node sets. In addition, each estimation of expected spread takes measurements on $r$ graphs, and each measurement needs $O(m)$ time. These lead to an $O(kmnr)$ total running time.

\subsection{Borgs et al.'s Method} \label{sec:prelim-ris}

The main reason for {\em Greedy}'s inefficiency is that it requires estimating the expected spread of $O(kn)$ node sets. Intuitively, most of those $O(kn)$ estimations are {\em wasted} since, in each iteration of {\em Greedy}, we are only interested in the node set with the largest expected spread. %As such, a natural idea is to improve {\em Greedy}'s performance is to avoid evaluating those node sets whose expected spreads are small. This idea, however, is
Yet, such wastes of computation are difficult to avoid under the framework of {\em Greedy}. To explain, consider the first iteration of {\em Greedy}, where we are to identify a single node in $G$ with the maximum expected spread. Without prior knowledge on the expected spread of each node, we would have to evaluate $\E[I(\{v\})]$ for each node $v$ in $G$. In that case, the overhead of the first iteration alone would be $O(mnr)$.

Borgs et al.\ \cite{Borgs14} avoid the limitation of {\em Greedy} and propose a drastically different method for influence maximization under the IC model. We refer to the method as {\em Reverse Influence Sampling (RIS)}. To explain how {\em RIS} works, we %first introduce the following concept:
first introduce two concepts:
%\begin{definition}[Transpose Graph $G^T$] \label{def:def-transpose}
%The transpose graph $G^T$ is constructed from $G$ by exchanging the starting and ending points for each edge $e$, without changing $e$'s propagation probability $p(e)$.
%\end{definition}
%In other words, for each edge $e = \langle v_1, v_2\rangle$ in $G$, there exists an edge $e^T = \langle v_2, v_1\rangle$ in $G^T$ with $p(e) = p(e^T)$.
\begin{definition}[Reverse Reachable Set] \label{def:def-rrset}
Let $v$ be a node in $G$, and $g$ be a graph obtained by removing each edge $e$ in $G$ with $1 - p(e)$ probability. The reverse reachable (RR) set for $v$ in $g$ is the set of nodes in $g$ that can reach $v$. (That is, for each node $u$ in the RR set, there is a directed path from $u$ to $v$ in $g$.)
%(We say that a node $u$ in $g$ can reach $v$, if there exists a directed path in $g$ that starts from $u$ and ends at $v$.)
\end{definition}
\begin{definition}[Random RR Set] \label{def:def-random-rrset}
Let $\G$ be the distribution of $g$ induced by the randomness in edge removals from $G$. A {\em random RR set} is an RR set generated on an instance of $g$ randomly sampled from $\G$, for a node selected uniformly at random from $g$.
\end{definition}
%Let $\G$ be the distribution of $g$ induced by the randomness in edge removals from $G$. We use the term {\em random RR set} to refer to an RR set generated on an instance of $g$ randomly sampled from $\G$, for a node selected uniformly at random from $g$.

%we write $R(v)$ as $R$ when $v$ is clear from the context, and

%a {\em transposed graph} $G^T$, which is constructed from $G$ by exchanging the starting and ending points for each edge $e$, without changing $e$'s propagation probability $p(e)$. That is, for each edge $e = \langle v_1, v_2\rangle$ in $G$, there exists an edge $e^T = \langle v_2, v_1\rangle$ in $G^T$ with $p(e) = p(e^T)$.

%Consider that we select a node $v$ from $G^T$, and then use $\{v\}$ as the seed set to run an influence propagation process on $G^T$ under the IC model (see Section~\ref{sec:prelim-def}). Let $A$ be the set of nodes activated in the process, including $v$ itself. Observe that, if a node $u$ appears in $A$, then there exists a directed path in $G$ that starts from $u$ and ends at $v$. Therefore, $u$ should have a chance to activate $v$ if we run an influence propagation process on $G$. More specifically, Borgs et al.\ show that if $u$ appears in $A$ with a probability $\rho$, then when we use $\{u\}$ to conduct an influence propagation process on $G$, we should have $\rho$ probability to activate $v$. Based on this result, Borgs et al.\ devise an influence maximization algorithm (which we refer to as {\em RIS}\footnote{Reverse Influence Sampling}) that runs in five steps as follows:

By definition, if a node $u$ appears in an RR set generated for a node $v$, then $u$ can reach $v$ via a certain path in $G$. As such, $u$ should have a chance to activate $v$ if we run an influence propagation process on $G$ using $\{u\}$ as the seed set. Borgs et al.\ show a result that is consistent with the above observation: If an RR set generated for $v$ has $\rho$ probability to overlap with a node set $S$, then when we use $S$ as the seed set to run an influence propagation process on $G$, we have $\rho$ probability to activate $v$ (See Lemma~\ref{lmm:tim-nodesel-rr}). Based on this result, Borgs et al.'s {\em RIS} algorithm runs in two steps

%three steps:
\begin{enumerate}
\item Generate a certain number of random RR sets from $G$.

\item Consider the {\em maximum coverage} problem \cite{Vazirani02} of selecting $k$ nodes to cover the maximum number of RR sets generated\footnote{We say that a node $v$ covers a set of nodes $S$ if and only if $v \in S$.}. Use the standard greedy algorithm \cite{Vazirani02} to derive a $(1 - 1/e)$-approximate solution $S^*_k$ for the problem. Return $S^*_k$ as the final result.
%
%Use a greedy algorithm to identify a size-$k$ node set $S_k$ that {\em covers}
%
%
%the largest number of activator sets. (We say that $S_k$ {\em covers} an activator set $A$, if $S_k \cap A \ne \emptyset$.)

%\item Return $S^*_k$ as the final result.
\end{enumerate}
The rationale of {\em RIS} is as follows: If a node $u$ appears in a large number of RR sets, then it should have a high probability to activate many nodes under the IC model; in that case, $u$'s expected spread should be large. By the same reasoning, if a size-$k$ node set $S^*_k$ covers most RR sets, then $S^*_k$ is likely to have the maximum expected spread among all size-$k$ node sets in $G$. In that case, $S^*_k$ should be a good solution to influence maximization. We illustrate {\em RIS} with an example.

%\begin{figure}[t]
%\centering
%\begin{small}
%\includegraphics[height=12mm]{./Figures/original.eps}
%%\vspace{-4mm}
%\figcapup
%\caption{Social network $G$.}
%\end{small}
%\label{fig:def-G}
%\figcapdown
%\end{figure}

\begin{example} \label{expl:def-RIS} \em
Consider that we invoke {\em RIS} on the social network $G$ in Figure~\ref{fig:def-G}, setting $k = 1$. {\em RIS} first generates a number of random RR sets, each of which is pertinent to (i) a node sampled uniformly at random from $G$ and (ii) a random graph obtained by removing each edge $e$ in $G$ with $1 - p(e)$ probability (see Definition~\ref{def:def-random-rrset}). Assume that the first RR set $R_1$ is pertinent to $v_1$ and the random graph $g_1$ in Figure~\ref{fig:def-random}. Then, we have $R_1 = \{v_1, v_4\}$, since $v_1$ and $v_4$ are the only two nodes in $g_1$ that can reach $v_1$.

Suppose that, besides $R_1$, {\em RIS} only constructs three other random RR sets $R_2$, $R_3$, and $R_4$, which are pertinent to three random graphs $g_2$, $g_3$, and $g_4$, respectively. For simplicity, assume that (i) $g_2$, $g_3$, and $g_4$ are identical to $g_1$, and (ii) the node that {\em RIS} samples from $g_i$ ($i \in [2, 4]$) is $v_i$. Then, we have $R_2 = \{v_2\}$, $R_3 = \{v_3\}$, and $R_4 = \{v_4\}$. In that case, $v_4$ is the node that covers the most number of RR sets, since it appears in two RR sets (i.e., $R_1$ and $R_4$), whereas any other node only covers one RR set. Consequently, {\em RIS} returns $S^*_k = \{v_4\}$ as the result. \done
\end{example}

Compared with {\em Greedy}, {\em RIS} can be more efficient as it avoids estimating the expected spreads of a large number of node sets. That said, we need to carefully control the number of random RR sets generated in Step 1 of {\em RIS}, so as to strike a balance between efficiency and accuracy. Towards this end, Borgs et al.\ propose a threshold-based approach: they allow {\em RIS} to keep generating RR sets, until the total number of nodes and edges examined during the generation process reaches a pre-defined threshold $\tau$. They show that when $\tau$ is set to $\Theta(k(m + n) \log n /\varepsilon^3)$, {\em RIS} runs in time linear to $\tau$, and it returns a $(1-1/e-\varepsilon)$-approximate solution to the influence maximization problem with at least a constant probability. They then provide an algorithm that amplifies the success probability to at least $1 - n^{-\ell}$, by increasing $\tau$ by a factor of $\ell$, and repeating {\em RIS} for $\Omega(\ell \log n)$ times.

%As we show in Section~\ref{sec:compare-RIS}, however, there is a gap in Borgs et al.'s analysis that invalidates the above asymptotic bounds of {\em RIS}. The reason can be intuitively explained as follows. Given that {\em RIS} sets a threshold $\tau$ on the total cost of Step 1, the RR sets sampled in Step 1 are correlated. Due to such correlations, some nodes in $G$ may appear in RR sets more frequently than normal\footnote{To demonstrate this phenomenon, imagine that we repeatedly sample from a Bernoulli distribution with $p = 0.5$, until the sum of samples reaches $1$. It can be verified that our sample set has $1/2$ probability to contain more $1$ than $0$, but only $1/4$ probability to contain more $0$ than $1$. In other words, $1$ is the most frequent number in the sample set with an abnormally high probability.}. In that case, even if we identify a node set that covers the most number of RR sets, it still may not be a good solution to the influence maximization problem. Borgs et al.\ attempt to mitigate the effects of correlations by setting $\tau$ to a relatively large number, but their approach turns out to be insufficient.

Despite of its near-linear time complexity, {\em RIS} still incurs significant computational overheads in practice, as we show in Section~\ref{sec:exp}. The reason can be intuitively explained as follows. Given that {\em RIS} sets a threshold $\tau$ on the total cost of Step 1, the RR sets sampled in Step 1 are correlated, due to which some nodes in $G$ may appear in RR sets more frequently than normal\footnote{To demonstrate this phenomenon, imagine that we repeatedly sample from a Bernoulli distribution with $p = 0.5$, until the sum of samples reaches $1$. It can be verified that our sample set has $1/2$ probability to contain more $1$ than $0$, but only $1/4$ probability to contain more $0$ than $1$. In other words, $1$ is the most frequent number in the sample set with an abnormally high probability.}. In that case, even if we identify a node set that covers the most number of RR sets, it still may not be a good solution to the influence maximization problem. Borgs et al.\ mitigate the effects of correlations by setting $\tau$ to a large number. However, this not only results in the $\varepsilon^{-3}$ term in {\em RIS}'s time complexity, but also renders {\em RIS}'s practical efficiency less than satisfactory.

%and a large hidden constant factor

%, due to the correlations among samples induced by the threshold on the sample sum.}.

%To remedy the deficiency of {\em RIS}, we present 
\section{Proposed Solution} \label{sec:tim}

This section presents {\em TIM}, an influence maximization method that borrows ideas from {\em RIS} but overcomes its limitations with a novel algorithm design. At a high level, {\em TIM} consists of two phases as follows:
\begin{enumerate}
\item {\bf Parameter Estimation.} This phase computes a lower-bound of the maximum expected spread among all size-$k$ node sets, and then uses the lower-bound to derive a parameter $\theta$.

\item {\bf Node Selection.} This phase samples $\theta$ random RR sets from $G$, and then derives a size-$k$ node set $S^*_k$ that covers a large number of RR sets. After that, it returns $S^*_k$ as the final result.
\end{enumerate}
The node selection phase of {\em TIM} is similar to {\em RIS}, except that it samples a {\em pre-decided} number (i.e., $\theta$) of random RR sets, instead of using a threshold on computation cost to indirectly control the number. This ensures that the RR sets generated by {\em TIM} are independent (given $\theta$), thus avoiding the correlation issue that plagues {\em RIS}. Meanwhile, the derivation of $\theta$ in the parameter estimation phase is non-trivial: As we shown in Section~\ref{sec:tim-nodesel}, $\theta$ needs to be larger than a certain threshold to ensure the correctness of {\em TIM}, but the threshold depends on the optimal result of influence maximization, which is unknown. To address this challenge, we compute a $\theta$ that is above the threshold but still small enough to ensure the overall efficiency of {\em TIM}.

In what follows, we first elaborate the node selection phase of {\em TIM}, and then detail the parameter estimation phase. For ease of reference, Table~\ref{tbl:tim-notation} lists the notations frequently used. Unless otherwise specified, all logarithms in this paper are to the base $e$.

\begin{table}[t]
\centering
\begin{small}
\renewcommand{\arraystretch}{1.3}
%\hspace{-5mm}
\begin{tabularx}{1.01\linewidth}{|@{\hspace{1mm}}c@{\hspace{1mm}}|X|} \hline
   {\bf Notation} & \multicolumn{1}{c|}{\bf Description} \\ \hline
   $G$, $G^T$                   & \multicolumn{1}{m{2.5in}|}{a social network $G$, and its {\em transpose} $G^T$ constructed by exchanging the starting and ending points of each edge in $G$} \\
   \hline
   $n$       & the number of nodes in $G$ (resp.\ $G^T$) \\
   \hline
   $m$       & the number of edges in $G$ (resp.\ $G^T$) \\
   \hline
   $k$       & the size of the seed set for influence maximization \\
   \hline
   $p(e)$    & the propagation probability of an edge $e$ \\
   \hline
%   $g$, $g^T$ & \multicolumn{1}{m{2.5in}|}{a graph $g$ obtained by removing each edge $e$ in $G$ with a probability of $1 - p(e)$, and its transposed version $g^T$} \\
%   \hline
%   $\G^T$ & \multicolumn{1}{m{2.5in}|}{the distribution of $g^T$ induced by the randomization in edge removals} \\
%   \hline
%   $\G$ (resp.\ $\G^T$) & \multicolumn{1}{m{2.5in}|}{the distribution of $g$ (resp.\ $g^T$) induced by the randomization in edge removals} \\
%   \hline
   $I(S)$ & \multicolumn{1}{m{2.5in}|}{the spread of a node set $S$ in an influence propagation process on $G$ (see Section~\ref{sec:prelim-greedy})} \\
   \hline
   $w(R)$ & \multicolumn{1}{m{2.5in}|}{the number of edges in $G^T$ that starts from the nodes in an RR set $R$ (see Equation~\ref{eqn:tim-nodesel-width})} \\
   \hline
   $\kappa(R)$ & \multicolumn{1}{m{2.5in}|}{see Equation~\ref{eqn:tim-para-omega}} \\
   \hline
   $\R$  & \multicolumn{1}{m{2.5in}|}{the set of all RR sets generated in Algorithm~\ref{alg:tim-nodesel}} \\
   \hline
   $F_{\R}(S)$  & \multicolumn{1}{m{2.5in}|}{the fraction of RR sets in $\R$ that are covered by a node set $S$} \\
   \hline
   $EPT$ & \multicolumn{1}{m{2.5in}|}{the expected width of a random RR set} \\
   \hline
   $OPT$ & \multicolumn{1}{m{2.5in}|}{the maximum $I(S)$ for any size-$k$ node set $S$} \\
   \hline
   $KPT$ & \multicolumn{1}{m{2.5in}|}{a lower-bound of $OPT$ established in Section~\ref{sec:tim-para}} \\
   \hline
   $\lambda$ & \multicolumn{1}{m{2.5in}|}{see Equation~\ref{eqn:tim-lambda}} \\
   \hline
%   $C_g(v)$ & \multicolumn{1}{m{2.5in}|}{the set of nodes in $g$ that are reachable from a node $v$ via a directed path starting from $v$} \\
%   \hline
%   $C_{g^T}(v)$ & \multicolumn{1}{m{2.5in}|}{the set of nodes in $g^T$ that are reachable from a node $v$ via a directed path starting from $v$} \\
%   \hline
%   $N(S)$ & \multicolumn{1}{m{2.5in}|}{the number of activator sets covered by a node set $S$ in Algorithm~\ref{alg:tim-nodesel}} \\
%   \hline
 \end{tabularx}
 \caption{Frequently used notations.}\label{tbl:tim-notation}
 \end{small}
  \tbldown
\end{table}

\subsection{Node Selection} \label{sec:tim-nodesel}

Algorithm~\ref{alg:tim-nodesel} presents the pseudo-code of {\em TIM}'s node selection phase. Given $G$, $k$, and a constant $\theta$, the algorithm first generates $\theta$ random RR sets, and inserts them into a set $\R$ (Lines 1-2). The subsequent part of the algorithm consists of $k$ iterations (Lines 3-7). In each iteration, the algorithm selects a node $v_j$ that covers the largest number of RR sets in $\R$, and then removes all those covered RR sets from $\R$. The $k$ selected nodes are put into a set $S^*_k$, which is returned as the final result.

\header
{\bf Implementation.} Lines 6-10 in Algorithm~\ref{alg:tim-nodesel} correspond to a standard greedy approach for a {\em maximum coverage} problem \cite{Vazirani02}, i.e., the problem of selecting $k$ nodes to cover the largest number of node sets. It is known that this greedy approach returns $(1 - 1/e)$-approximate solutions, and has a linear-time implementation. For brevity, we omit the description of the implementation and refer interested readers to \cite{Borgs14} for details.

Meanwhile, the generation of each RR set in Algorithm~\ref{alg:tim-nodesel} is implemented as a randomized breath-first search (BFS) on $G$. Given a node $v$ in $G$, we first create an empty queue, and then flip a coin for each {\em incoming} edge $e$ of $v$; with $p(e)$ probability, we retrieve the node $u$ from which $e$ starts, and we put $u$ into the queue. Subsequently, we iteratively extract the node $v'$ at the top of the queue, and examine each incoming edge $e'$ of $v$; if $e'$ starts from an unvisited node $u'$, we add $u'$ into the queue with $p(e')$ probability. This iterative process terminates when the queue becomes empty. Finally, we collect all nodes visited during the process (including $v$), and use them to form an RR set.

\header
{\bf Performance Bounds.} We define the {\em width} of an RR set $R$, denoted as $w(R)$, as the number of directed edges in $G$ whose point to the nodes in $R$. That is
\begin{equation} \label{eqn:tim-nodesel-width}
w(R) = \sum_{v \in R} \left(\textrm{the indegree of $v$ in $G$}\right).
\end{equation}
Observe that if an edge is examined in the generation of $R$, then it must point to a node in $R$. Let $EPT$ be the expected width of a random RR set. It can be verified that Algorithm~\ref{alg:tim-nodesel} runs in $O(\theta \cdot EPT)$ time. In the following, we analyze how $\theta$ should be set to minimize the expected running time while ensuring solution quality. Our analysis frequently uses the Chernoff bounds \cite{MotwaniR95}:
%and requires $O(\theta \cdot EPT)$ space
\begin{lemma} \label{lmm:tim-chernoff}
Let $X$ be the sum of $c$ i.i.d.\ random variables sampled from a distribution on $[0, 1]$ with a mean $\mu$. For any $\delta > 0$,
\begin{align*}
Pr\Big[X - c\mu \ge \delta \cdot c\mu\Big] & \le  \exp\left(-\frac{\delta^2}{2+\delta} c\mu\right), \\
Pr\Big[X - c\mu \le -\delta \cdot c\mu\Big] & \le  \exp\left(-\frac{\delta^2}{2} c\mu\right).
\end{align*}
\end{lemma}

In addition, we utilize the following lemma from \cite{Borgs14} that establishes the connection between RR sets and the influence propagation process on $G$:
\begin{lemma}\label{lmm:tim-nodesel-rr}
Let $S$ be a fixed set of nodes, and $v$ be a fixed node. Suppose that we generate an RR set $R$ for $v$ on a graph $g$ that is constructed from $G$ by removing each edge $e$ with $1 - p(e)$ probability. Let $\rho_1$ be the probability that $S$ overlaps with $R$, and $\rho_2$ be the probability that $S$, when used as a seed set, can activate $v$ in an influence propagation process on $G$. Then, $\rho_1 = \rho_2$.
\end{lemma}
The proofs of all theorems, lemmas, and corollaries in Section~\ref{sec:tim} are included in the appendix.

\begin{algorithm}[t]
\begin{small}
\caption{NodeSelection ($G$, $k$, $\theta$) \label{alg:tim-nodesel}}
\begin{algorithmic}[1]
 \State Initialize a set $\R = \emptyset$.
 \State Generate $\theta$ random RR sets and insert them into $\R$.
 \State Initialize a node set $S^*_k = \emptyset$.
 \For{$j = 1$ to $k$}
    \State Identify the node $v_j$ that covers the most RR sets in $\R$.
    \State Add $v_j$ into $S^*_k$.
    \State Remove from $\R$ all RR sets that are covered by $v_j$.
 \EndFor
 \State \Return{$S^*_k$}
\end{algorithmic}
\end{small}
\end{algorithm}

Let $\R$ be the set of all RR sets generated in Algorithm~\ref{alg:tim-nodesel}. For any node set $S$, let $F_{\R}(S)$ be fraction of RR sets in $\R$ covered by $S$. Then, based on Lemma~\ref{lmm:tim-nodesel-rr}, we can prove that the expected value of $n \cdot F_{\R}(S)$ equals the expected spread of $S$ in $G$:
\begin{corollary} \label{coro:tim-fraction}
$\E[n \cdot F_{\R}(S)] = \E[I(S)]$.
\end{corollary}

Let $OPT$ be the maximum expected spread of any size-$k$ node set in $G$. Using the Chernoff bounds, we show that $n \cdot F_{\R}(S)$ is an accurate estimator of any node set $S$'s expected spread, when $\theta$ is sufficiently large:
\begin{lemma} \label{lmm:tim-theta}
Suppose that $\theta$ satisfies
\begin{equation} \label{eqn:tim-theta-1}
\theta \ge (8+2\varepsilon) n \cdot \frac{\ell \log n + \log{n \choose k} + \log 2 }{OPT \cdot \varepsilon^2}.
\end{equation}
Then, for any set $S$ of at most $k$ nodes, the following inequality holds with at least $1 - n^{-\ell}/ {n \choose k}$ probability:
\begin{equation} \label{eqn:tim-theta-2}
\Big|n\cdot F_{\R}(S) - \E[I(S)]\Big| < \frac{\varepsilon}{2} \cdot OPT.
\end{equation}
\end{lemma}

Based on Lemma~\ref{lmm:tim-theta}, we prove that when Equation~\ref{eqn:tim-theta-1} holds, Algorithm~\ref{alg:tim-nodesel} returns a $(1 - 1/e - \varepsilon)$-approximate solution with high probability:
\begin{theorem} \label{thrm:tim-approx}
Given a $\theta$ that satisfies Equation~\ref{eqn:tim-theta-1}, Algorithm~\ref{alg:tim-nodesel} returns a $(1 - 1/e - \varepsilon)$-approximate solution with at least $1 - n^{-\ell}$ probability.
\end{theorem}

Notice that it is difficult to set $\theta$ directly based on Equation~\ref{eqn:tim-theta-1}, since $OPT$ is unknown. We address this issue in Section~\ref{sec:tim-para}, by presenting an algorithm that returns a $\theta$ which not only satisfies Equation~\ref{eqn:tim-theta-1}, but also leads to an $O((k+\ell)(m+n)\log n /\varepsilon^2)$ expected time complexity for Algorithm~\ref{alg:tim-nodesel}. For simplicity, we define
\begin{equation} \label{eqn:tim-lambda}
\lambda = (8+2\varepsilon) n \cdot \big(\ell \log n + \log{\textstyle {n \choose k}} + \log 2\big) \cdot \varepsilon^{-2},
\end{equation}
and we rewrite Equation~\ref{eqn:tim-theta-1} as
\begin{equation} \label{eqn:tim-theta-3}
\theta \ge \lambda/OPT.
\end{equation}
\subsection{Parameter Estimation} \label{sec:tim-para}

Recall that the expected time complexity of Algorithm~\ref{alg:tim-nodesel} is ${O(\theta \cdot EPT)}$, where $EPT$ is the expected number of coin tosses required to generate an RR set for a randomly selected node in $G$. Our objective is to identify an $\theta$ that makes $\theta \cdot EPT$ reasonably small, while still ensuring $\theta \ge \lambda / OPT$. Towards this end, we first define a probability distribution $\V$ over the nodes in $G$, such that the probability mass for each node is proportional to its in-degree in $G$. Let $v^*$ be a random variable following $\V$. We have the following lemma:
\begin{lemma} \label{lmm:tim-para-ept}
$\frac{n}{m} EPT = \E[I(\{v^*\})]$, where the expectation of $I(\{v^*\})$ is taken over the randomness in $v^*$ and the influence propagation process.
\end{lemma}
In other words, if we randomly sample a node from $\V$ and calculate its expected spread $s$, then on average we have $s = \frac{n}{m} EPT$. This implies that $\frac{n}{m}EPT \le OPT$, since $OPT$ equals the maximum expected spread of any size-$k$ node set.

Suppose that we are able to identify a number $t$ such that $t = \Omega(\frac{n}{m}EPT)$ and $t \le OPT$. Then, by setting $\theta = \lambda / t$, we can guarantee that Algorithm~\ref{alg:tim-nodesel} is correct and has an expected time complexity of
\begin{equation} \label{eqn:tim-para-time}
O(\theta \cdot EPT) = O\left(\frac{m}{n} \lambda\right) = O\left((k+\ell)(m+n)\log n /\varepsilon^2\right).
\end{equation}

\header
{\bf Choices of $\boldsymbol{t}$.} An intuitive choice of $t$ is $t = \frac{n}{m}EPT$, since (i) both $n$ and $m$ are known and (ii) $EPT$ can be estimated by measuring the average width of RR sets. However, we observe that when $k \gg 1$, $t = \frac{n}{m}EPT$ renders $\theta = \lambda/t$ unnecessarily large, which in turn leads to inferior efficiency. To explain, recall that $\frac{n}{m}EPT$ equals the mean of the expected spread of a node $v^*$ sampled from $\V$, and hence, it is independent of $k$. In contrast, $OPT$ increases monotonically with $k$. Therefore, the difference between $\frac{n}{m}EPT$ and $OPT$ increases with $k$, which makes $t = \frac{n}{m}EPT$ an unfavorable choice of $t$ when $k$ is large. To tackle this problem, we replace $\frac{n}{m}EPT$ with a closer approximation of $OPT$ that increases with $k$, as explained in the following.

Suppose that we take $k$ samples from $\V$, and use them to form a node set $S^*$. (Note that $S^*$ may contain fewer than $k$ nodes due to the elimination of duplicate samples.) Let $KPT$ be the mean of the expected spread of $S^*$ (over the randomness in $S^*$ and the influence propagation process). It can be verified that
\begin{equation} \label{eqn:tim-para-kpt}
{\textstyle \frac{n}{m}EPT \le KPT \le OPT},
\end{equation}
and that $KPT$ increases with $k$. We also have the following lemma:
\begin{lemma} \label{lmm:tim-para-kpt}
Let $R$ be a random RR set and $w(R)$ be the width of $R$. Define
\begin{equation} \label{eqn:tim-para-omega}
\kappa(R) = 1 - \left(1 - \frac{w(R)}{m}\right)^k. %{\textstyle 1 - (1 - \frac{m_A}{m})^k}
\end{equation}
Then, $KPT = n \cdot \E\left[\kappa(R)\right]$, where the expectation is taken over the random choices of $R$.
\end{lemma}

By Lemma~\ref{lmm:tim-para-kpt}, we can estimate $KPT$ by first measuring $n \cdot \kappa(R)$ on a set of random RR sets, and then taking the average of the measurements. But how many measurements should we taken? By the Chernoff bounds, if we are to obtain an estimate of $KPT$ with $\delta \in (0, 1)$ relative error with at least $1 - n^{-\ell}$ probability, then the number of measurements should be $\Omega( \ell n \log n \cdot \delta^{-2} / KPT)$. In other words, the number of measurements required depends on $KPT$, whereas $KPT$ is exactly the subject being measured. We resolve this dilemma with an adaptive sampling approach that dynamically adjusts the number of measurements based on the observed samples of RR sets.

\header
{\bf Estimation of $\boldsymbol{KPT}$.} Algorithm~\ref{alg:tim-kpt} presents our sampling approach for estimating $KPT$. The high level idea of the algorithm is as follows. We first generate a relatively small number of RR sets, and use them to derive an estimation of $KPT$ with a bounded absolute error. If the estimated value of $KPT$ is much larger than the error bound, we infer that the estimation is accurate enough, and we terminate the algorithm. On the other hand, if the estimated value of $KPT$ is not large compared with the error bound, then we generate more RR sets to obtain a new estimation of $KPT$ with a reduced absolute error. After that, we re-evaluate the accuracy of our estimation, and if necessary, we further increase the number of RR sets, until a precise estimation of $KPT$ is computed.

\begin{algorithm}[t]
\begin{small}
\caption{KptEstimation ($G$, $k$) \label{alg:tim-kpt}}
\begin{algorithmic}[1]
 \For{$i = 1$ to $\log_2 n - 1$}
    \State Let $c_i = \left(6\ell \log n + 6\log(\log_2 n)\right)\cdot 2^i$.
    \State Let $sum = 0$.
    \For{$j = 1$ to $c_i$}
        \State Generate a random RR set $R$.
        \State $\kappa(R) = 1 - \left(1 - \frac{w(R)}{m}\right)^k$
        \State $sum = sum + \kappa(R)$.
    \EndFor
    \If{$sum/c_i > 1/2^{i}$}
        \State \Return{$KPT^* = n \cdot sum/(2 \cdot c_i)$}
    \EndIf
 \EndFor
 \State \Return{$KPT^* = 1$}
\end{algorithmic}
\end{small}
\end{algorithm}

More specifically, Algorithm~\ref{alg:tim-kpt} runs in at most $\log_2 n - 1$ iterations. In the $i$-th iteration, it samples $c_i$ RR sets from $G$ (Lines 2-7), where
\begin{equation} \label{eqn:tim-kpt-ci}
c_i = \big(6\ell \log n + 6\log\left(\log_2 n\right)\big)\cdot 2^i.
\end{equation}
Then, it measures $\kappa(R)$ on each RR set $R$, and computes the average value of $\kappa(R)$. Our choice of $c_i$ ensures that if this average value is larger than $2^{-i}$, then with a high probability, $\E[\kappa(R)]$ is at least half of the average value; in that case, the algorithm terminates by returning a $KPT^*$ that equals the average value times $n/2$ (Lines 8-9). Meanwhile, if the average value is no more than $2^{-i}$, then the algorithm proceeds to the ($i+1$)-th iteration.

On the other hand, if the average value is smaller than $2^{-i}$ in all $\log_2 n - 1$ iterations, then the algorithm returns $KPT^* = 1$, which equals the smallest possible $KPT$ (since each node in the seed set can always activate itself). As we show shortly, $\E[\frac{1}{KPT^*}] = O(\frac{1}{KPT})$, and $KPT^* \in [KPT/4, OPT]$ holds with a high probability. Hence, setting $\theta = \lambda / KPT^*$ ensures that Algorithm~\ref{alg:tim-nodesel} is correct and achieves the expected time complexity in Equation~\ref{eqn:tim-para-time}.

\header
%\noindent
{\bf Theoretical Analysis.} Although Algorithm~\ref{alg:tim-kpt} is conceptually simple, proving its correctness and effectiveness is non-trivial as it requires a careful analysis of the algorithm's behavior in each iteration. In what follows, we present a few supporting lemmas, and then use them to establish Algorithm~\ref{alg:tim-kpt}'s performance guarantees.

Let $\K$ be the distribution of $\kappa(R)$ over random RR sets in $G$. Then, $\K$ has a domain $[0, 1]$. Let $\mu = KPT/n$, and $s_i$ be the sum of $c_i$ i.i.d.\ samples from $\K$, where $c_i$ is as defined in Equation~\ref{eqn:tim-kpt-ci}. By the Chernoff bounds, we have the following result:
\begin{lemma} \label{lmm:tim-para-stopup}
If $\mu \le 2^{-j}$, then for any $i \in  [1, j - 1]$,
\begin{equation*}
\Pr\left[\frac{s_i}{c_i} > \frac{1}{2^i}\right] < \frac{1}{n^{\ell}\cdot \log_2 n}.
\end{equation*}
\end{lemma}
By Lemma~\ref{lmm:tim-para-stopup}, if $KPT \le 2^{-j}$, then Algorithm~\ref{alg:tim-kpt} is very unlikely to terminate in any of the first $j - 1$ iterations. This prevents the algorithm from outputting a $KPT^*$ too much larger than $KPT$.

\begin{lemma} \label{lmm:tim-para-stopdown}
If $\mu \ge 2^{-j}$, then for any $i \ge j + 1$,
\begin{equation*}
\Pr\left[\frac{s_i}{c_i} > \frac{1}{2^i}\right] > 1 - n^{-\ell \cdot 2^{i-j-1}}/\log_2 n.
\end{equation*}
\end{lemma}
By Lemma~\ref{lmm:tim-para-stopdown}, if $KPT \le 2^{-j}$ and Algorithm~\ref{alg:tim-kpt} happens to enter its $i > j+1$ iteration, then it will almost surely terminate in the $i$-th iteration. This ensures that the algorithm would not output a $KPT^*$ that is considerably smaller than $KPT$.

Based on Lemmas \ref{lmm:tim-para-stopup} and \ref{lmm:tim-para-stopdown}, we prove the following theorem on the accuracy and expected time complexity of Algorithm~\ref{alg:tim-kpt}:
\begin{theorem} \label{thrm:tim-para-kpt}
When $n \ge 2$ and $\ell \ge 1/2$, Algorithm~\ref{alg:tim-kpt} returns $KPT^* \in [KPT/4, OPT]$ with at least $1 - n^{-\ell}$ probability, and runs in $O(\ell (m+n) \log n)$ expected time. Furthermore, $\E\left[\frac{1}{KPT^*}\right] < \frac{12}{KPT}$.
\end{theorem} 
\subsection{Putting It Together} \label{sec:tim-together}

In summary, our {\em TIM} algorithm works as follows. Given $G$, $k$, and two parameters $\varepsilon$ and $\ell$, {\em TIM} first feeds $G$ and $k$ as input to Algorithm~\ref{alg:tim-kpt}, and obtains a number $KPT^*$ in return. After that, {\em TIM} computes $\theta = \lambda / KPT^*$, where $\lambda$ is as defined in Equation~\ref{eqn:tim-lambda} and is a function of $k$, $\ell$, $n$, and $\varepsilon$. Finally, {\em TIM} gives $G$, $k$, and $\theta$ as input to Algorithm~\ref{alg:tim-nodesel}, whose output $S^*_k$ is the final result of influence maximization.

By Theorems~\ref{thrm:tim-approx} and \ref{thrm:tim-para-kpt}, Equation~\ref{eqn:tim-para-time}, and the union bound, {\em TIM} runs in $O((k+\ell)(m+n)\log n /\varepsilon^2)$ expected time, and returns a ${(1 - 1/e - \varepsilon)}$-approximate solution with at least $1 - 2\cdot n^{-\ell}$ probability. This success probability can easily increased to $1 - n^{-\ell}$, by scaling $\ell$ up by a factor of $1 + \log 2/\log n$. Finally, we note that the time complexity of {\em TIM} is near-optimal under the IC model, as it is only a $\log n$ factor larger than the $\Omega(m + n)$ lower-bound proved by Borgs et al.\ \cite{Borgs14} (for fixed $k$, $\ell$, and $\varepsilon$).

\section{Extensions} \label{sec:ext}

In this section, we present a heuristic method for improving the practical performance of {\em TIM} (without affecting its asymptotic guarantees), and extend {\em TIM} to an influence propagation model more general than the IC model.

\subsection{Improved Parameter Estimation} \label{sec:ext-faster}

The efficiency of {\em TIM} highly depends on the output $KPT^*$ of Algorithm~\ref{alg:tim-kpt}. If $KPT^*$ is close to $OPT$, then $\theta = \lambda / KPT^*$ is small; in that case, Algorithm~\ref{alg:tim-nodesel} only needs to generate a relatively small number of RR sets, thus reducing computation overheads. However, we observe that $KPT^*$ is often much smaller than $OPT$ on real datasets, which severely degrades the efficiency of Algorithm~\ref{alg:tim-nodesel} and the overall performance of {\em TIM}.

\begin{algorithm}[t]
\begin{small}
\caption{RefineKPT ($G$, $k$, $KPT^*$, $\varepsilon'$) \label{alg:ext-refine}}
\begin{algorithmic}[1]
 \State Let $\R'$ be the set of all RR sets generated in the last iteration of Algorithm~\ref{alg:tim-kpt}.
 \State Initialize a node set $S'_k = \emptyset$.
 \For{$j = 1$ to $k$}
    \State Identify the node $v_j$ that covers the most RR sets in $\R'$.
    \State Add $v_j$ into $S'_k$.
    \State Remove from $\R'$ all RR sets that are covered by $v_j$.
 \EndFor
 \State Let $\lambda' = (2+\varepsilon') \ell n \log n \cdot (\varepsilon')^{-2}$.
 \State Let $\theta' = \lambda'/KPT^*$.
 \State Generate $\theta'$ random RR sets; put them into a set $\R''$.
 \State Let $f$ be the fraction of the RR sets in $\R''$ that is covered by $S'_k$.
 \State Let $KPT' = f \cdot n /(1+\varepsilon')$
 \State \Return $KPT^+ = \max\{KPT', KPT^*\}$
\end{algorithmic}
\end{small}
\end{algorithm}

Our solution to the above problem is to add an intermediate step between Algorithms \ref{alg:tim-nodesel} and \ref{alg:tim-kpt} to refine $KPT^*$ into a (potentially) much tighter lower-bound of $OPT$. Algorithm~\ref{alg:ext-refine} shows the pseudo-code of the intermediate step. The algorithm first retrieves the set $\R'$ of all RR sets created in the last iteration of Algorithm~\ref{alg:tim-kpt}, i.e., the RR sets that from which $KPT^*$ is computed. Then, it invokes the greedy approach (for the maximum coverage problem) on $\R'$, and obtains a size-$k$ node set $S'_k$ that covers a large number of RR sets in $\R'$ (Lines 2-6 in Algorithm~\ref{alg:ext-refine}).

Intuitively, $S'_k$ should have a large expected spread, and thus, if we can estimate $\E[I(S'_k)]$ to a reasonable accuracy, then we may use the estimation to derive a good lower-bound for $OPT$. Towards this end, Algorithm~\ref{alg:ext-refine} generates a number $\theta'$ of random RR sets, and examine the fraction $f$ of RR sets that are covered by $S'_k$ (Lines 7-10). By Corollary~\ref{coro:tim-fraction}, $f \cdot n$ is an unbiased estimation of $\E[I(S'_k)]$. We set $\theta'$ to a reasonably large number to ensure that $f \cdot n < (1+\varepsilon')\cdot \E[I(S'_k)]$ occurs with at most $1 - n^{-\ell}$ probability. Based on this, Algorithm~\ref{alg:ext-refine} computes $KPT' = f \cdot n / (1 + \varepsilon')$, which scales $f\cdot n$ down by a factor of $1 + \varepsilon'$ to ensure that $KPT' \le \E[I(S'_k)] \le OPT$. The final output of Algorithm~\ref{alg:ext-refine} is $KPT^+ = \max\{KPT', KPT^*\}$, i.e., we choose the larger one between $KPT'$ and $KPT^*$ as the new lower-bound for $OPT$. The following lemma shows the theoretical guarantees of Algorithm~\ref{alg:ext-refine}:
\begin{lemma} \label{lmm:ext-refine}
Given that $\E[\frac{1}{KPT^*}] < \frac{12}{EPT}$, Algorithm~\ref{alg:ext-refine} runs in $O(\ell (m + n) \log n /(\varepsilon')^{2})$ expected time. In addition, it returns $KPT^+ \in [KPT^*, OPT]$ with at least $1 - n^{-\ell}$ probability, if $KPT^* \in [KPT/4, OPT]$.
\end{lemma}
Note that the time complexity of Algorithm~\ref{alg:ext-refine} is smaller than that of Algorithm~\ref{alg:tim-nodesel} by a factor of $k$, since the former only needs to accurately estimate the expected spread of one node set (i.e., $S'_k$), whereas the latter needs to ensure accurate estimations for ${n \choose k}$ node sets simultaneously.

We integrate Algorithm~\ref{alg:ext-refine} into {\em TIM} and obtain an improved solution (referred to as {\em TIM$^+$}) as follows. Given $G$, $k$, $\varepsilon$, and $\ell$, we first invoke Algorithm~\ref{alg:tim-kpt} to derive $KPT^*$. After that, we feed $G$, $k$, $KPT^*$, and a parameter $\varepsilon'$ to Algorithm~\ref{alg:ext-refine}, and obtain $KPT^+$ in return. Then, we compute $\theta = \lambda/KPT^+$. Finally, we run Algorithm~\ref{alg:tim-nodesel} with $G$, $k$, and $\theta$ as the input, and get the final result of influence maximization. It can be verified that when $\varepsilon' \ge \varepsilon/\sqrt{k}$, {\em TIM$^+$} has the same time complexity with {\em TIM}, and it returns a $(1-1/e-\varepsilon)$-approximate solution with at least $1 - 3n^{-\ell}$ probability. The success probability can be raised to $1 - n^{-\ell}$ by increasing $\ell$ by a factor of $1 + \log 3 / \log n$.

Finally, we discuss the choice of $\varepsilon'$. Ideally, we should set $\varepsilon'$ to a value that minimizes the total number $\beta$ of RR sets generated in Algorithms \ref{alg:tim-nodesel} and \ref{alg:ext-refine}. However, $\beta$ is difficult to estimate as it depends on unknown variables such as $KPT^*$ and $KPT^+$. In our implementation of {\em TIM$^+$}, we set
\begin{equation*}
\varepsilon' = 5 \cdot \sqrt[3]{\ell \cdot \varepsilon^2/(k + \ell)}
\end{equation*}
 for any $\varepsilon \le 1$. This is obtained by using a function of $\varepsilon'$ to roughly approximate $\beta$, and then taking the minimizer of the function.

\subsection{Generalization to the Triggering Model} \label{sec:ext-general}

The {\em triggering model} \cite{KempeKT03} is a influence propagation model that generalizes the IC model. It assumes that each node $v$ is associated with a {\em triggering distribution} $\T(v)$ over the power set of $v$'s incoming neighbors, i.e., each sample from $\T(v)$ is a subset of the nodes that has an outgoing edge to $v$.

Given a seed set $S$, an influence propagation process under the triggering model works as follows. First, for each node $v$, we take a sample from $\T(v)$, and define the sample as the {\em triggering set} of $v$. After that, at timestamp 1, we activate the nodes in $S$. Then, at subsequent timestamp $i$, if an activated node appears in the triggering set of an inactive node $v$, then $v$ becomes activated at timestamp $i+1$. The propagation process terminates when no more nodes can be activated.

The influence maximization problem under the triggering model asks for a size-$k$ seed set $S$ that can activate the largest number of nodes in expectation. To understand why the triggering model captures the IC model as a special case, consider that we assign a triggering distribution to each node $v$, such that each of $v$'s incoming neighbors independently appears in $v$'s trigger set with $p(e)$ probability, where $e$ is the edge that goes from the neighbor to $v$. It can be verified that influence maximization under this distribution is equivalent to that under the IC model.

Interestingly, our solutions can be easily extended to support the triggering model. To explain, observe that Algorithms \ref{alg:tim-nodesel}, \ref{alg:tim-kpt}, and \ref{alg:ext-refine} do not rely on anything specific to the IC model, except that they require a subroutine to generate random RR sets, whereas RR sets are defined under the IC model only. To address this issue, we revise the definition of RR sets to accommodate the triggering model, as explained in the following.  %For that purpose, we refine several related concepts as follows.

Suppose that we generate random graphs $g$ from $G$, by first sampling a node set $T$ for each node $v$ from its triggering distribution $\T(v)$, and then removing any outgoing edge of $v$ that does not point to a node in $T$. Let $\G$ be the distribution of $g$ induced by the random choices of triggering sets. We refer to $\G$ as the {\em triggering graph distribution} for $G$. For any given node $u$ and a graph $g$ sampled from $G$, we define the reverse reachable (RR) set for $u$ in $g$ as the set of nodes that can reach $u$ in $g$. In addition, we define a {\em random RR set} as one that is generated on an instance of $g$ randomly sampled from $\G$, for a node selected from $g$ uniformly at random.

To construct random RR sets defined above, we employ a randomized BFS algorithm as follows. Let $v$ be a randomly selected node. Given $v$, we first take a sample $T$ from $v$'s triggering distribution $\T(v)$, and then put all nodes in $T$ into a queue. After that, we iteratively extract the node at the top of the queue; for each node $u$ extracted, we sample a set $T'$ from $u$'s triggering distribution, and we insert any unvisited node in $T'$ into the queue. When the queue becomes empty, we terminate the process, and form a random RR set with the nodes visited during the process. The expected cost of the whole process is $O(EPT)$, where $EPT$ denotes the expected number of edges in $G$ that point to the nodes in a random RR set. This expected time complexity is the same as that of the algorithm for generating random RR sets under the IC model.

By incorporating the above BFS approach into Algorithms \ref{alg:tim-nodesel}, \ref{alg:tim-kpt}, and \ref{alg:ext-refine}, our solutions can readily support the triggering model. Our next step is to show that the revised solution retains the performance guarantees of {\em TIM} and {\em TIM$^+$}. For this purpose, we first present an extended version of Lemma~\ref{lmm:tim-nodesel-rr} for the triggering model. (The proof of the lemma is almost identical to that of Lemma~\ref{lmm:tim-nodesel-rr}.)
\begin{lemma}\label{lmm:ext-rr}
Let $S$ be a fixed set of nodes, $v$ be a fixed node, and $\G$ be the triggering graph distribution for $G$. Suppose that we generate an RR set $R$ for $v$ on a graph $g$ sampled from $\G$. Let $\rho_1$ be the probability that $S$ overlaps with $R$, and $\rho_2$ be the probability that $S$ (as a seed set) can activate $v$ in an influence propagation process on $G$ under the triggering model. Then, $\rho_1 = \rho_2$.
\end{lemma}

Next, we note that all of our theoretical analysis of {\em TIM} and {\em TIM$^+$} is based on the Chernoff bounds and Lemma~\ref{lmm:tim-nodesel-rr}, without relying on any other results specific to the IC model. Therefore, once we establish Lemma~\ref{lmm:ext-rr}, it is straightforward to combine it with the Chernoff bounds to show that, under the triggering model, both {\em TIM} and {\em TIM$^+$} provide the same performance guarantees as in the case of the IC model. Thus, we have the following theorem:
\begin{theorem}\label{thrm:ext-performance}
Under the triggering model, { TIM} (resp. { TIM$^+$}) runs in $O((k+\ell)(m+n)\log n /\varepsilon^2)$ expected time, and returns a ${(1 - 1/e - \varepsilon)}$-approximate solution with at least $1 - 2 \cdot n^{-\ell}$ probability (resp.\ $1 - 3 \cdot n^{-\ell}$ probability).
\end{theorem}

\section{Theoretical Comparisons} \label{sec:compare}

%This section compares the asymptotic performance of our solutions with that of {\em RIS} and {\em Greedy}.

\vspace{-1mm}
\header
%\noindent
{\bf Comparison with {\bf \em RIS}.} Borgs et al.\ \cite{Borgs14} show that, under the IC model, {\em RIS} can derive a $(1 - 1/e-\varepsilon)$-approximate solution for the influence maximization problem, with ${O(k \ell^2 (m + n) \log^2 n /\varepsilon^3)}$ running time and at least $1-n^{-\ell}$ success probability. The time complexity of {\em RIS} is larger than the expected time complexity of {\em TIM} and {\em TIM$^+$} by a factor of $\ell \log n / \varepsilon$. Therefore, both {\em TIM} and {\em TIM$^+$} are superior to {\em RIS} in terms of asymptotic performance.

%comparable to that of our solutions. However, there exists a gap in Borgs et al.'s analysis that invalidates the above asymptotic bounds, as explained in the following.

%\subsection{Comparison with {\large \bf \em Greedy}} \label{sec:compare-greedy}

\header
{\bf Comparison with {\bf \em Greedy}.} As mentioned in Section~\ref{sec:prelim-greedy}, {\em Greedy} runs in $O(kmnr)$ time, where $r$ is the number of Monte Carlo samples used to estimate the expected spread of each node set. Kempe et al.\ do not provide a formal result on how $r$ should be set to achieve a $(1 - 1/e - \varepsilon)$-approximation ratio; instead, they only point out that when each estimation of expected spread has $\varepsilon$ related error, {\em Greedy} returns a $(1 - 1/e - \varepsilon')$-approximate solution for a certain $\varepsilon'$ \cite{KempeKT05}.

We present a more detailed characterization on the relationship between $r$ and {\em Greedy}'s approximation ratio:
\begin{lemma} \label{lmm:compare-greedy}
{\em Greedy} returns a $(1 - 1/e - \varepsilon)$-approximate solution with at least $1 - n^{-\ell}$ probability, if
\begin{eqnarray} \label{eqn:compare-r}
r &\ge& (8k^2 + 2k\varepsilon) \cdot n \cdot \frac{(\ell+1)\log n + \log k}{\varepsilon^2 \cdot OPT}.
\end{eqnarray}
\end{lemma}
Assume that we know $OPT$ in advance and set $r$ to the smallest value satisfying the above inequality, in {\em Greedy}'s favor. In that case, the time complexity of {\em Greedy} is $O(k^3 \ell m n^2 \varepsilon^{-2} \log n/ OPT)$. Given that $OPT \le n$, this complexity is much worse than the expected time complexity of {\em TIM} and {\em TIM$^+$}.

\section{Additional Related Work} \label{sec:related}

There has been a large body of literature on influence maximization over the past decade (see \cite{KimKY13,WangCW12,ChenYZ10,GoyalBL11,JungHC12,ChenWY09,ChenWW10,Wang2010community,leskovec2007cost,ChenLZ12,BharathiKS07,SeemanS13,KempeKT03,KempeKT05,Borgs14} and the references therein). Besides {\em Greedy} \cite{KempeKT03} and {\em RIS} \cite{Borgs14}, the work most related to ours is by Leskovec et al.\ \cite{leskovec2007cost}, Chen et al.\ \cite{ChenWY09}, and Goyal et al.\ \cite{GoyalLL11Celf}. In particular, Leskovec et al.\ \cite{leskovec2007cost} propose an algorithmic optimization of {\em Greedy} that avoids evaluating the expected spreads of a large number of node sets. This optimization reduces the computation cost of {\em Greedy} by up to $700$-fold, without affecting its approximation guarantees. Subsequently, Chen et al.\ \cite{ChenWY09} and Goyal et al.\ \cite{GoyalLL11Celf} further enhance Leskovec et al.'s approach, and achieve up to $50\%$ additional improvements in terms of efficiency.

Meanwhile, there also exist a plethora of algorithms \cite{KimKY13,WangCW12,ChenYZ10,GoyalBL11,JungHC12,ChenWY09,ChenWW10,Wang2010community} that rely on heuristics to efficiently derive solutions for influence maximization. For example, Chen et al.\ \cite{ChenWW10} propose to reduce computation costs by omitting the social network paths with low propagation probabilities; Wang et al.\ \cite{Wang2010community} propose to divide the social network into smaller communities, and then identify influential nodes from each community individually; Goyal et al.\ \cite{GoyalLL11} propose to estimate the expected spread of each node set $S$ only based on the nodes that are close to $S$. In general, existing heuristic solutions are shown to be much more efficient than {\em Greedy} (and its aforementioned variants \cite{leskovec2007cost,GoyalLL11Celf,ChenWY09}), but they fail to retain the $(1 - 1/e - \varepsilon)$-approximation ratio. As a consequence, they tend to produce less accurate results, as shown in the experiments in \cite{KimKY13,WangCW12,ChenYZ10,GoyalBL11,JungHC12,ChenWY09,ChenWW10,Wang2010community}.

Considerable research has also been done to extend Kempe et al.'s formulation of influence maximization \cite{KempeKT03} to various new settings, e.g., when the influence propagation process follows a different model \cite{Li0WZ13,GoyalBL11}, when there are multiple parties that compete with each other for social influence \cite{BharathiKS07,LuB0L13}, or when the influence propagation process terminates at a predefined timestamp \cite{ChenLZ12}. The solutions derived for those scenarios are inapplicable under our setting, due to the differences in problem formulations. Finally, there is recent research on learning the parameters of influence propagation model (e.g., the propagation probability on each edge) from observed data \cite{GoyalBL10,KutzkovBBG13,SaitoMIGM08}. This line of research complements (and is orthogonal to) the existing studies on influence maximization.

\begin{table}[t]
\centering
\begin{small}
\renewcommand{\arraystretch}{1.3}
%\hspace{-5mm}
\begin{tabular}{|c|r|r|c|c|}
    \hline
    {\bf Name} & \multicolumn{1}{c|}{$\boldsymbol{n}$} & \multicolumn{1}{c|}{$\boldsymbol{m}$} & {\bf Type} & {\bf Average degree}\\
    \hline
    {\em NetHEPT}   &15K    &31K          & undirected & $\,\,\,$4.1 \\
    \hline
    {\em Epinions}  &76K     &509K        &directed & 13.4 \\
    \hline
    {\em DBLP}      &655K    &2M      &undirected & $\,\,\,$6.1 \\
    \hline
    {\em LiveJournal} &4.8M &69M    & directed & 28.5 \\
    \hline
    {\em Twitter}   & 41.6M &1.5G & directed & 70.5 \\
    \hline
\end{tabular}
\caption{Dataset characteristics.} \label{tbl:exp-data}
\end{small}
\tbldown \vspace{-1mm}
\end{table}

\section{Experiments} \label{sec:exp}

This section experimentally evaluates {\em TIM} and {\em TIM$^+$}. Our experiments are conducted on a machine with an Intel Xeon $2.4$GHz CPU and $48$GB memory, running 64bit Ubuntu 13.10. All algorithms tested are implemented in C++ and compiled with g++ 4.8.1.

\subsection{Experimental Settings} \label{sec:exp-setting}

\vspace{-1mm}
\header
{\bf Datasets.} Table~\ref{tbl:exp-data} shows the datasets used in our experiments. Among them, {\em NetHEPT}, {\em Epinions}, {\em DBLP}, and {\em LiveJournal} are benchmarks in the literature of influence maximization \cite{KimKY13}. Meanwhile, {\em Twitter} contains a social network crawled from Twitter.com in July 2009, and it is publicly available from \cite{TWITTER}. Note that {\em Twitter} is significantly larger than the other four datasets.

\header
{\bf Propagation Models.} We consider two influence propagation models, namely, the IC model (see Section~\ref{sec:prelim-def}) and the {\em linear threshold (LT)} model \cite{KempeKT03}. Specifically, the LT model is a special case of the triggering model, such that for each node $v$, any sample from $v$'s triggering distribution $\T(v)$ is either $\emptyset$ or a singleton containing an incoming neighbor of $v$. Following previous work \cite{ChenYZ10}, we construct $\T(v)$ for each node $v$, by first assigning a random probability in $[0, 1]$ to each of $v$'s incoming neighbors, and then normalizing the probabilities so that they sum up to $1$. As for the IC model, we set the propagation probability of each edge $e$ as follows: we first identify the node $v$ that $e$ points to, and then set $p(e) = 1/i$, where $i$ denotes the in-degree of $v$. This setting of $p(e)$ is widely adopted in prior work \cite{WangCW12,ChenWW10,GoyalBL11,JungHC12}.

\begin{figure}[t]
\begin{small}
\centering
\begin{tabular}{cc}
\multicolumn{2}{c}{\hspace{-4mm} \includegraphics[height=2.9mm]{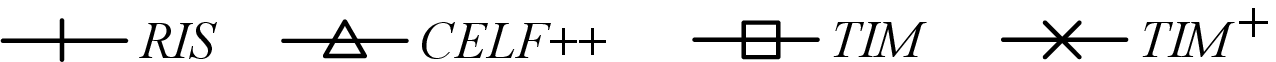}} \vspace{0mm} \\
\hspace{-6.2mm}\includegraphics[height=36mm]{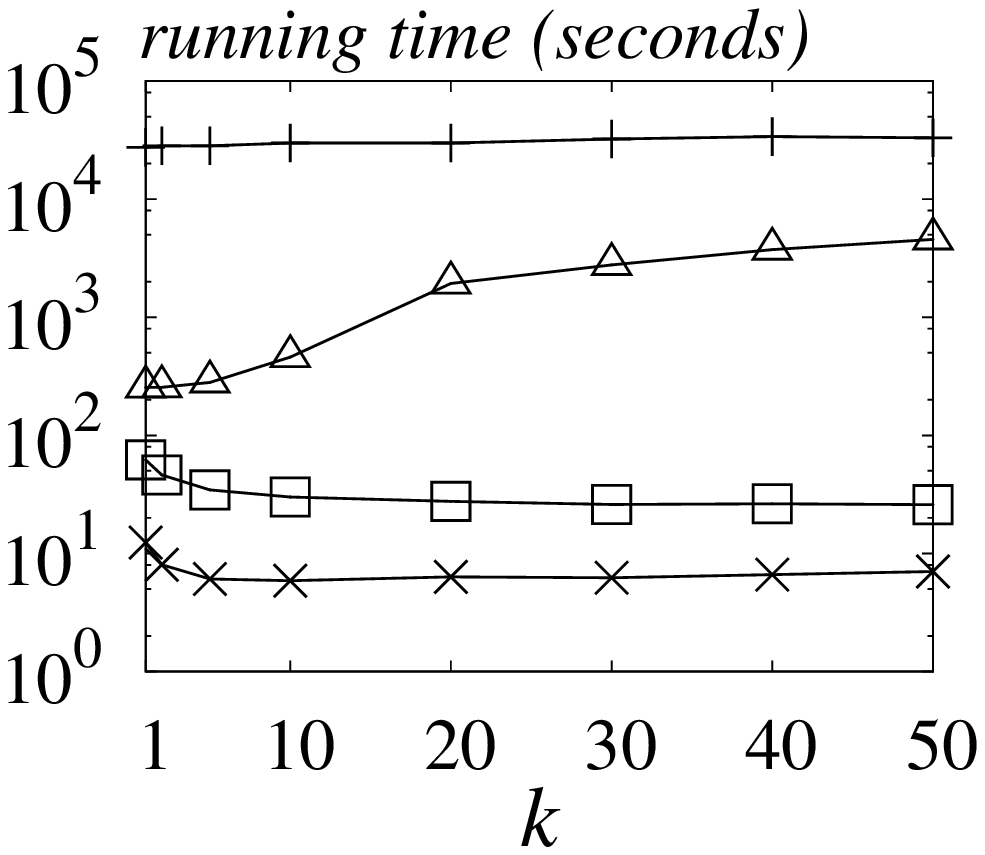}
&
\hspace{-12mm}\includegraphics[height=36mm]{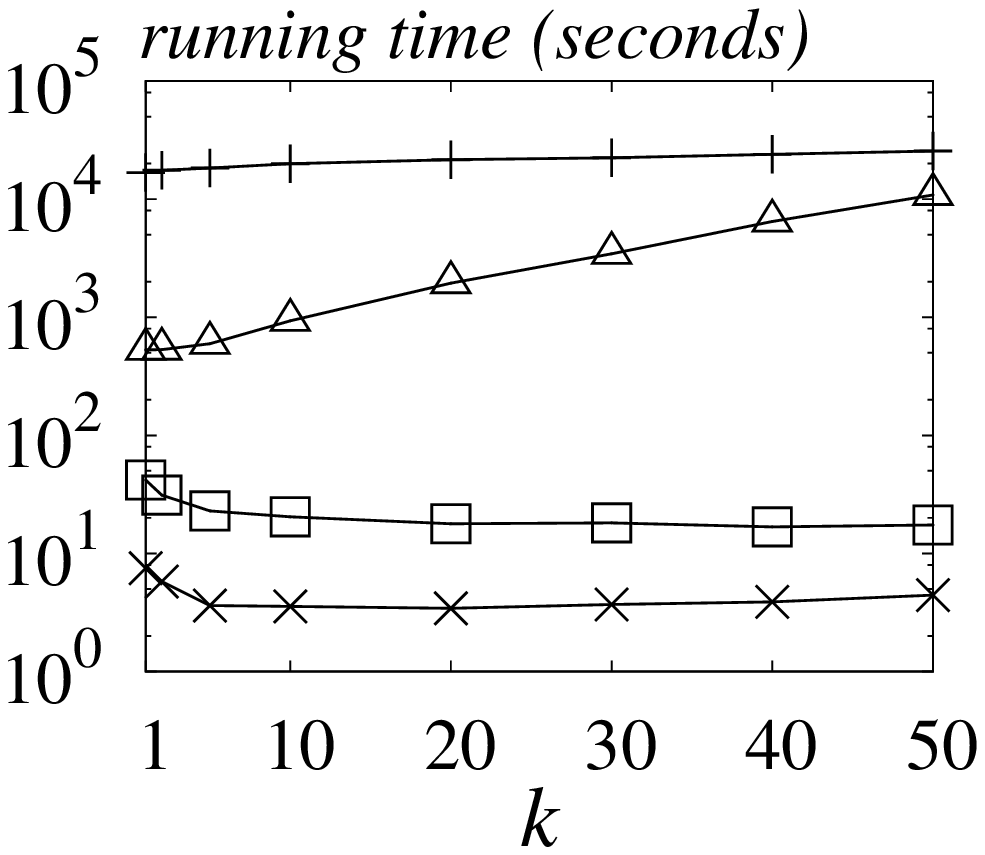}
\\
\hspace{-6.2mm}(a) The IC model. & \hspace{-12mm}(b) The LT model.
\end{tabular}
\figcapup  \vspace{-1mm} \caption{Computation time vs.\ $k$ on {\em NetHEPT}.} %\figcapdown %\vspace{-2mm}  % %
\label{fig:exp-nethept-time}
\end{small}
\end{figure}

\begin{figure}[t]
\begin{small}
\centering
\begin{tabular}{cc}
\multicolumn{2}{c}{\hspace{-4mm} \includegraphics[height=2.7mm]{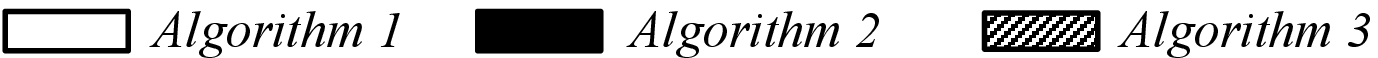}} \vspace{0mm} \\
\hspace{-7mm}\includegraphics[height=35mm]{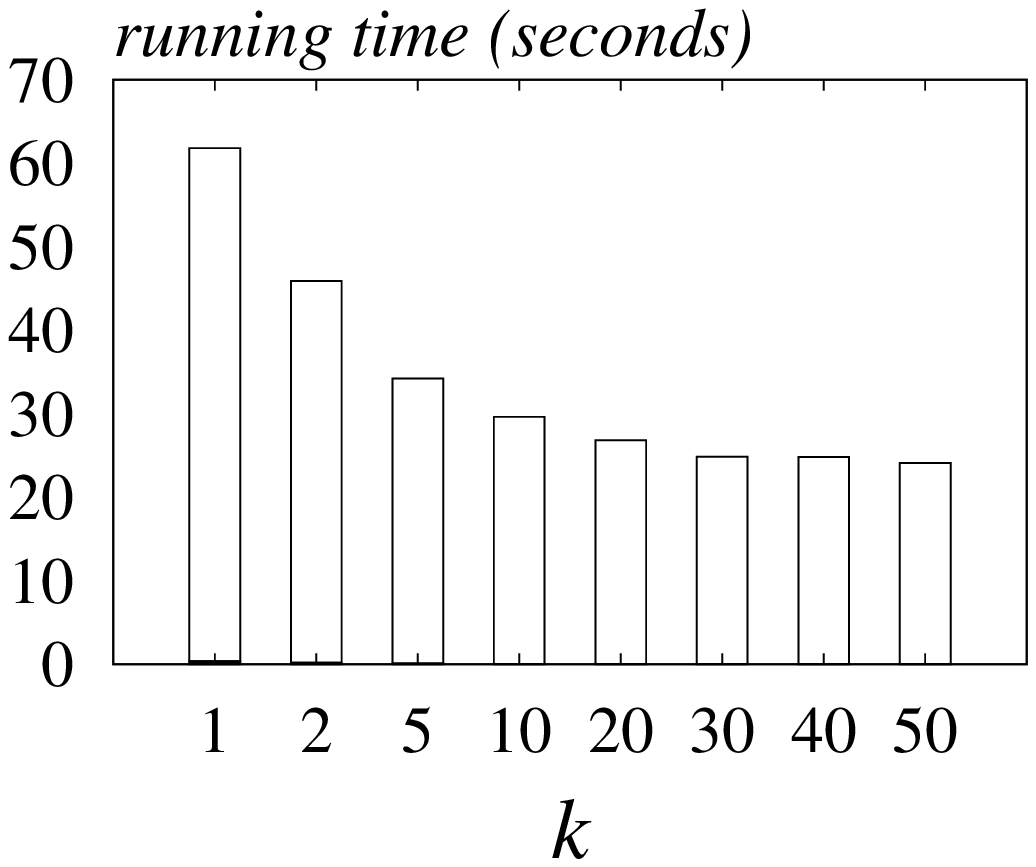}
&
\hspace{-10mm}\includegraphics[height=35mm]{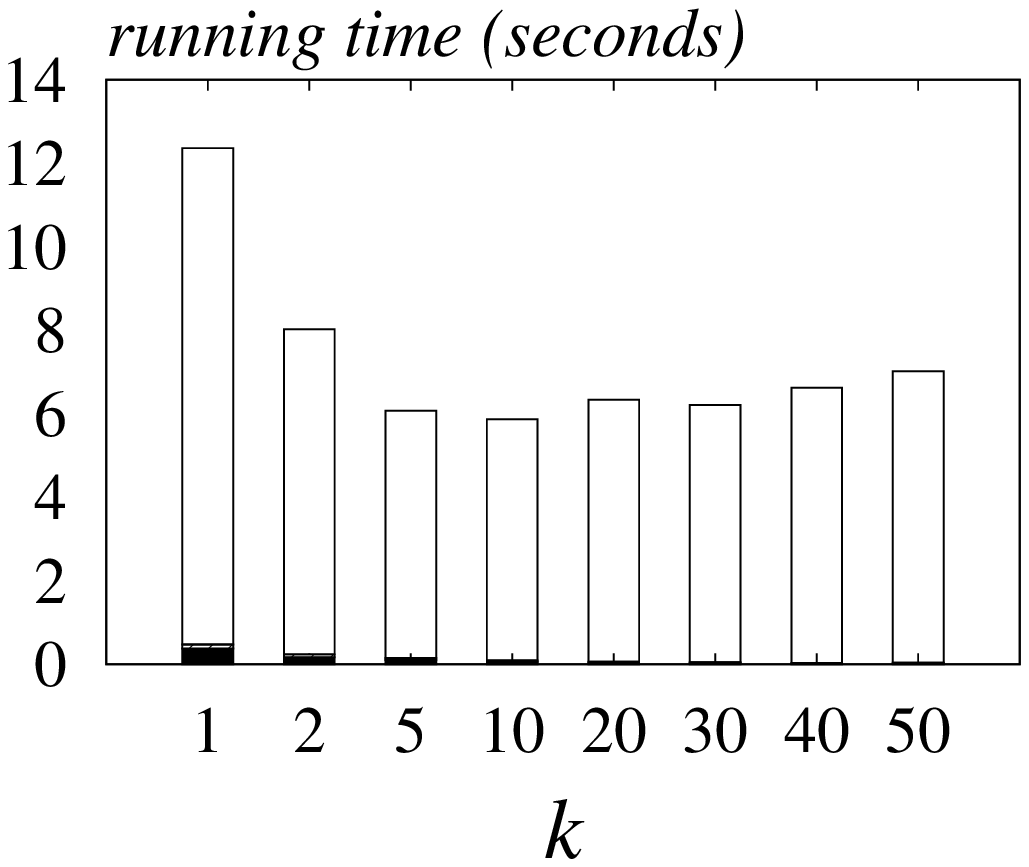}
\\
\hspace{-5mm}(a) {\em TIM} (the IC model). & \hspace{-10mm}(b) {\em TIM$^+$} (the IC model).
\end{tabular}
\figcapup \vspace{-1mm}  \caption{Breakdown of computation time on {\em NetHEPT}.} \figcapdown \vspace{1mm}
\label{fig:exp-nethept-bar}
\end{small}
\end{figure}

\begin{figure}[t]
\begin{small}
\centering
\begin{tabular}{cc}
\multicolumn{2}{c}{\hspace{-8mm} \includegraphics[height=3mm]{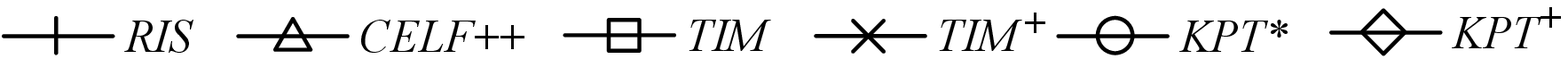}} \vspace{0mm} \\
\hspace{-8.5mm}\includegraphics[height=36mm]{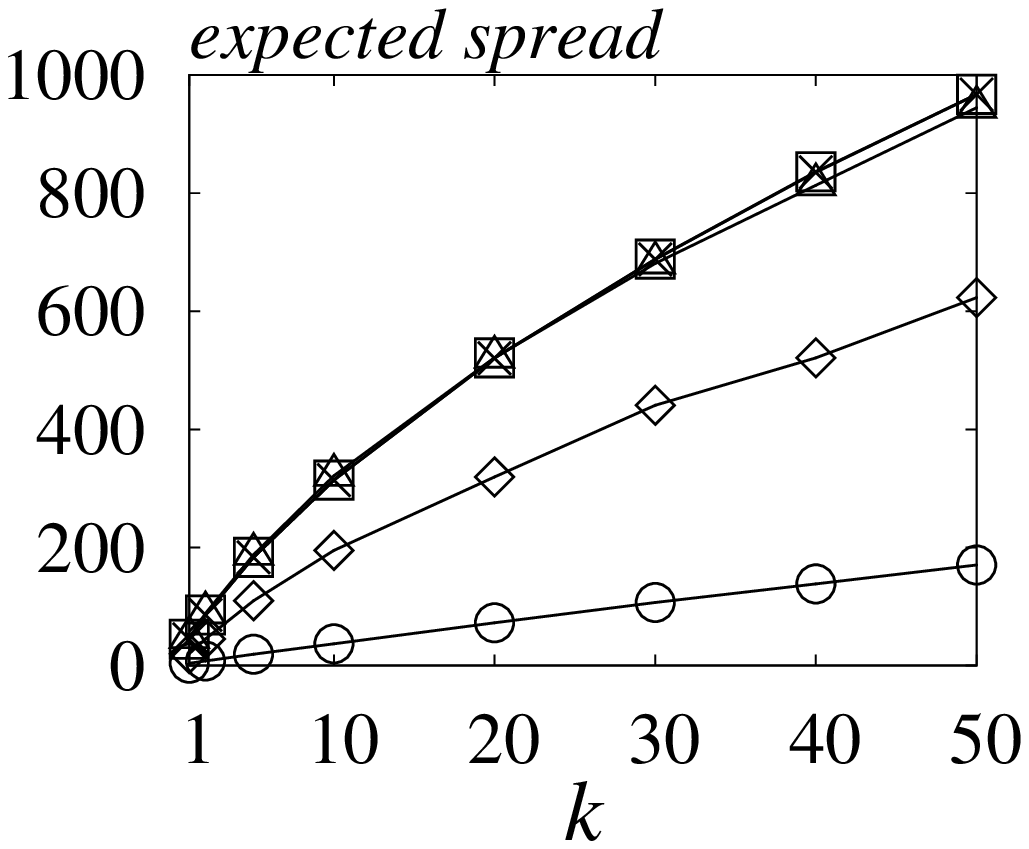}
&
\hspace{-12mm}\includegraphics[height=36mm]{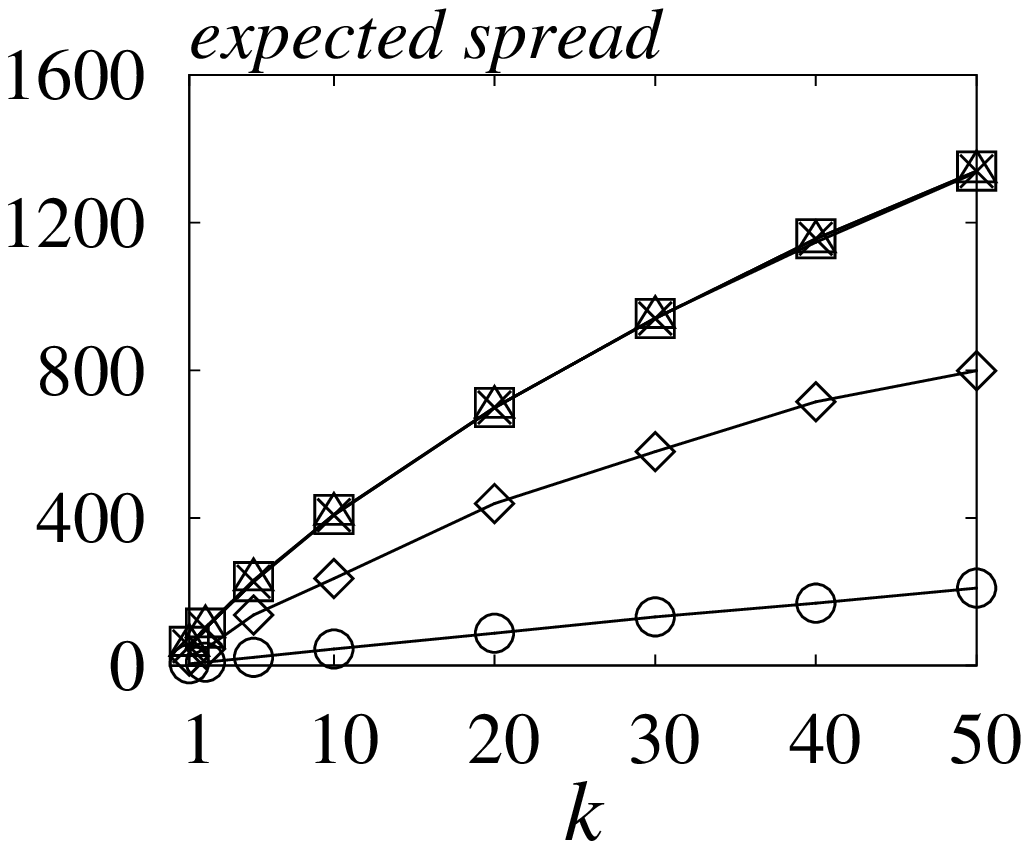}
\\
\hspace{-2.5mm}(a) The IC model. & \hspace{-4mm}(b) The LT model.
\end{tabular}
\figcapup \vspace{-1mm}  \caption{Expected spreads, $KPT^*$, and $KPT^+$ on {\em NetHEPT}.} \figcapdown \vspace{1mm}
\label{fig:exp-nethept-spread}
\end{small}
\end{figure}

\header
{\bf Algorithms.} We compare our solutions with four methods, namely, {\em RIS} \cite{Borgs14}, {\em CELF++} \cite{GoyalLL11Celf}, {\em IRIE} \cite{JungHC12}, and {\em SIMPATH} \cite{GoyalLL11}. In particular, {\em CELF++} is a state-of-the-art variant of {\em Greedy} that considerably improves the efficiency of {\em Greedy} without affecting its theoretical guarantees, while {\em IRIE} and {\em SIMPATH} are the most advanced heuristic methods under the IC and LT models, respectively. We adopt the C++ implementations of {\em CELF++}, {\em IRIE}, and {\em SIMPATH} made available by their inventors, and we implement {\em RIS} and our solutions in C++. Note that {\em RIS} is designed under the IC model only, but we incorporate the techniques in Section~\ref{sec:ext-general} into {\em RIS} and extend it to the LT model.

\header
{\bf Parameters.} Unless otherwise specified, we set $\varepsilon = 0.1$ and $k = 50$ in our experiments. For {\em RIS} and our solutions, we set $\ell$ in a way that ensures a success probability of $1 - 1/n$. For {\em CELF++}, we set the number of Monte Carlo steps to $r = 10000$, following the standard practice in the literature. Note that this choice of $r$ is to the advantage of {\em CELF++} because, by Lemma~\ref{lmm:compare-greedy}, the value of $r$ required in our experiments is always larger than $10000$. In each of our experiments, we repeat each method three times and report the average result.

\begin{figure*}[t]
\begin{small}
\centering
\begin{tabular}{cccc}
\multicolumn{4}{c}{\hspace{-4mm} \includegraphics[height=3.5mm]{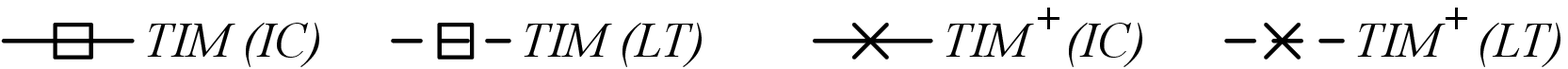}} \vspace{0mm} \\
\hspace{-4mm}\includegraphics[height=36mm]{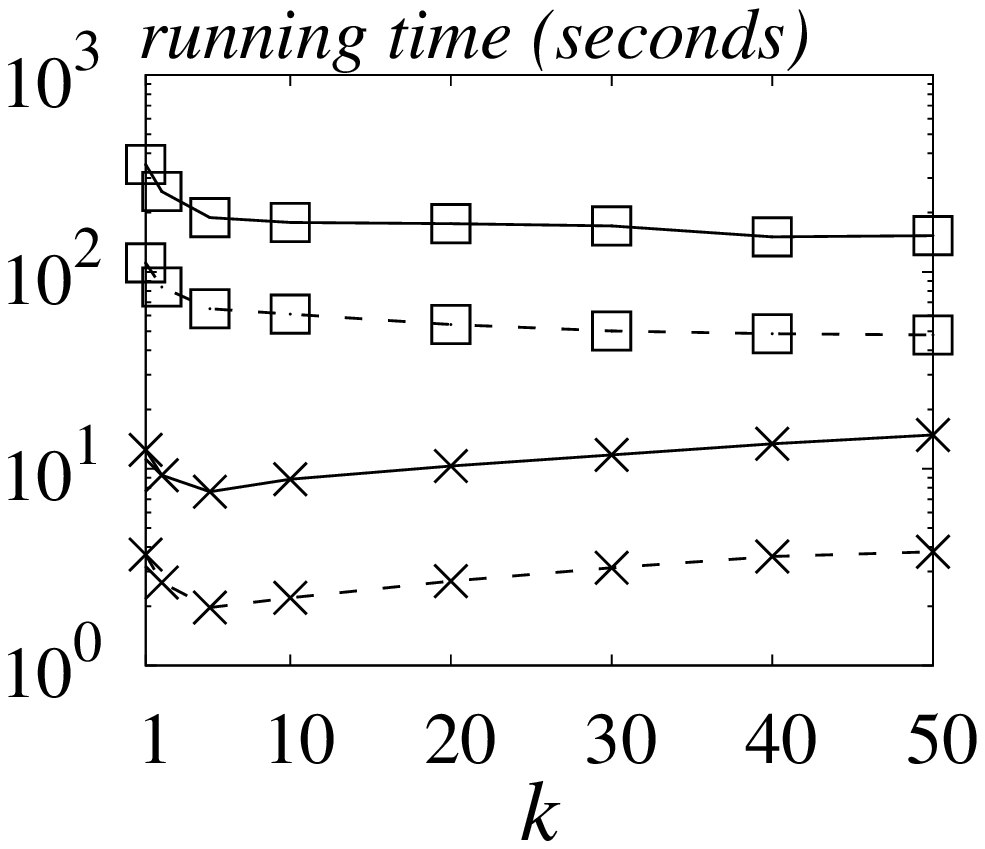}
&
\hspace{-11mm}\includegraphics[height=36mm]{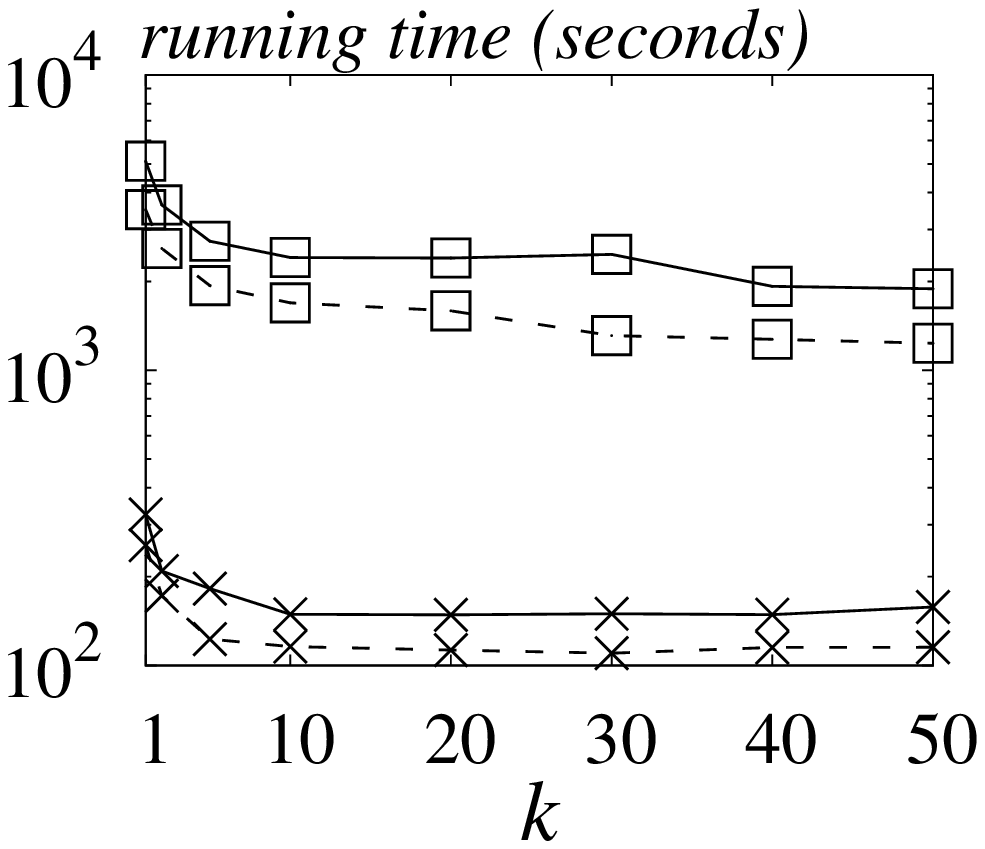}
&
\hspace{-11mm}\includegraphics[height=36mm]{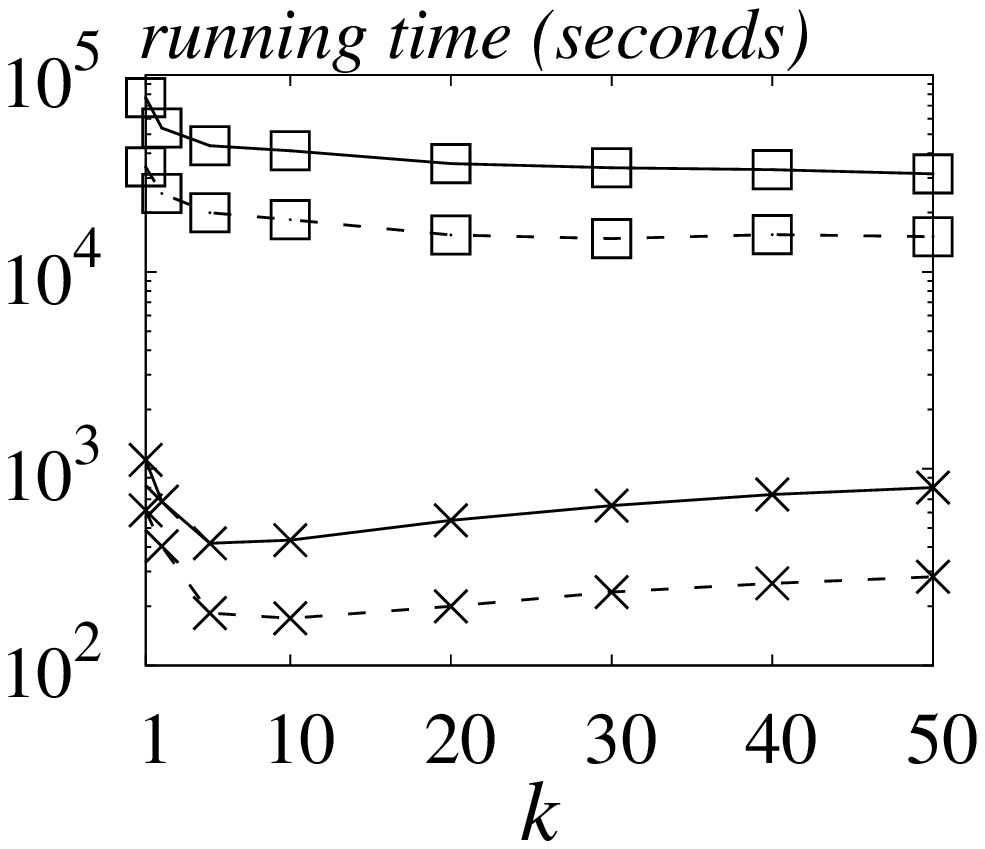}
&
\hspace{-11mm}\includegraphics[height=36mm]{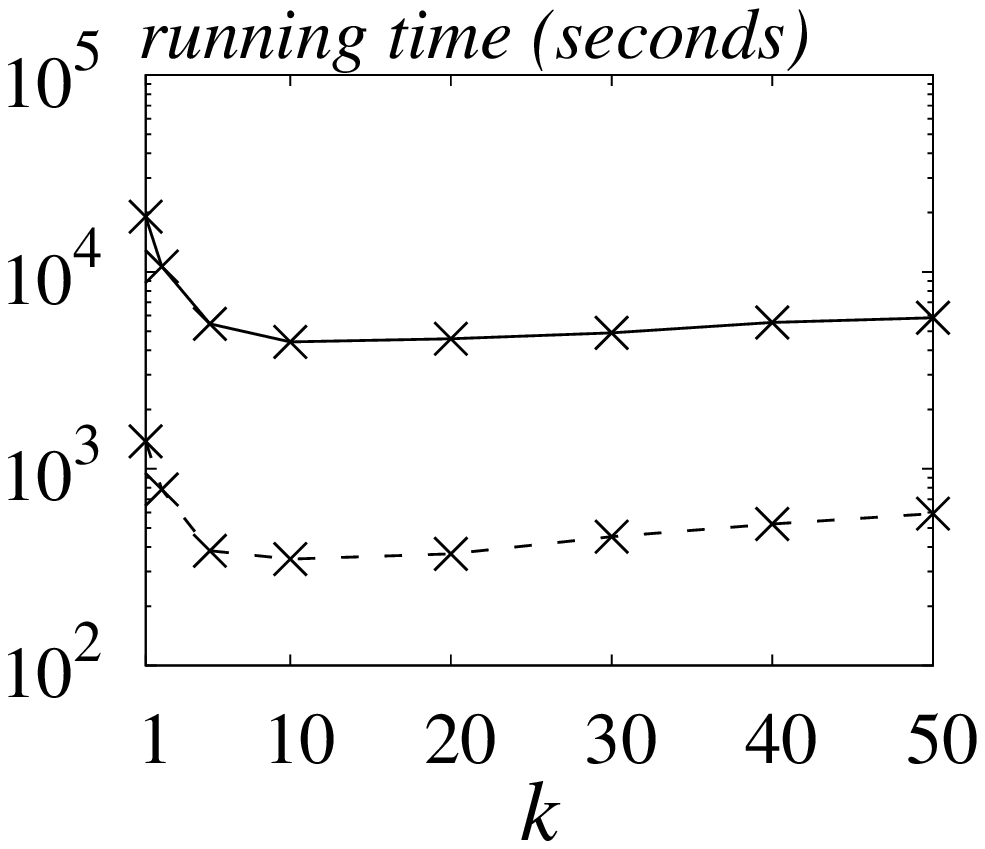}
\\
\hspace{-4mm}(a) {\em Epinions} & \hspace{-11mm}(b) {\em DBLP} &
\hspace{-11mm}(c) {\em LiveJournal} & \hspace{-11mm} (d)  {\em Twitter}
\end{tabular}
\figcapup  \vspace{-1mm} \caption{Running time vs.\ $k$ on large datasets.} %\vspace{-1mm} %\figcapdown
\label{fig:exp-large-time}
\end{small}
\end{figure*}

\begin{figure*}[t]
\begin{small}
\centering
\begin{tabular}{cccc}
\multicolumn{4}{c}{\hspace{-4mm} \includegraphics[height=3.5mm]{./figures/figure67.eps}} \vspace{0mm} \\
\hspace{-4mm}\includegraphics[height=36mm]{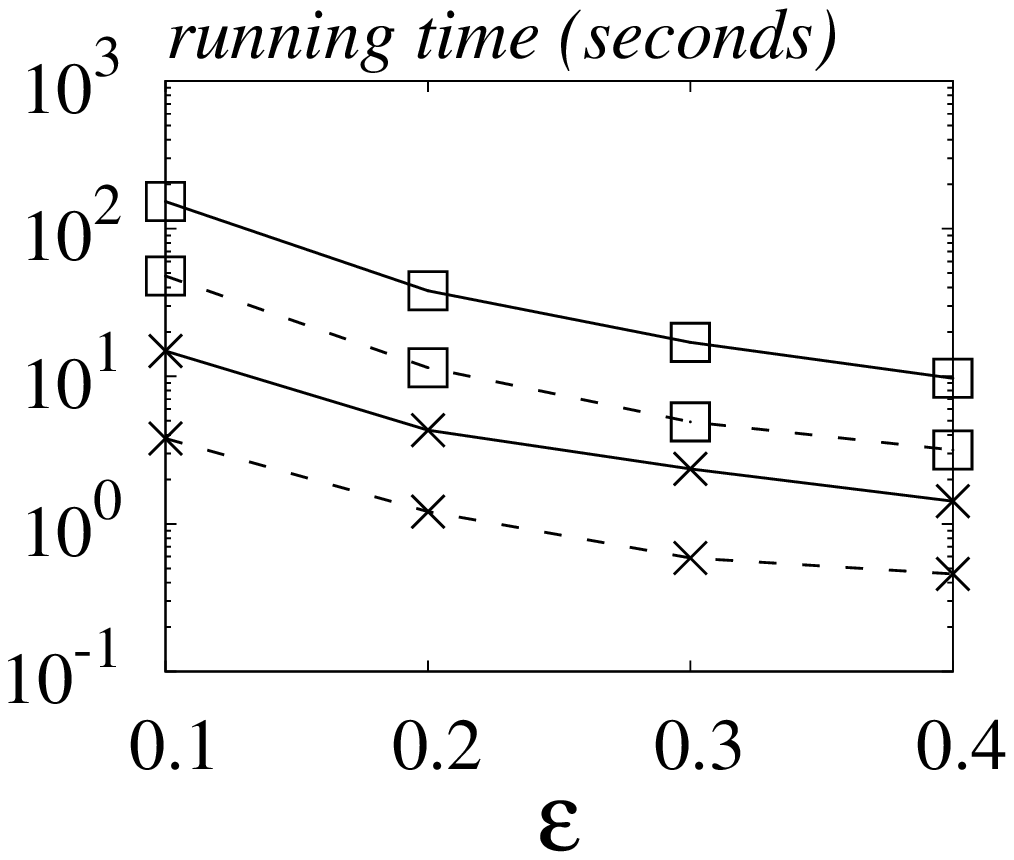}
&
\hspace{-11mm}\includegraphics[height=36mm]{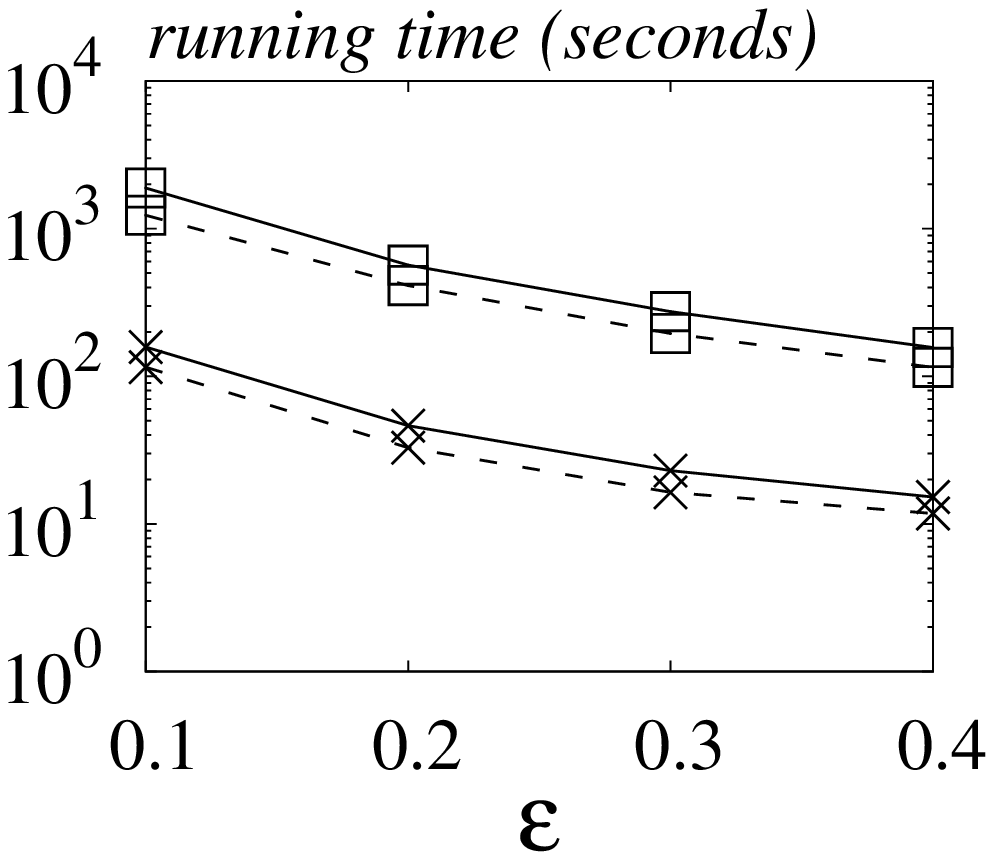}
&
\hspace{-11mm}\includegraphics[height=36mm]{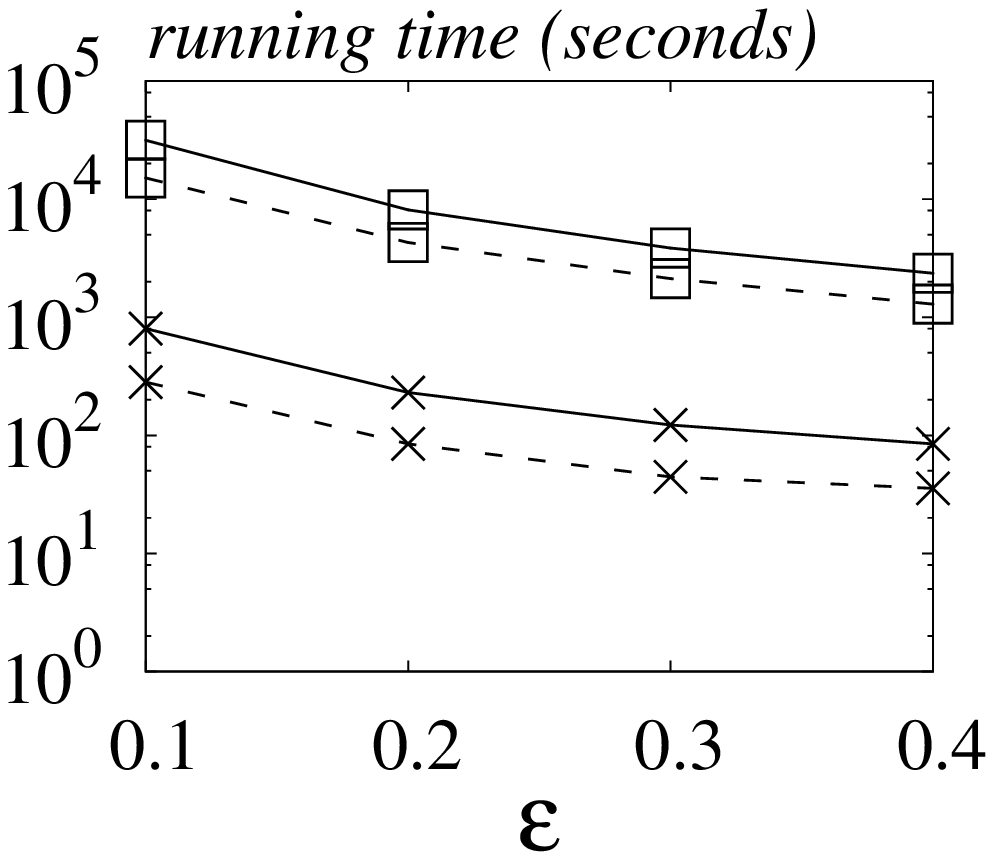}
&
\hspace{-11mm}\includegraphics[height=36mm]{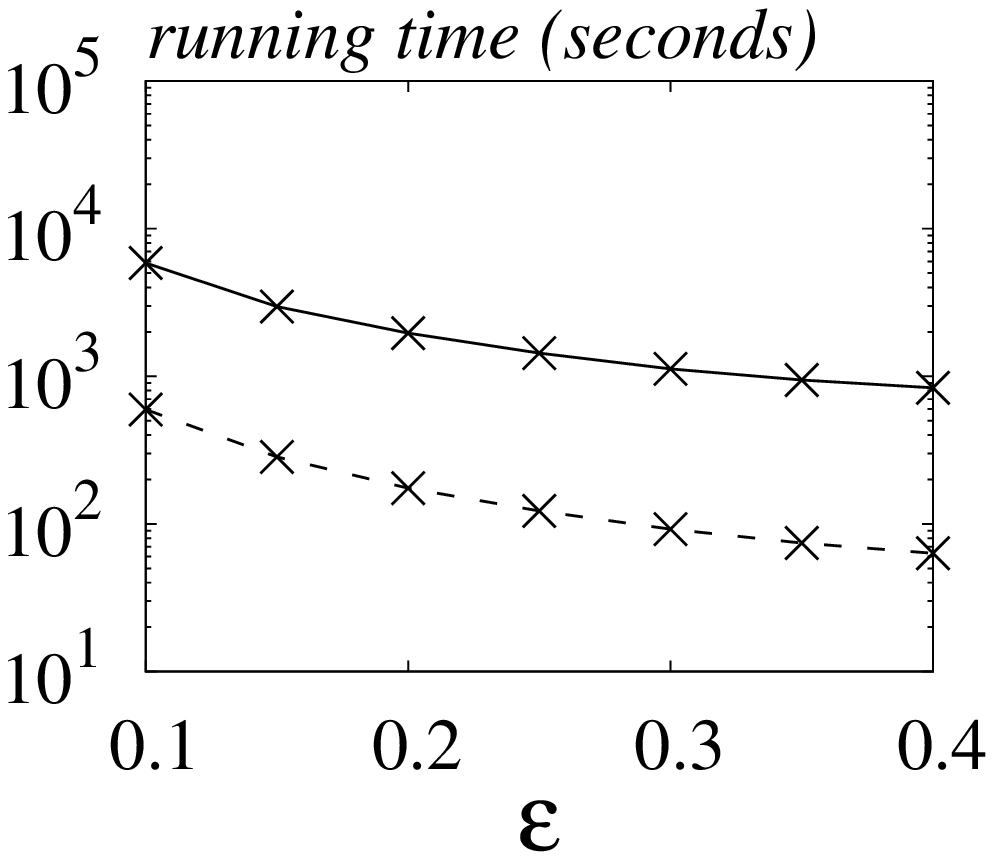}
\\
\hspace{-4mm}(a) {\em Epinions} & \hspace{-11mm}(b) {\em DBLP} &
\hspace{-11mm}(c) {\em LiveJournal} & \hspace{-11mm}  (d) {\em Twitter}
\end{tabular}
\figcapup \vspace{-1mm} \caption{Running time vs.\ $\varepsilon$ on large datasets.} \figcapdown %\vspace{-1mm} %
\label{fig:exp-large-eps}
\end{small}
\end{figure*}

\subsection{Comparison with {\large \bf \em  CELF++} and {\large \bf \em RIS}} \label{sec:exp-result}

Our first set of experiments compares our solutions with {\em CELF++} and {\em RIS}, i.e., the state of the arts among the solutions that provide non-trivial approximation guarantees.

\header
{\bf Results on {\bf \em NetHEPT}. } Figure~\ref{fig:exp-nethept-time} shows the computation cost of each method on the {\em NetHEPT} dataset, varying $k$ from $1$ to $50$. Observe that {\em TIM$^+$} consistently outperforms {\em TIM}, while {\em TIM} is up to two orders of magnitude faster than {\em CELF++} and {\em RIS}. In particular, when $k = 50$, {\em CELF++} requires more than an hour to return a solution, whereas {\em TIM$^+$} terminates within ten seconds. These results are consistent with our theoretical analysis (in Section~\ref{sec:compare}) that {\em Greedy}'s time complexity is much higher than those of {\em TIM} and {\em TIM$^+$}. On the other hand, {\em RIS} is the slowest method in all cases despite of its near-linear time complexity, because of the $\varepsilon^{-3}$ term and the large hidden constant factor in its performance bound. One may improve the empirical efficiency of {\em RIS} by reducing the threshold $\tau$ on its running time (see Section~\ref{sec:prelim-ris}), but in that case, the worst-case quality guarantee of {\em RIS} is not necessarily retained.

%On the other hand, {\em RIS} is the slowest method in all cases despite of its near-linear time complexity, because of the $\varepsilon^{-3}$ term and the large hidden constant factor in its performance bound. Its running time is insensitive to $k$, since its threshold $\tau$ on computation cost is independent of $k$ (see Equation~\ref{eqn:compare-tau-2}). In contrast, the computation time of {\em CELF++} increases with $k$, because a larger $k$ requires {\em CELF++} to evaluate the expected spread of an increased number of node sets.

The computation overheads of {\em RIS} and {\em CELF++} increase with $k$, because (i) {\em RIS}'s threshold $\tau$ on running time is linear to $k$, while (ii) a larger $k$ requires {\em CELF++} to evaluate the expected spread of an increased number of node sets. Surprisingly, when $k$ increases, the running time of {\em TIM} and {\em TIM$^+$} tends to decrease. To understand this, we show, in Figure~\ref{fig:exp-nethept-bar}, a breakdown of {\em TIM} and {\em TIM$^+$}'s computation overheads under the IC model. Evidently, both algorithms' overheads are mainly incurred by Algorithm~\ref{alg:tim-nodesel}, i.e., the node selection phase. Meanwhile, the computation cost of Algorithm~\ref{alg:tim-nodesel} is mostly decided by the number $\theta$ of RR sets that it needs to generate. For {\em TIM}, we have $\theta = \lambda/KPT^*$, where $\lambda$ is as defined in Equation~\ref{eqn:tim-lambda}, and $KPT^*$ is a lower-bound of $OPT$ produced by Algorithm~\ref{alg:tim-kpt}. Both $\lambda$ and $KPT^*$ increase with $k$, and it happens that, on {\em NetHEPT}, the increase of $KPT^*$ is more pronounced than that of $\lambda$, which leads to the decrease in {\em TIM}'s running time. Similar observations can be made on {\em TIM$^+$} and on the case of the LT model.

From Figure~\ref{fig:exp-nethept-bar}, we can also observe that the computation cost of Algorithm~\ref{alg:ext-refine} (i.e, the intermediate step) is negligible compared with the total cost of {\em TIM$^+$}. Yet, Algorithm~\ref{alg:ext-refine} is so effective that it reduces {\em TIM$^+$}'s running time to at most $1/3$ of {\em TIM}'s. This indicates that Algorithm~\ref{alg:ext-refine} returns a much tighter lower-bound of $OPT$ than Algorithm~\ref{alg:tim-kpt} (i.e., the parameter estimation phase) does. To support this argument, Figure~\ref{fig:exp-nethept-spread} illustrates the lower-bounds $KPT^*$ and $KPT^+$ produced by Algorithms \ref{alg:tim-kpt} and \ref{alg:ext-refine}, respectively. Observe that $KPT^+$ is at least three times $KPT^*$ in all cases, which is consistent with {\em TIM$^+$}'s $3$-fold efficiency improvement over {\em TIM}.

In addition, Figure~\ref{fig:exp-nethept-spread} also shows the expected spreads of the node sets selected by each method on {\em NetHEPT}. (We estimate the expected spread of a node set by taking the average of $10^5$ Monte Carlo measurements.) There is no significant difference among the expected spreads pertinent to different methods.

\header
{\bf Results on Large Datasets.} Next, we experiment with the four larger datasets, i.e., {\em Epinion}, {\em DBLP}, {\em LiveJournal}, and {\em Twitter}. As {\em RIS} and {\em CELF++} incur prohibitive overheads on those four datasets, we omit them from the experiments. Figure~\ref{fig:exp-large-time} shows the running time of {\em TIM} and {\em TIM$^+$} on each dataset. Observe that {\em TIM$^+$} outperforms {\em TIM} in all cases, by up to two orders of magnitude in terms of running time. Furthermore, even in the most adversarial case when $k = 1$, {\em TIM$^+$} terminates within four hours under both the IC and LT models. ({\em TIM} is omitted from Figure~\ref{fig:exp-large-time}d due to its excessive computation cost on {\em Twitter}.)

Interestingly, both {\em TIM} and {\em TIM$^+$} are more efficient under the LT model than the IC model. This is caused by the fact that we use different methods to generate RR sets under the two models. Specifically, under the IC model, we construct each RR set with a randomized BFS on $G$; for each incoming edge that we encounter during the BFS, we need to generate a random number to decide whether the edge should be ignored. In contrast, when we perform a randomized BFS on $G$ to create an RR set under the LT model, we generate a random number $x$ for each node $v$ that we visit, and we use $x$ to pick an incoming edge of $v$ to traverse. In other words, the number of random numbers required under the IC (resp.\ LT) model is proportional to the number of edges (resp.\ nodes) examined. Given that each of our datasets contains much more edges than nodes, it is not surprising that our solutions perform better under the LT model.

Finally, Figure~\ref{fig:exp-large-eps} shows the running time of {\em TIM} and {\em TIM$^+$} as a function of $\varepsilon$. The performance of both algorithms significantly improves with the increase of $\varepsilon$, since a larger $\varepsilon$ leads to a less stringent requirement on the number of RR sets. In particular, when $\varepsilon \ge 0.2$, {\em TIM$^+$} requires less than $1$ hour to process {\em Twitter} under both the IC and LT models.

\subsection{Comparison with {\large \bf \em  IRIE} and {\large \bf \em SIMPATH}} \label{sec:exp-additional}

Our second set of experiments compares {\em TIM$^+$} with {\em IRIE} \cite{JungHC12} and {\em SIMPATH} \cite{GoyalLL11}, namely, the state-of-the-art heuristic methods under the IC and LT models, respectively. (We omit {\em TIM} as it performs consistently worse than {\em TIM$^+$}.) Both {\em IRIE} and {\em SIMPATH} have two internal parameters that control the trade-off between computation cost and result accuracy. In our experiments, we set those parameters according to the recommendations in \cite{JungHC12,GoyalLL11}. Specifically, we set {\em IRIE} parameters $\alpha$ and $\theta$ to $0.7$ and $1/320$, respectively, and {\em SIMPATH}'s parameters $\eta$ and $\ell$ to $10^{-3}$ and $4$, respectively. For {\em TIM$^+$}, we set $\varepsilon = \ell = 1$, in which case {\em TIM$^+$} provides weak theoretical guarantees but high empirical efficiency. We evaluate the algorithms on all datasets except {\em Twitter}, as the memory consumptions of {\em IRIE} and {\em SIMPATH} exceed the size of the memory on our testing machine (i.e., $48$GB).

\begin{figure*}[t]
\begin{small}
\centering
\begin{tabular}{cccc}
\multicolumn{4}{c}{\hspace{-4mm} \includegraphics[height=3.5mm]{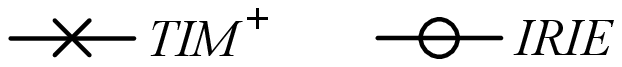}} \vspace{0mm} \\
\hspace{-5.5mm}\includegraphics[height=36mm]{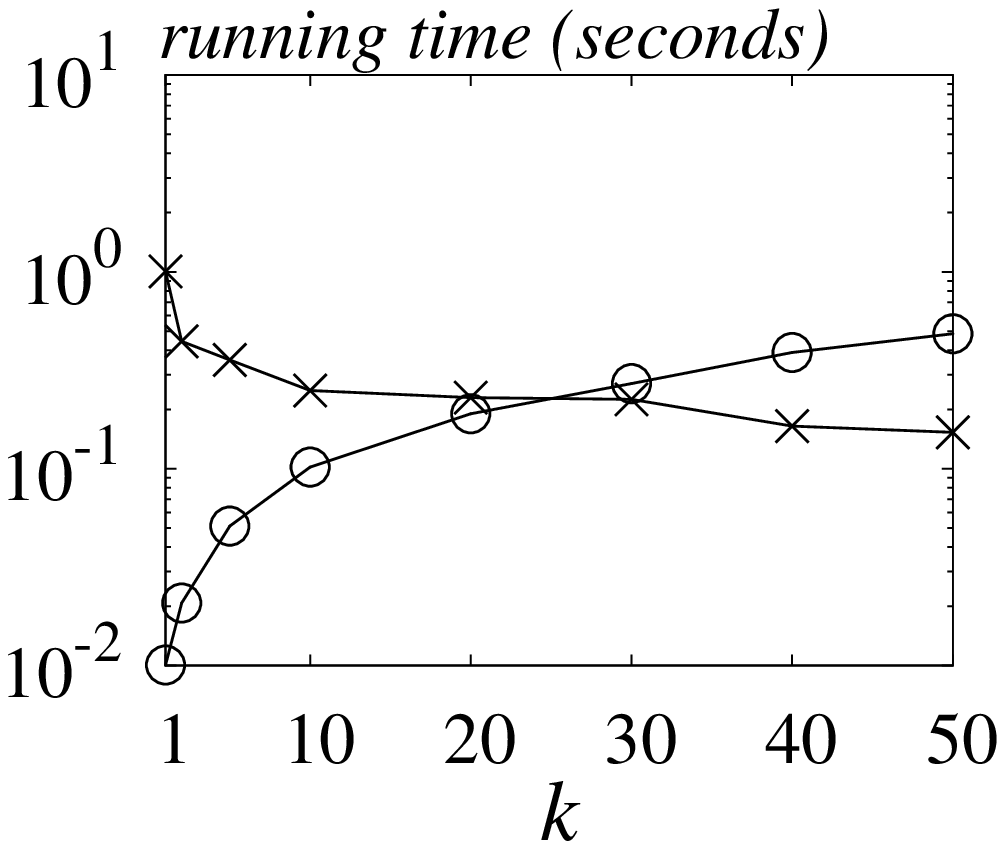}
&
\hspace{-10mm}\includegraphics[height=36mm]{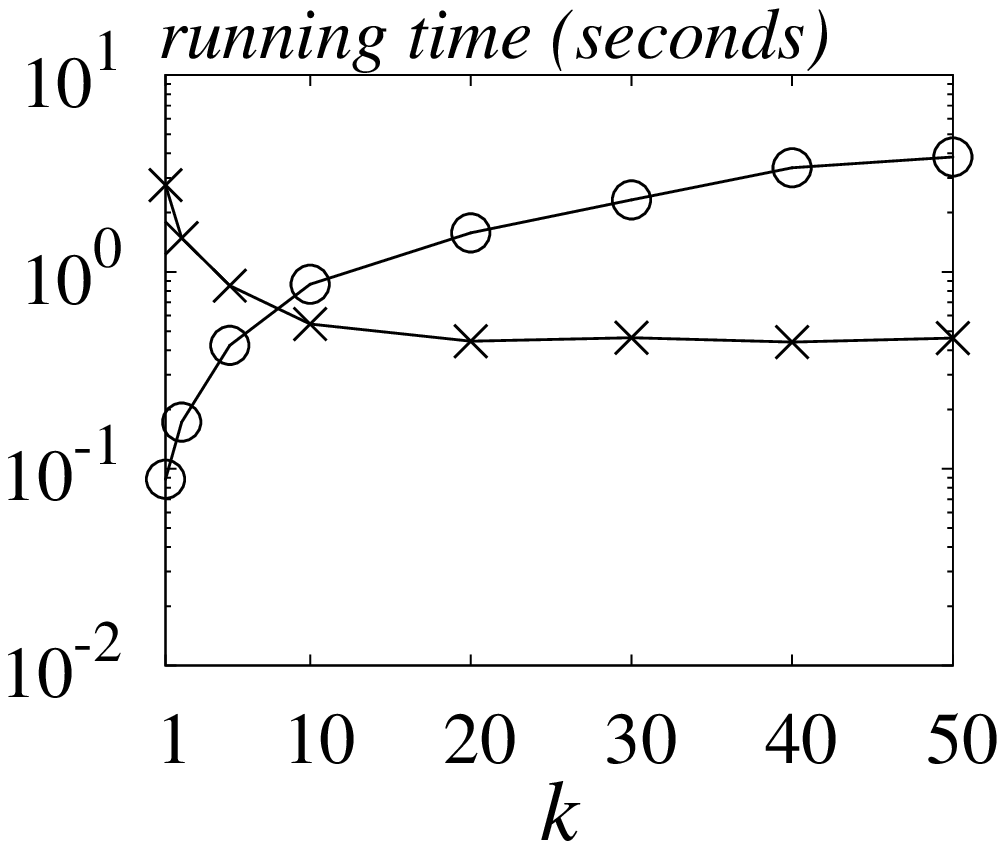}
&
\hspace{-10mm}\includegraphics[height=36mm]{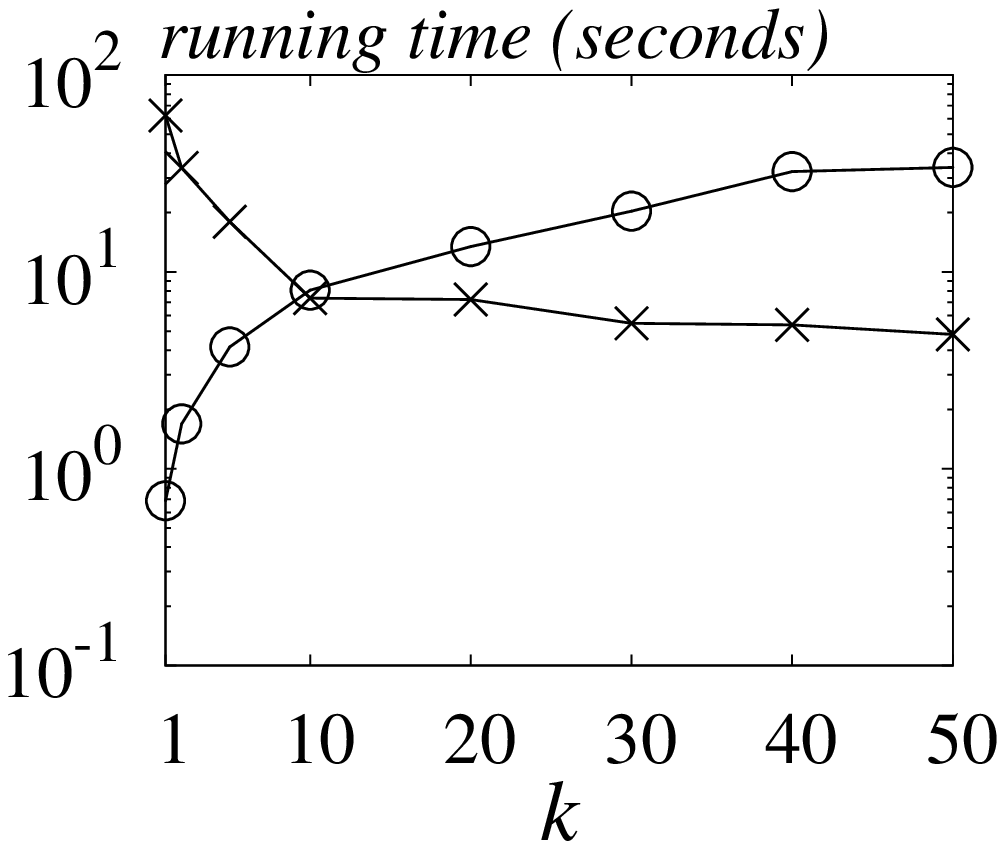}
&
\hspace{-10mm}\includegraphics[height=36mm]{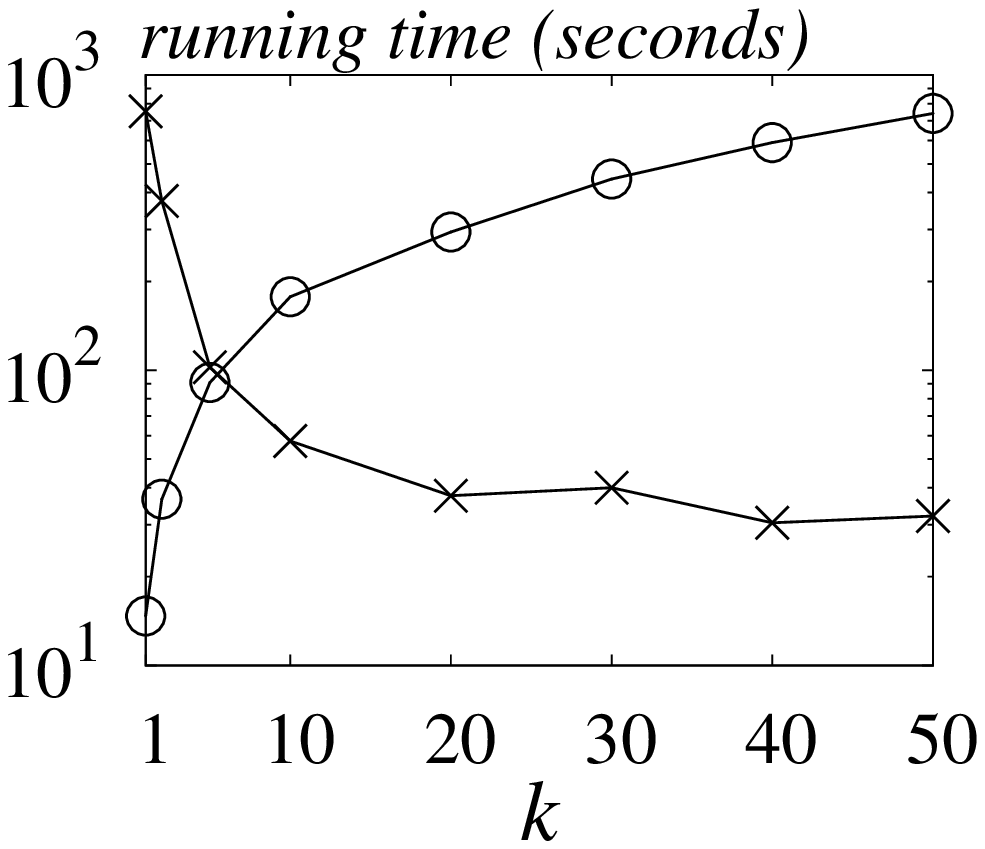}
\\
\hspace{-5.5mm}(a) {\em NetHEPT} & \hspace{-10mm}(b) {\em Epinions} &
\hspace{-10mm}(c) {\em DBLP} & \hspace{-10mm} (d)  {\em LiveJournal}
\end{tabular}
\figcapup  \vspace{-1mm} \caption{Running time vs.\ $k$ under the IC model.} %\vspace{-1mm} %\figcapdown
\label{fig:exp-compare-ic-time}
\end{small}
\end{figure*}

\begin{figure*}[t]
\begin{small}
\centering
\begin{tabular}{cccc}
\multicolumn{4}{c}{\hspace{-4mm} \includegraphics[height=3.5mm]{./figures/compare_ic_legend.eps}} \vspace{0mm} \\
\hspace{-6mm}\includegraphics[height=36mm]{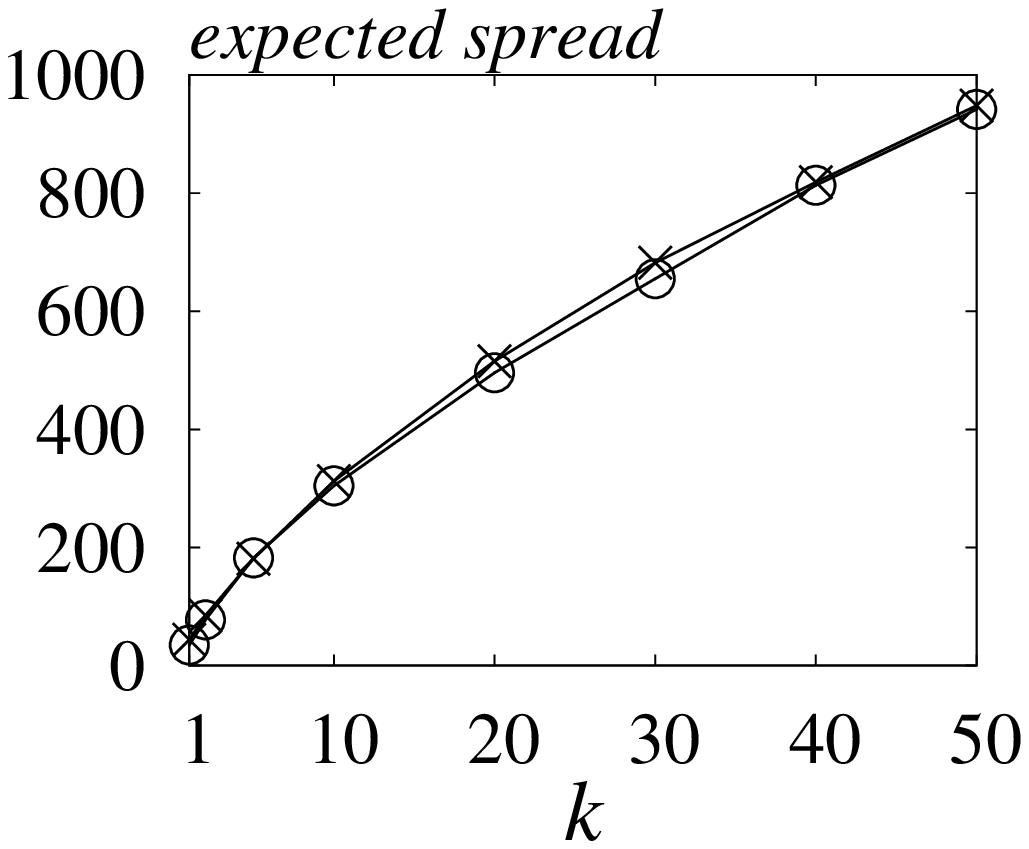}
&
\hspace{-10.5mm}\includegraphics[height=36mm]{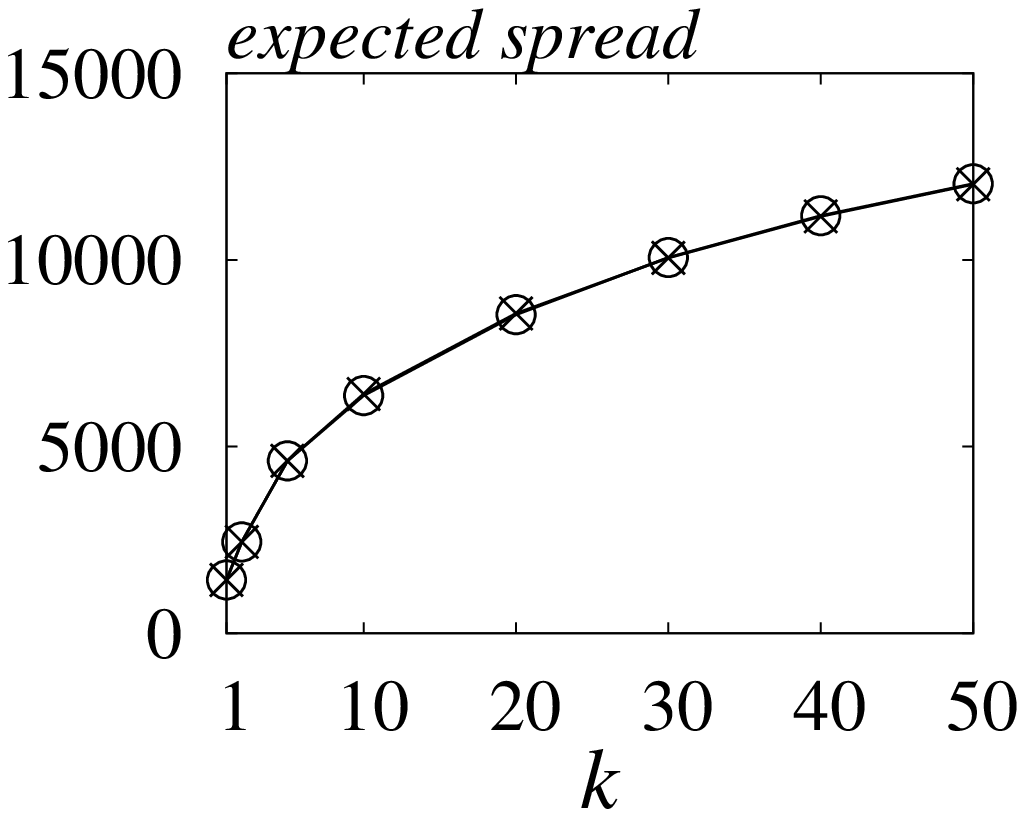}
&
\hspace{-10.5mm}\includegraphics[height=36mm]{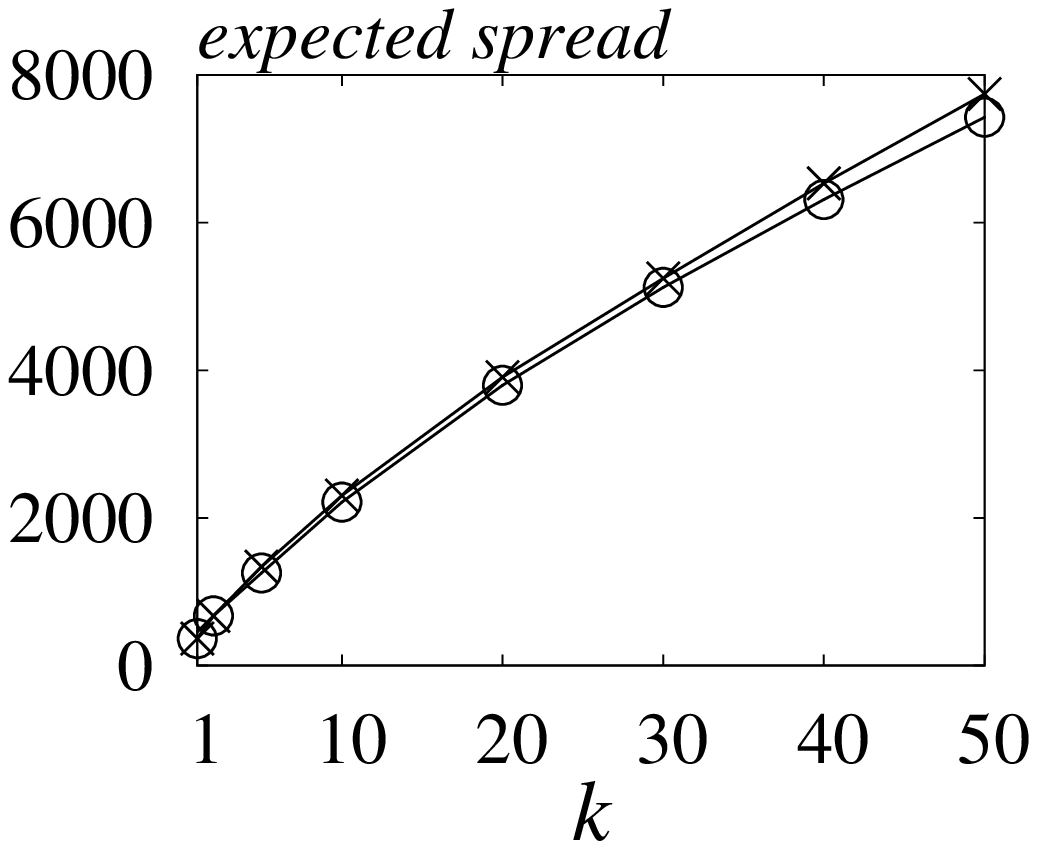}
&
\hspace{-10.5mm}\includegraphics[height=36mm]{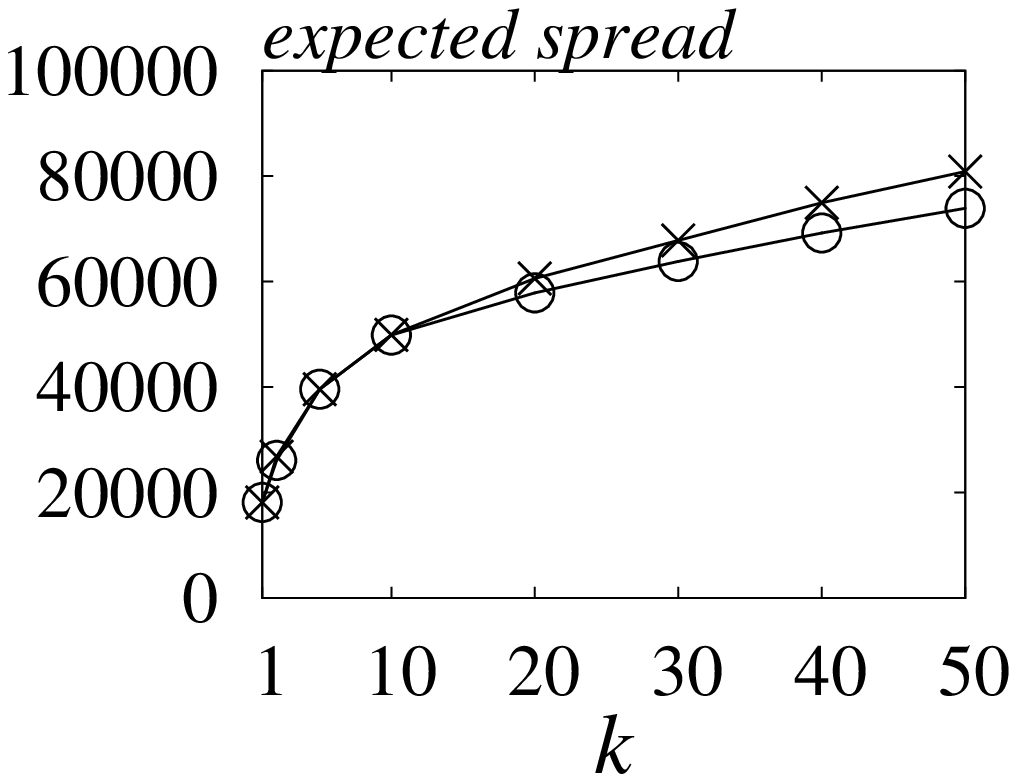}
\\
\hspace{-3mm}(a) {\em NetHEPT} & \hspace{-5mm}(b) {\em Epinions} &
\hspace{-5mm}(c) {\em DBLP} & \hspace{-3mm} (d)  {\em LiveJournal}
\end{tabular}
\figcapup  \vspace{-1mm} \caption{Expected spreads vs.\ $k$ under the IC model.} %\vspace{-1mm} %\figcapdown
\label{fig:exp-compare-ic-inf}
\end{small}
\end{figure*}

Figure~\ref{fig:exp-compare-ic-time} shows the running time of {\em TIM$^+$} and {\em IRIE} under the IC model, varying $k$ from $1$ to $50$. The computation cost of {\em TIM$^+$} tends to decrease with the increase of $k$, as a result of the subtle interplay among several variables (e.g., $\lambda$, $KPT^*$, and $KPT^+$) that decide the number of random RR sets required in {\em TIM$^+$}. Meanwhile, {\em IRIE}'s computation time increases with $k$, since (i) it adopts a greedy approach to iteratively select $k$ nodes from the input graph $G$, and (ii) a larger $k$ results in more iterations in {\em IRIE}, which leads to a higher processing cost. Overall, {\em TIM$^+$} is not as efficient as {\em IRIE} when $k$ is small, but it clearly outperforms {\em IRIE} on all datasets when $k > 20$. In particular, when $k = 50$, {\em TIM$^+$}'s computation time on {\em LiveJournal} is less than $5\%$ of {\em IRIE}'s.

Figure~\ref{fig:exp-compare-ic-inf} illustrates the expected spreads of the node sets returned by {\em TIM$^+$} and {\em IRIE}. Compared with {\em IRIE}, {\em TIM$^+$} have (i) noticeably higher expected spreads on {\em DBLP} and {\em LiveJournal}, and (ii) similar expected spreads on {\em NetHEPT} and {\em Epinion}. This indicates that {\em TIM$^+$} generally provides more accurate results than {\em IRIE} does, even when we set $\varepsilon = \ell = 1$ for {\em TIM$^+$}.

\begin{figure*}[t]
\begin{small}
\centering
\begin{tabular}{cccc}
\multicolumn{4}{c}{\hspace{-4mm} \includegraphics[height=3.5mm]{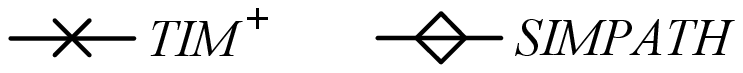}} \vspace{0mm} \\
\hspace{-5.5mm}\includegraphics[height=36mm]{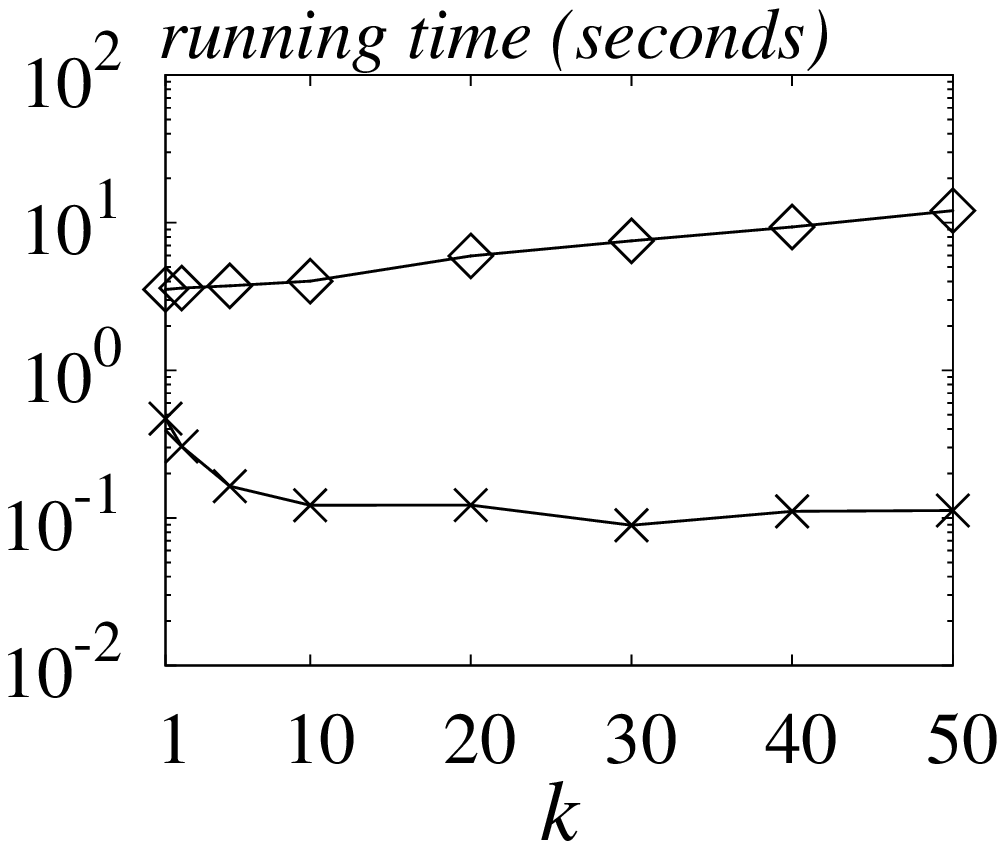}
&
\hspace{-10mm}\includegraphics[height=36mm]{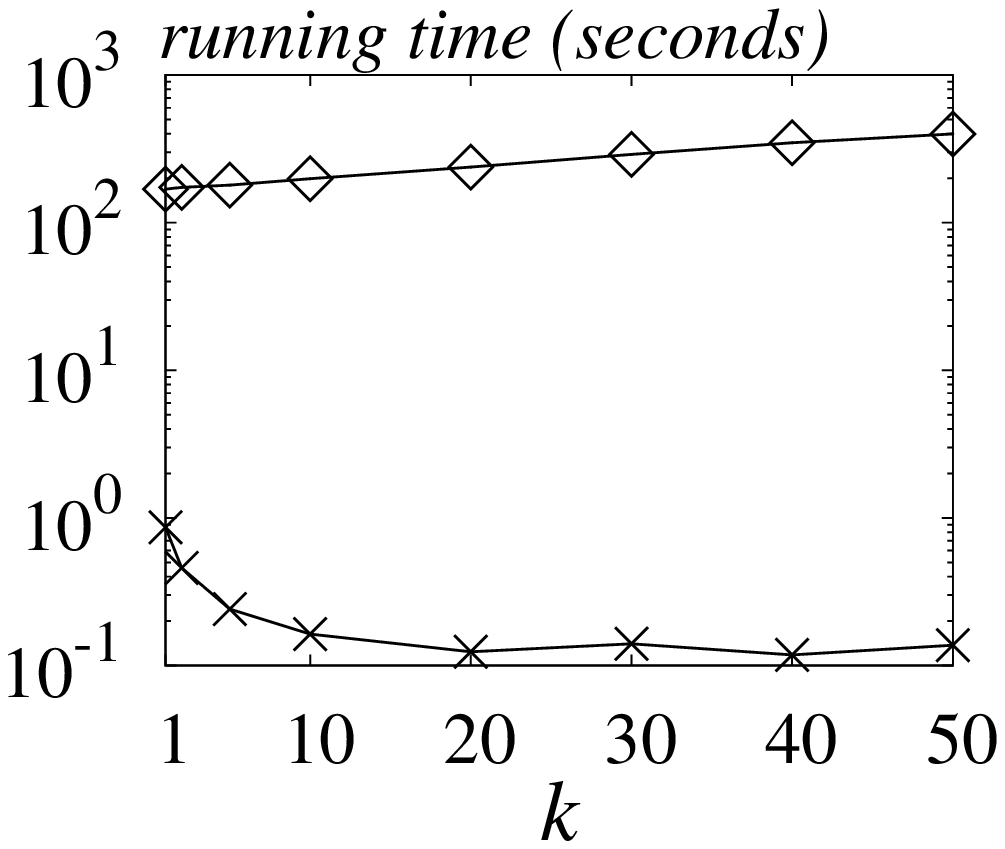}
&
\hspace{-10mm}\includegraphics[height=36mm]{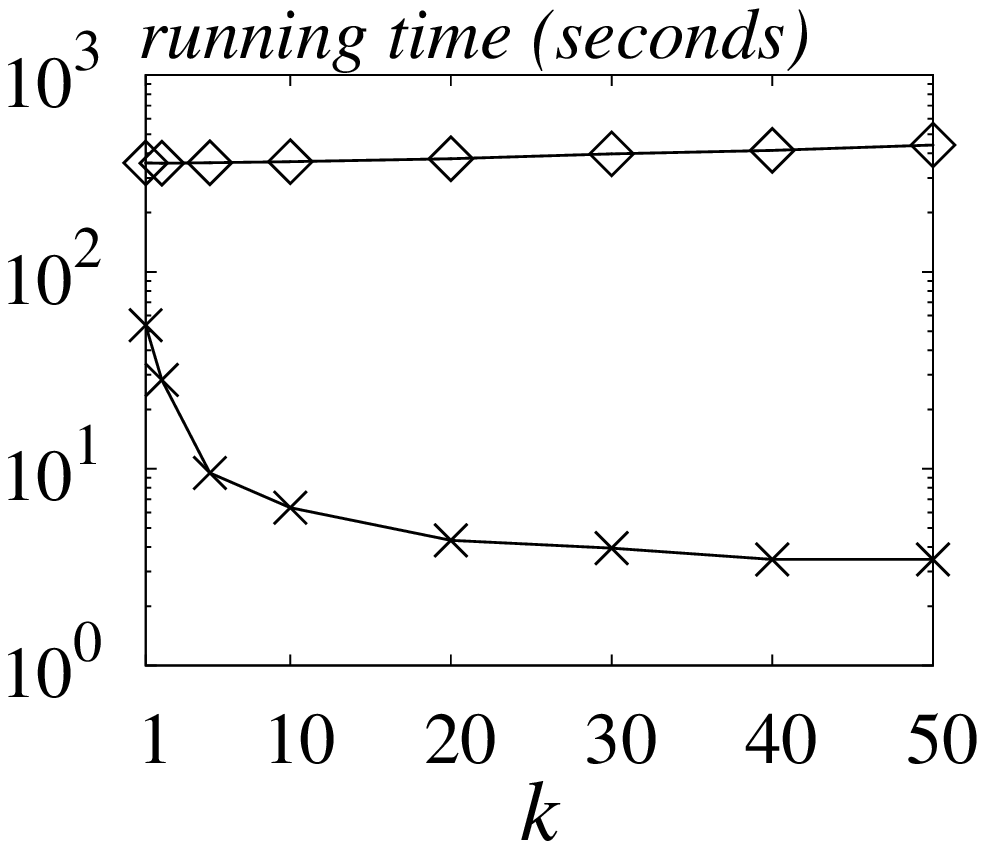}
&
\hspace{-10mm}\includegraphics[height=36mm]{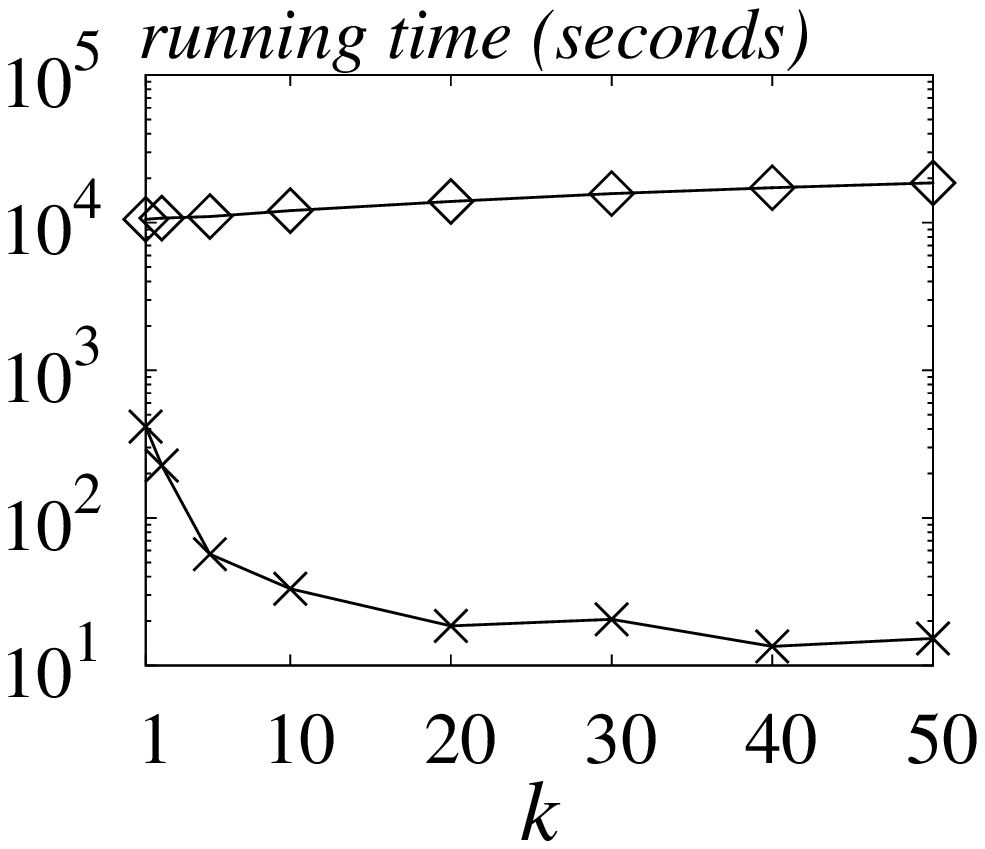}
\\
\hspace{-5.5mm}(a) {\em NetHEPT} & \hspace{-10mm}(b) {\em Epinions} &
\hspace{-10mm}(c) {\em DBLP} & \hspace{-10mm} (d)  {\em LiveJournal}
\end{tabular}
\figcapup  \vspace{-1mm} \caption{Running time vs.\ $k$ under the LT model.} %\vspace{-1mm} %\figcapdown
\label{fig:exp-compare-lt-time}
\end{small}
\end{figure*}

\begin{figure*}[t]
\begin{small}
\centering
\begin{tabular}{cccc}
\multicolumn{4}{c}{\hspace{-4mm} \includegraphics[height=3.5mm]{./figures/compare_lt_legend.eps}} \vspace{0mm} \\
\hspace{-6mm}\includegraphics[height=36mm]{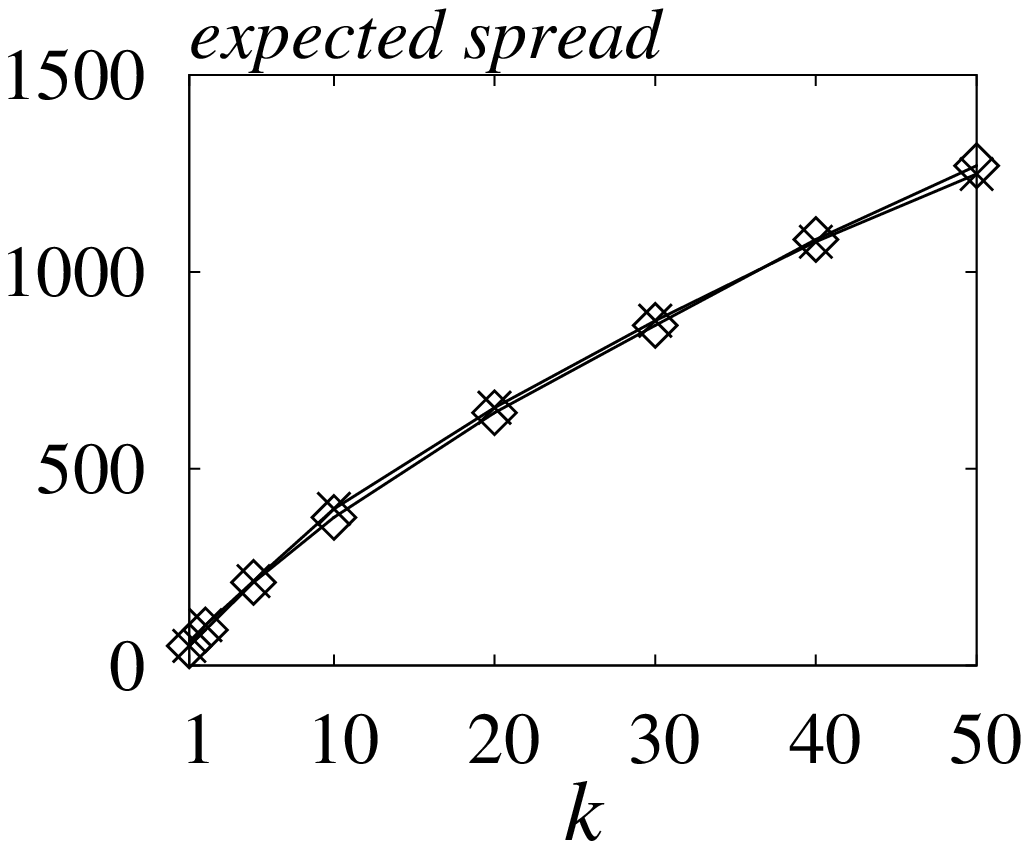}
&
\hspace{-11mm}\includegraphics[height=36mm]{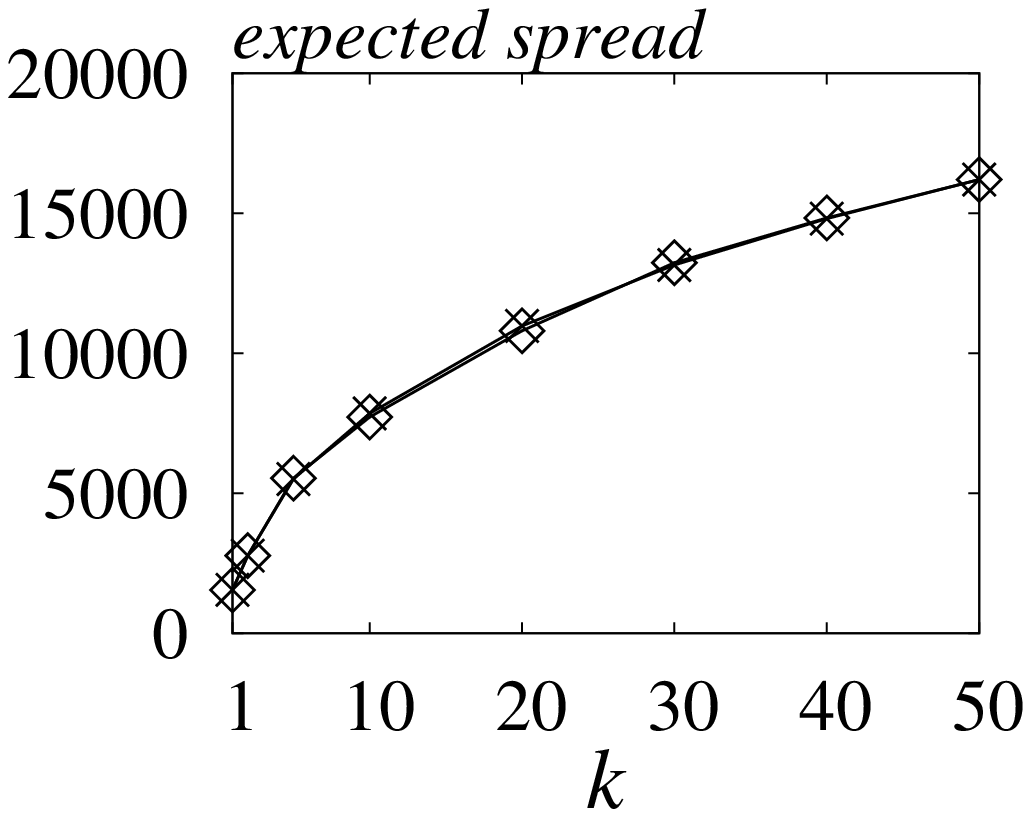}
&
\hspace{-11mm}\includegraphics[height=36mm]{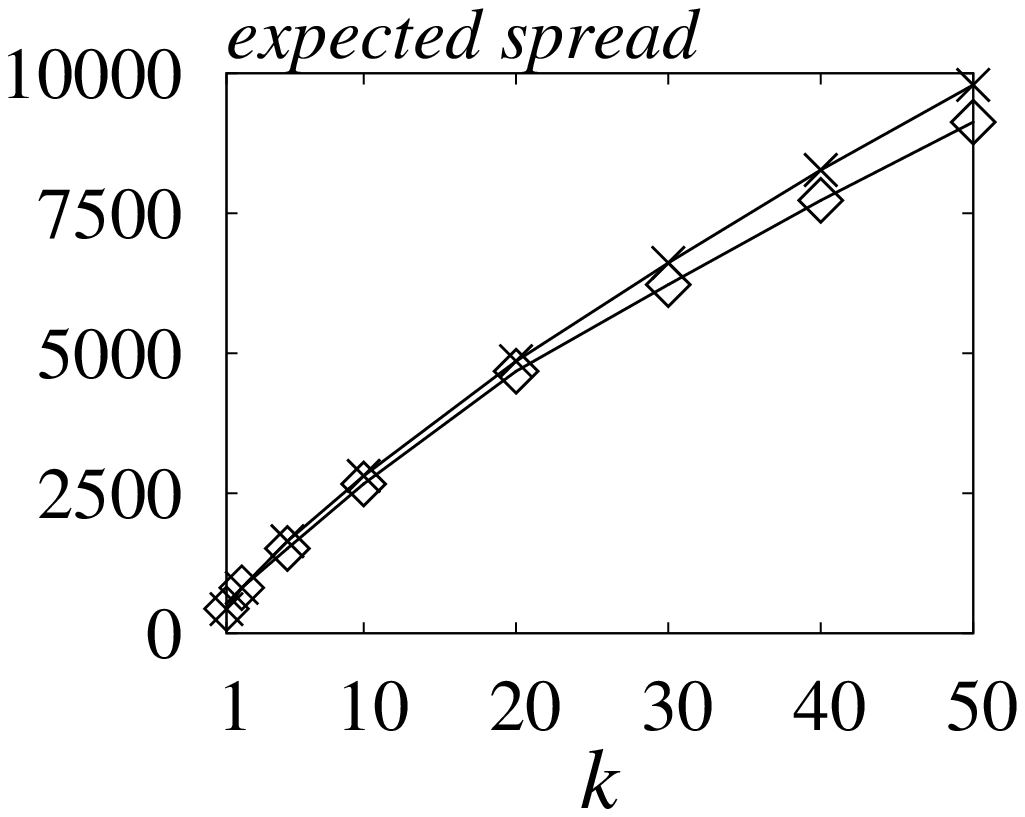}
&
\hspace{-11mm} \includegraphics[height=36mm]{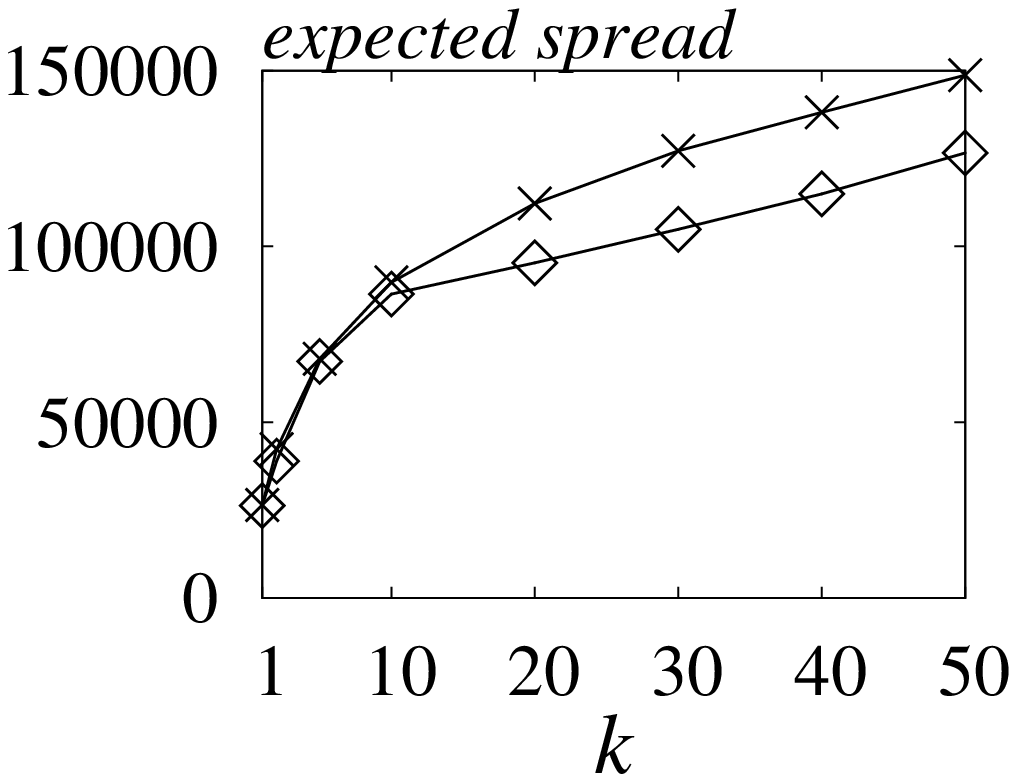}
\\
\hspace{-3mm}(a) {\em NetHEPT} & \hspace{-5mm}(b) {\em Epinions} &
\hspace{-5mm}(c) {\em DBLP} & \hspace{-3mm} (d)  {\em LiveJournal}
\end{tabular}
\figcapup  \vspace{-1mm} \caption{Expected spreads vs.\ $k$ under the LT model.} %\vspace{-1mm} %\figcapdown
\label{fig:exp-compare-lt-inf}
\end{small}
\end{figure*}

\begin{figure*}[!t]
\begin{small}
\centering
\begin{tabular}{ccc}
\multicolumn{3}{c}{\hspace{-0mm} \includegraphics[height=3.8mm]{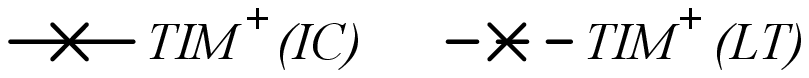}} \vspace{0mm} \\
\hspace{0mm}\includegraphics[height=36mm]{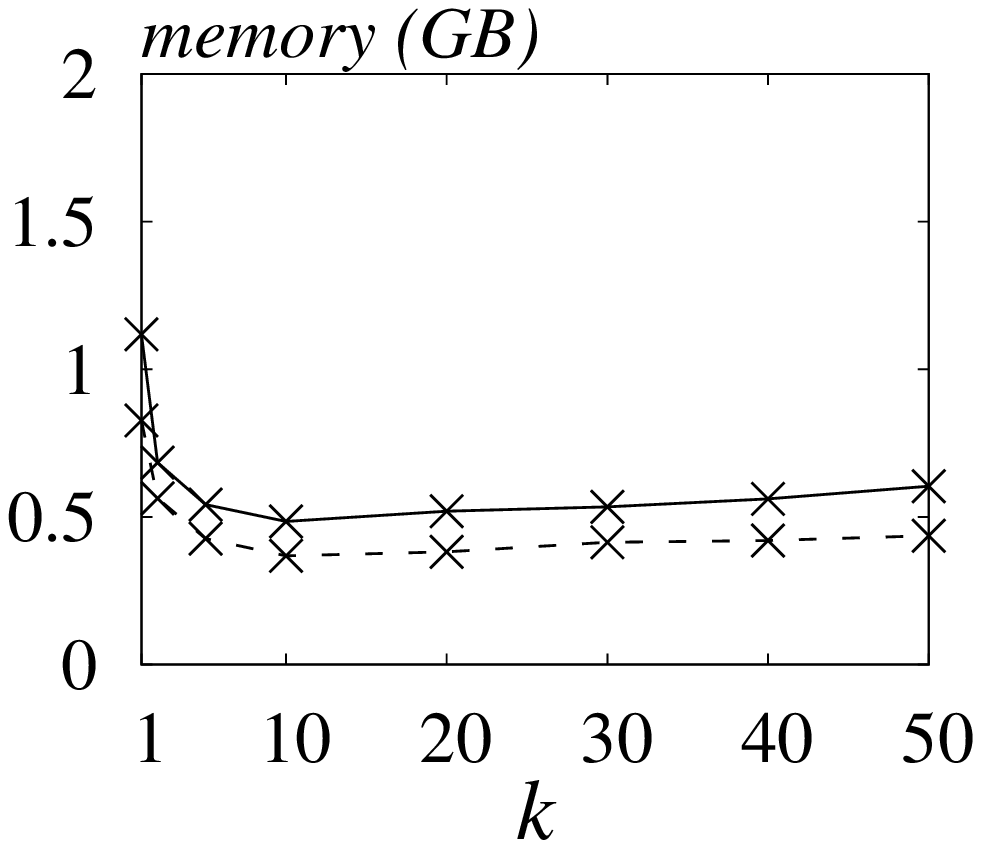}
&
\hspace{0mm}\includegraphics[height=36mm]{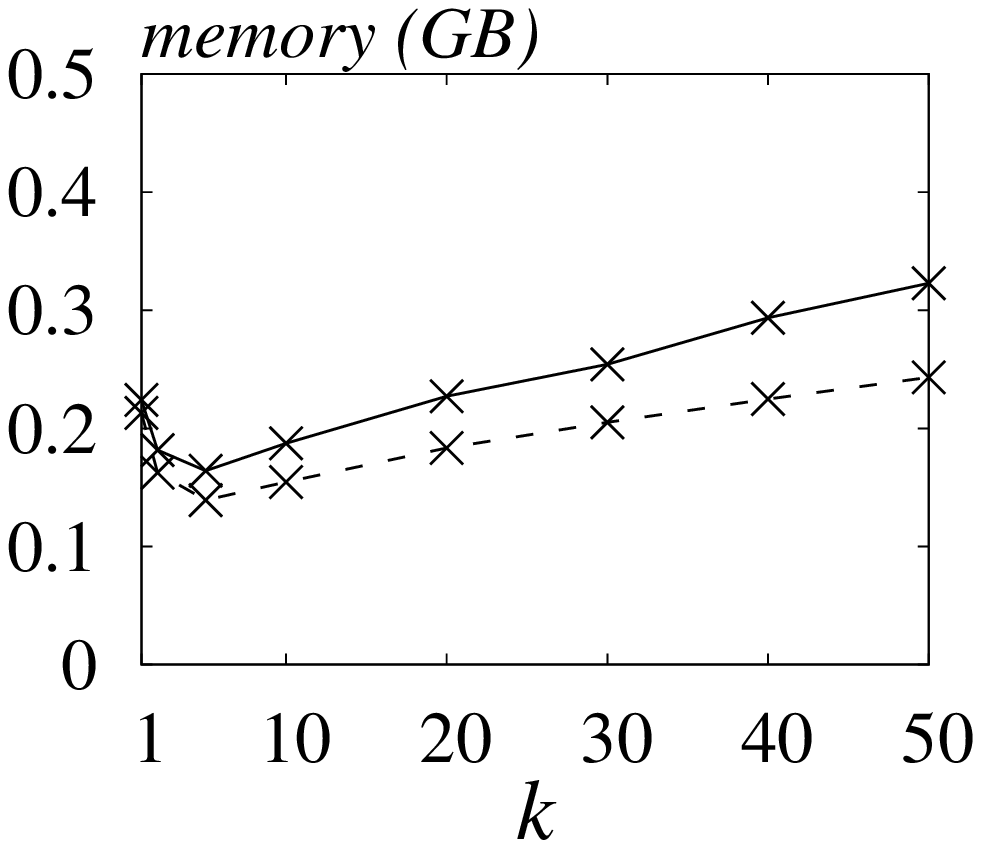}
&
\hspace{0mm}\includegraphics[height=36mm]{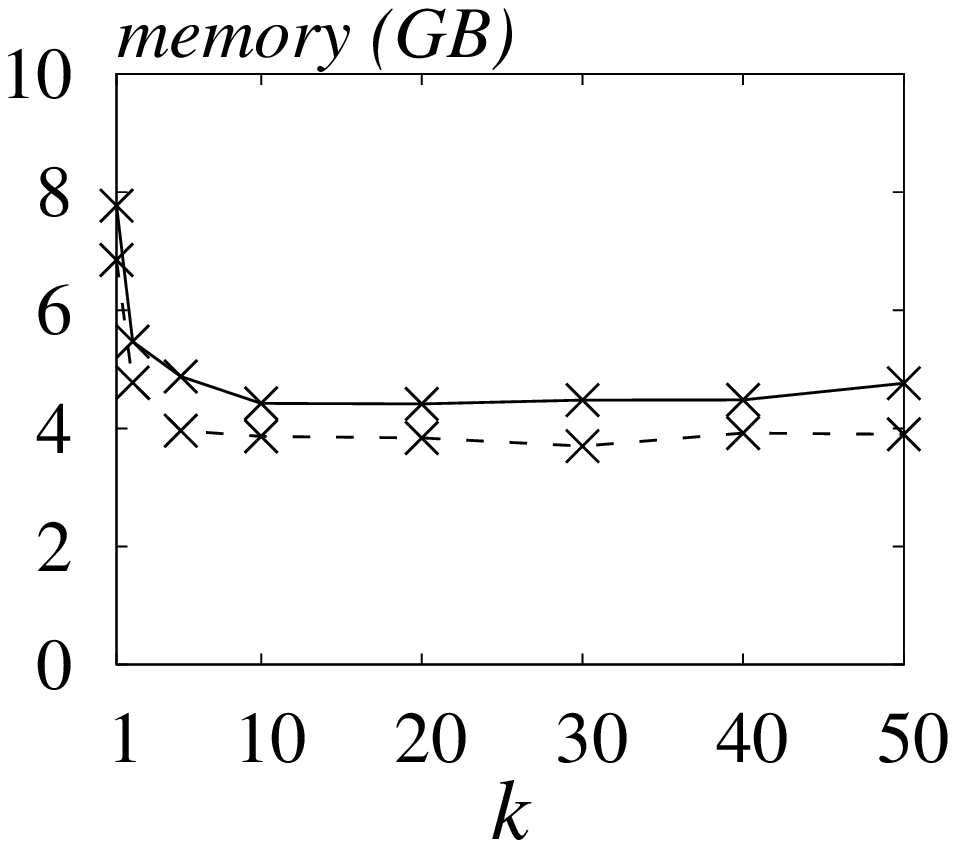}
\\
\hspace{2mm}(a) {\em NetHEPT} & \hspace{3mm}(b) {\em Epinions} &
\hspace{2mm}(c) {\em DBLP}
\end{tabular}
\begin{tabular}{cc}
\hspace{0mm} \includegraphics[height=36mm]{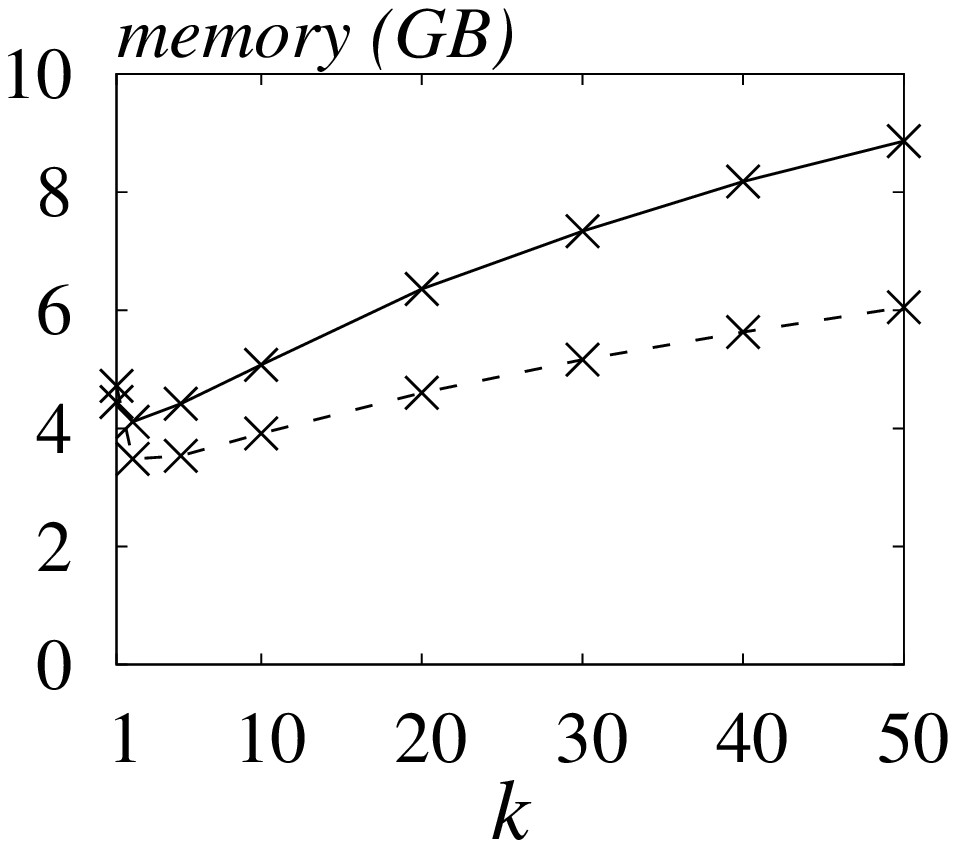}
&
\hspace{0mm} \includegraphics[height=36mm]{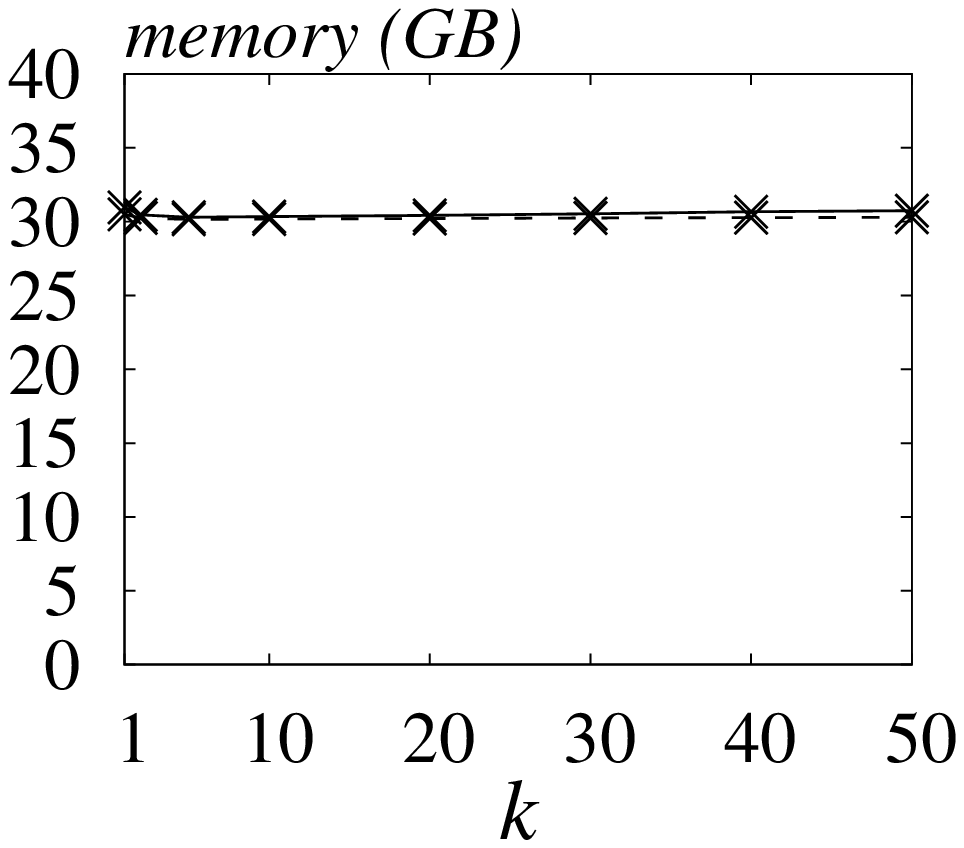} \\
\hspace{2mm} (d)  {\em LiveJournal} & \hspace{0mm} (e)  {\em Twitter}
\end{tabular}
\figcapup  \vspace{-0mm} \caption{Memory consumptions of {\em TIM$^+$} vs.\ $k$.} %\vspace{-1mm} %\figcapdown
\label{fig:exp-memory}
\end{small}
\end{figure*}

Figure~\ref{fig:exp-compare-lt-time} compares the computation efficiency of {\em TIM$^+$} and {\em SIMPATH} under the LT model, when $k$ varies. Observe that {\em TIM$^+$} consistently outperforms {\em SIMPATH} by large margins. In particular, when $k = 50$, the former's running time on {\em LiveJournal} is lower than the latter's by three orders of magnitude. Furthermore, as shown in Figure~\ref{fig:exp-compare-lt-inf}, {\em TIM$^+$}'s expected spreads are significantly higher than {\em SIMPATH}'s on {\em LiveJournal}, and are no worse on the other three datasets. Therefore, {\em TIM$^+$} is clearly more preferable than {\em SIMPATH} for influence maximization under the LT model.

\subsection{Memory Consumptions} \label{sec:exp-space}

Our last set of experiments evaluates {\em TIM$^+$}'s memory consumptions, setting $\varepsilon = 0.1$ and $\ell = 1 + \log 3 / \log n$ (i.e., we ensure a success probability of at least $1 - 1/n$). Note that $\varepsilon = 0.1$ is adversarial to {\em TIM$^+$}, due to the following reasons:
\begin{enumerate}
\item The memory costs of {\em TIM$^+$} is mainly incurred by the set $\R$ of random RR sets generated in Algorithm~\ref{alg:tim-nodesel};

\item {\em TIM$^+$} sets the size of $\R$ to $\lambda / KPT^+$, where $\lambda$ is as defined in Equation~\ref{eqn:tim-lambda} and $KPT^+$ is a lowerbound of $OPT$ generated by Algorithm~\ref{alg:ext-refine};

\item $\lambda$ is inverse proportional to $\varepsilon^2$, i.e., a smaller $\varepsilon$ leads to a larger $\R$, which results in a higher space overhead.
\end{enumerate}
Figure~\ref{fig:exp-memory} shows the memory costs of {\em TIM$^+$} on each dataset under the IC and LT models. In all cases, {\em TIM$^+$} requires more memory under the IC model than under the LT model. The reason is that $\R$'s size is inverse proportional to $KPT^+$, while $KPT^+$ tends to be larger under the LT model (see Figure~\ref{fig:exp-nethept-spread} for example). The memory consumption of {\em TIM$^+$} tends to be larger when the dataset size increases, since $\R = \lambda / KPT^+$, while $\lambda$ increases with $n$, i.e., the number of nodes in the dataset. But interestingly, {\em TIM$^+$} incurs a higher space overhead on {\em NetHEPT} than on {\em Epinion}, even though the latter has a larger number of nodes. To explain, observe from Figures \ref{fig:exp-compare-ic-inf} and \ref{fig:exp-compare-lt-inf} that nodes in {\em Epinion} tend to have much higher expected spreads than those in {\em NetHEPT}. As a consequence, {\em TIM$^+$} obtains a considerably larger $KPT^+$ from {\em Epinion} than from {\em NetHEPT}. This pronounced increase in $KPT^+$ renders $\R = \lambda / KPT^+$ smaller on {\em Epinion} than on {\em NetHEPT}, despite of the fact that $\lambda$ is smaller on the latter.

\section{Conclusion} \label{sec:conclu}

This paper presents {\em TIM}, an influence maximization algorithm that supports the triggering model by Kempe et al.\ \cite{KempeKT03}. The algorithm runs in $O((k+\ell)(n+m) \log n / \varepsilon^2)$ expected time, and returns $(1-1/e-\varepsilon)$-approximate solutions with at least $1 - n^{-\ell}$ probability. In addition, it incorporates heuristic optimizations that lead to up to $100$-fold improvements in empirical efficiency. Our experiments show that, when $k = 50$, $\varepsilon = 0.2$, and $\ell = 1$, the algorithm can process a billion-edge graph on a commodity machine within an hour. Such practical efficiency is unmatched by any existing solutions that provide non-trivial approximation guarantees for the influence maximization problem. For future work, we plan to investigate how we can turn {\em TIM} into a distributed algorithm, so as to handle massive graphs that do not fit in the main memory of a single machine. In addition, we plan to extend {\em TIM} to other formulations of the influence maximization problem, e.g., {\em competitive influence maximization} \cite{{BharathiKS07,LuB0L13}}.

\bibliographystyle{abbrv}
\bibliography{ref}
\appendix
%\section{Proofs} \label{sec:proofs}

%\begin{proof}[of Lemma~\ref{lmm:tim-fraction}]
%We first present a supporting lemma from \cite{Borgs14} that we frequently use. The proof of the lemma is included for completeness.

{\bf Proof of Lemma~\ref{lmm:tim-nodesel-rr}.}
Let $g$ be a graph constructed from $G$ by removing each edge $e$ with $1 - p(e)$ probability. Then, $\rho_2$ equals the probability that $v$ is reachable from $S$ in $g$. Meanwhile, by Definition~\ref{def:def-rrset}, $\rho_1$ equals the probability that $g$ contains a directed path that ends at $v$ and starts at a node in $S$. It follows that $\rho_1 = \rho_2$. \done
%
%We write this as
%\begin{equation} \label{eqn:appen-act-1}
%\rho_2 = {\Pr}_{g}\big[S \rightarrow \{v\} \textrm{ in } g\big],
%\end{equation}
%where $A \rightarrow B$ denotes that there exists a directed path from some node in a set $A$ to some node in a set $B$.
%
%Let $g'$ be a graph obtained from $G^T$ by eliminating each edge $e^T$ with $1 - p(e^T)$ probability. Then,
%\begin{equation} \label{eqn:appen-act-2}
%{\Pr}_{g}\big[S \rightarrow \{v\} \textrm{ in } g\big] = {\Pr}_{g'}\left[\{v\} \rightarrow S \textrm{ in } g'\right]
%\end{equation}
%Observe that $\Pr_{g'}\left[\{v\} \rightarrow S \textrm{ in } g'\right]$ equals the probability that $S$ overlaps with an RR set generated for $v$, which is exactly $\rho_1$. Combining this with Equations \ref{eqn:appen-act-1} and \ref{eqn:appen-act-2}, we have $\rho_1 = \rho_2$. \done

%\section{Proofs}

\pheader
{\bf Proof of Corollary~\ref{coro:tim-fraction}.} Observe that $\E[F_{\R}(S)]$ equals the probability that $S$ intersects a random RR set, while $\E[I(S)]/n$ equals the probability that a randomly selected node can be activated by $S$ in an influence propagation process on $G$. By Lemma~\ref{lmm:tim-nodesel-rr}, the two probabilities are equal, leading to $\E[n\cdot F_{\R}(S)] = \E[I(S)]$. \done

\pheader
{\bf Proof of Lemma~\ref{lmm:tim-theta}.} Let $\rho$ be the probability that $S$ overlaps with a random RR set. Then, $\theta \cdot F_{\R}(S)$ can be regarded as the sum of $\theta$ i.i.d.\ Bernoulli variables with a mean $\rho$. By Corollary~\ref{coro:tim-fraction},
\begin{equation*}
\rho = \E[F_{\R}(S)] = \E[I(S)]/n.
\end{equation*}
Then, we have
\begin{eqnarray} \label{eqn:tim-theta-proof-1}
\lefteqn{\Pr\Big[\big|n \cdot F_{\R}(S) - \E[I(S)]\big| \ge \frac{\varepsilon}{2} \cdot OPT \Big]} \nonumber \\
& = & \Pr\Big[\big|\theta \cdot F_{\R}(S) - \rho \theta \big| \ge \frac{\varepsilon\theta}{2n} \cdot OPT \Big] \nonumber \\
%& = & \Pr\Big[\big|\theta \cdot F_{\R}(S) - \rho \theta \big| \ge \frac{\varepsilon \cdot OPT}{2n \rho} \cdot \rho \theta \Big] \nonumber \\
& = & \Pr\Big[\big|\theta \cdot F_{\R}(S) - \rho \theta \big| \ge \frac{\varepsilon \cdot OPT}{2n \rho} \cdot \rho \theta \Big].
%& \le & Pr\Big[\big|F_{\R}(S) - \rho \theta \big| > \varepsilon \rho \theta/2 \Big] \nonumber \\
%& \le & \exp(-\rho \theta \varepsilon^2/12) + \exp(-\rho \theta \varepsilon^2/8) \nonumber \\
%& \le & 2\exp(-\rho \theta \varepsilon^2/12)
\end{eqnarray}
Let $\delta = \varepsilon \cdot OPT/(2n \rho)$. By the Chernoff bounds, Equation~\ref{eqn:tim-theta-1}, and the fact that $\rho = \E[I(S)]/n \le OPT/n$, we have
\begin{align*}
\textrm{r.h.s.\ of Eqn.~\ref{eqn:tim-theta-proof-1}} & <  2\exp\left(-\frac{\delta^2}{2 + \delta} \cdot \rho\theta\right) \\
& =  2\exp\left(-\frac{\varepsilon^2\cdot OPT^2}{8 n^2 \rho + 2 \varepsilon n \cdot OPT}\cdot \theta\right) \\
& \le  2\exp\left(-\frac{\varepsilon^2\cdot OPT^2}{8 n \cdot OPT + 2 \varepsilon n \cdot OPT}\cdot \theta\right) \\
& =  2\exp\left(-\frac{\varepsilon^2\cdot OPT}{(8 + 2 \varepsilon)\cdot n}\cdot \theta\right) \;\; \le \frac{1}{{n \choose k} \cdot n^{l}}.
%& \le \frac{1}{{n \choose k} \cdot n^{l}}.
\end{align*}
Therefore, the lemma is proved. \done

\pheader
{\bf Proof of Theorem~\ref{thrm:tim-approx}.} Let $S_k$ be the node set returned by Algorithm~\ref{alg:tim-nodesel}, and $S^+_k$ be the size-$k$ node set that maximizes $F_{\R}(S^+_k)$ (i.e., $S^+_k$ covers the largest number of RR sets in $\R$). As $S_k$ is derived from $\R$ using a $(1 - 1/e)$-approximate algorithm for the maximum coverage problem, we have $F_{\R}(S_k) \ge (1-1/e) \cdot F_{\R}(S^+_k)$. Let $S^\circ_k$ be the optimal solution for the influence maximization problem on $G$, i.e., $\E[I(S^\circ_k)] = OPT$. We have $F_{\R}(S^+_k) \ge F_{\R}(S^\circ_k)$, which leads to $F_{\R}(S_k) \ge (1-1/e) \cdot F_{\R}(S^\circ_k)$.

Assume that $\theta$ satisfies Equation~\ref{eqn:tim-theta-1}. By Lemma~\ref{lmm:tim-theta}, Equation~\ref{eqn:tim-theta-2} holds with at least $1 - n^{-\ell}/{n \choose k}$ probability for any given size-$k$ node set $S$. The, by the union bound, Equation~\ref{eqn:tim-theta-2} should hold simultaneously for all size-$k$ node sets with at least $1 - n^{-\ell}$ probability. In that case, we have
\begin{eqnarray*}
\E[I(S_k)] &>& n \cdot F_{\R}(S_k) - \varepsilon/2\cdot OPT \\
&\ge& (1 - 1/e) \cdot n \cdot F_{\R}(S^+_k) - \varepsilon/2\cdot OPT \\
&\ge& (1 - 1/e) \cdot n \cdot F_{\R}(S^\circ_k) - \varepsilon/2\cdot OPT \\
&\ge& (1 - 1/e) \cdot (1 - \varepsilon/2) \cdot OPT - \varepsilon/2\cdot OPT \\
&>& (1 - 1/e - \varepsilon ) \cdot OPT.
\end{eqnarray*}
Thus, the theorem is proved. \done

\pheader
{\bf Proof of Lemma~\ref{lmm:tim-para-ept}.} Let $R$ be a random RR set, $p_R$ be the probability that a randomly selected edge from $G$ points to a node in $R$. Then,
%\vspace{-3mm}
%\begin{equation*}
%%\vspace{-1mm}
%EPT = \E[p_R \cdot m].
%\end{equation*}
$EPT = \E[p_R \cdot m]$, where the expectation is taken over the random choices of $R$.

Let $v^*$ be a sample from $\V$, and $b(v^*, R)$ be a boolean function that returns $1$ if $v^* \in R$, and $0$ otherwise. Then, for any fixed $R$,
\begin{equation*}
p_R = \sum_{v^*} \big(\Pr[v^*] \cdot b(v^*, R) \big).
\end{equation*}
Now consider that we fix $v^*$ and vary $R$. Define
\begin{equation*}
p_{v^*} = \sum_{R} \big(\Pr[R] \cdot b(v^*, R) \big).
\end{equation*}
By Lemma~\ref{lmm:tim-nodesel-rr}, $p_{v^*}$ equals the probability that a randomly selected node can be activated in an influence propagation process when $\{v^*\}$ is used as the seed set. Therefore, $\E[p_{v^*}] = \E[I(\{v^*\})]/n$. This leads to
\begin{align*}
EPT/m &= \E[p_R] = \sum_{R} \big( \Pr[R] \cdot p_R \big)\\
& = \sum_{R} \Big( \Pr[R] \cdot \sum_{v^*} \big(\Pr[v^*] \cdot b(v^*, R) \big) \Big) \\
& = \sum_{v^*} \Big( \Pr[v^*] \cdot \sum_{R} \big(\Pr[R] \cdot b(v^*, R) \big) \Big) \\
& = \sum_{v^*} \big( \Pr[v^*] \cdot p_{v^*} \big)  = \E[p_{v^*}] = \E[I(\{v^*\})]/n.
\end{align*}
Thus, the lemma is proved. \done

\pheader
{\bf Proof of Lemma~\ref{lmm:tim-para-kpt}.} Let $S^*$ be a node set formed by $k$ samples from $\V$, with duplicates removed. Let $R$ be a random RR set, and $\alpha_R$ be the probability that $S^*$ overlaps with $R$. Then, by Corollary~\ref{coro:tim-fraction},
\begin{equation*}
KPT = \E[I(S^*)] = \E[n \cdot \alpha_R].
\end{equation*}

Consider that we sample $k$ times over a uniform distribution on the edges in $G$. Let $E^*$ be the set of edges sampled, with duplicates removed. Let $\alpha'_R$ be the probability that one of the edges in $E^*$ points to a node in $R$. It can be verified that $\alpha'_R = \alpha_R$. Furthermore, given that there are $w(R)$ edges in $G$ that point to nodes in $R$, $\alpha'_R = 1 - (1 - w(R)/m)^k = \kappa(R)$. Therefore,
\begin{equation*}
KPT = \E[n \cdot \alpha_R] = \E[n \cdot \alpha'_R] = \E\left[n \cdot \kappa(R)\right],
\end{equation*}
which proves the lemma. \done

\pheader
{\bf Proof of Lemma~\ref{lmm:tim-para-stopup}.} Let $\delta = (2^{-i} - \mu)/\mu$. By the Chernoff bounds,
\begin{eqnarray*}
\Pr\left[\frac{s_i}{c_i} > 2^{-i}\right] &\le& \exp\left(-\frac{\delta^2}{2 + \delta} \cdot c_i \cdot \mu \right) \\
 &=& \exp\left(- c_i \cdot (2^{-i} - \mu)^2/(2^{-i} + \mu) \right)  \\
 &\le& \exp\left(- c_i \cdot 2^{-i-1}/3 \right) \quad = \; \frac{1}{n^{\ell}\cdot \log_2 n}.
% &=& \frac{1}{n^{\ell}\cdot \log_2 n}.
\end{eqnarray*}
This completes the proof. \done

\pheader
{\bf Proof of Lemma~\ref{lmm:tim-para-stopdown}.} Let $\delta = (\mu - 2^{-i})/\mu$. By the Chernoff bounds,
\begin{eqnarray*}
\Pr\left[\frac{s_i}{c_i} \le 2^{-i}\right] &\le& \exp\left(-\frac{\delta^2}{2} \cdot c_i \cdot \mu \right) \\
 &=& \exp\left(- c_i \cdot (\mu - 2^{-i})^2/(2 \cdot \mu)\right)  \\
 &\le& \exp\left(- c_i \cdot \mu/8 \right)  \quad < \; n^{-\ell \cdot 2^{i-j-1}}/\log_2 n.
% &<& n^{-\ell \cdot 2^{i-j-1}}/\log_2 n.
\end{eqnarray*}
This completes the proof. \done

\pheader
{\bf Proof of Theorem~\ref{thrm:tim-para-kpt}}. Assume that $KPT/n \in [2^{-j}, 2^{-j+1}]$. We first prove the accuracy of the $KPT^*$ returned by Algorithm~\ref{alg:tim-kpt}.

By Lemma~\ref{lmm:tim-para-stopup} and the union bound, Algorithm~\ref{alg:tim-kpt} terminates in or before the $(j - 2)$-th iteration with less than $n^{-\ell}(j-2)/\log_2 n$ probability. On the other hand, if Algorithm~\ref{alg:tim-kpt} reaches the $(j+1)$-th iteration, then by Lemma~\ref{lmm:tim-para-stopdown}, it terminates in the $(j+1)$-th iteration with at least $1 - n^{-\ell}/\log_2 n$ probability. Given the union bound and the fact that Algorithm~\ref{alg:tim-kpt} has at most $\log_2 n - 1$ iterations, Algorithm~\ref{alg:tim-kpt} should terminate in the $(j-1)$-th, $j$-th, or $(j+1)$-th iteration with a probability at least $1 - n^{-\ell}(\log_2 n - 2)/\log_2 n$. In that case, $KPT^*$ must be larger than $n/2 \cdot 2^{-j-1}$, which leads to $KPT^* > KPT/4$. Furthermore, $KPT^*$ should be $n/2$ times the average of at least $c_{j-1}$ i.i.d.\ samples from $\K$. By the Chernoff bounds, it can be verified that
\begin{equation*}
\Pr\left[KPT^* \ge KPT\right] \le n^{-\ell}/\log_2 n.
\end{equation*}
By the union bound, Algorithm~\ref{alg:tim-kpt} returns, with at least $1 - n^{-\ell}$ probability, $KPT^* \in [KPT/4, KPT] \subseteq [KPT/4, OPT]$.

Next, we analyze the expected running time of Algorithm~\ref{alg:tim-kpt}. Recall that the $i$-th iteration of the algorithm generates $c_i$ RR sets, and each RR sets takes $O(EPT)$ expected time. Given that $c_{i+1} = 2 \cdot c_{i}$ for any $i$, the first $j+1$ iterations generate less than $2 * c_{j+1}$ RR sets in total. Meanwhile, for any $i' \ge j+2$, Lemma~\ref{lmm:tim-para-stopdown} shows that Algorithm~\ref{alg:tim-kpt} has at most $n^{-\ell \cdot 2^{i'-j-1}}/\log_2 n$ probability to reach the $i'$-th iteration. Therefore, when $n \ge 2$ and $\ell \ge 1/2$, the expected number of RR sets generated after the first $j+1$ iterations is less than
\begin{eqnarray*}
{\textstyle \sum_{i' = j + 2}^{\log_2 n - 1} \left(c_{i'} \cdot n^{-\ell \cdot 2^{i' - j - 1}}/\log_2 n\right)} &<& c_{j+2}.
\end{eqnarray*}
Hence, the expected total number of RR sets generated by Algorithm~\ref{alg:tim-kpt} is less than $2 c_{j+1} + c_{j+2} = 2 c_{j+2}$. Therefore, the expected time complexity of the algorithm is
\begin{align*}
O(c_{j+2} & \cdot EPT)  =  O(2^j \ell \log n \cdot EPT) \\
& =  O(2^j \ell \log n \cdot (1 + \frac{m}{n}) \cdot KPT) \\
& =  O(2^j \ell \log n \cdot (m + n) \cdot 2^{-j}) = O(\ell (m+n) \log n ).
\end{align*}

Finally, we show that $\E[1/KPT^*] < 12/KPT$. Observe that if Algorithm~\ref{alg:tim-kpt} terminates in the $i$-th iteration, it returns $KPT^* \ge n \cdot 2^{-i-1}$. Let $\zeta_i$ denote the event that Algorithm~\ref{alg:tim-kpt} stops in the $i$-th iteration. By Lemma~\ref{lmm:tim-para-stopdown}, when $n \ge 2$ and $\ell \ge 1/2$, we have
\begin{align*}
\E[1/&KPT^*] = {\textstyle \sum_{i = 1}^{\log_2 n - 1} \Big( 2^{i+1}/n \cdot \Pr[\zeta_i]   \Big)} \\
&< {\textstyle \sum_{i = j+2}^{\log_2 n - 1} \Big( 2^{i+1}/n \cdot \left(n^{-\ell \cdot 2^{i - j - 1}}/\log_2 n \right) \Big) + 2^{j+2}/n} \\
&< (2^{j+3} + 2^{j+2})/n \quad \le \; 12/KPT.
\end{align*}
This completes the proof. \done

\pheader
{\bf Proof of Lemma~\ref{lmm:ext-refine}.}
We first analyze the expected time complexity of Algorithm~\ref{alg:ext-refine}. Observe that Lines 1-6 in Algorithm~\ref{alg:ext-refine} run in time linear the total size of the RR sets in $\mathcal{R}'$, i.e., the set of all RR sets generated in the last iteration of Algorithm~\ref{alg:tim-kpt}. Given that Algorithm~\ref{alg:tim-kpt} has an $O(\ell(m+n)\log n)$ expected time complexity (see Theorem~\ref{thrm:tim-para-kpt}), the expected total size of the RR sets in $\mathcal{R}'$ should be no more than $O(\ell(m+n)\log n)$. Therefore, Lines 1-6 of Algorithm~\ref{alg:ext-refine} have an expected time complexity $O(\ell(m+n)\log n)$.

On the other hand, the expected time complexity of Lines 7-12 of Algorithm~\ref{alg:ext-refine} is $O\left(\E\left[\frac{\lambda'}{KPT^*}\right] \cdot EPT\right)$, since they generate $\frac{\lambda'}{KPT^*}$ random RR sets, each of which takes $O(EPT)$ expected time. By Theorem~\ref{thrm:tim-para-kpt}, $\E[\frac{1}{KPT^*}] < \frac{12}{KPT}$. In addition, by Equation~\ref{eqn:tim-para-kpt}, $EPT \le \frac{m}{n} KPT$. Therefore,
\begin{eqnarray*}
{\textstyle O\left(\E\left[\frac{\lambda'}{KPT^*}\right] \cdot EPT\right)} & = & {\textstyle O\left(\frac{\lambda'}{KPT} \cdot EPT\right)} \\
& = & {\textstyle O\left(\frac{\lambda'}{KPT} \cdot (1 + \frac{m}{n}) \cdot KPT\right)} \\
& = & O\left(\ell(m+n)\log n/(\varepsilon')^2\right).
\end{eqnarray*}
Therefore, the expected time complexity of Algorithm~\ref{alg:ext-refine} is $O\left(\ell(m+n)\log n/(\varepsilon')^2\right)$.

Next, we prove that Algorithm~\ref{alg:ext-refine} returns $KPT^+ \in [KPT^*, OPT]$ with a high probability. First, observe that $KPT^+ \ge KPT^*$ trivially holds, as Algorithm~\ref{alg:ext-refine} sets $KPT^+ = \max\{KPT', KPT^*\}$, where $KPT'$ is derived in Line 11 of Algorithm~\ref{alg:ext-refine}. To show that $KPT^+ \in [KPT^*, OPT]$, it suffices to prove that $KPT' \le OPT$.

By Line 11 of Algorithm~\ref{alg:ext-refine}, $KPT' = f \cdot n / (1 + \varepsilon')$, where $f$ is the fraction of RR sets in $\mathcal{R}''$ that is covered by $S'_k$, while $\mathcal{R}''$ is a set of $\theta'$ random RR sets, and $S'_k$ is a size-$k$ node set generated from Lines 1-6 in Algorithm~\ref{alg:ext-refine}. Therefore, $KPT' \le OPT$ if and only if $f \cdot n \le (1 + \varepsilon') \cdot OPT$.

Let $\rho'$ be the probability that a random RR set is covered by $S'_k$. By Corollary~\ref{coro:tim-fraction}, $\rho'=\E[I(S_k')]/n$. In addition, $f \cdot \theta'$ can be regarded as the sum of $\theta'$ i.i.d.\ Bernoulli variables with a mean $\rho'$. Therefore, we have
\begin{eqnarray} \label{eqn:appen-refine-proof}
\lefteqn{\Pr\left[ f\cdot n >　(1+\varepsilon')\cdot OPT\right]} \nonumber \\
&\le& \Pr\Big[n\cdot f -\E\left[I(S'_k)\right]>\varepsilon'\cdot OPT\Big] \nonumber \\
&=& \Pr\left[\theta'\cdot f - \theta' \cdot \rho' > \frac{\theta'}{n} \cdot \varepsilon' \cdot OPT\right] \nonumber \\
&=& \Pr\left[\theta' \cdot f- \theta' \cdot \rho' > \frac{\varepsilon' \cdot OPT}{n \cdot \rho'}\cdot \theta' \cdot \rho'\right]
\end{eqnarray}
let $\delta=\varepsilon' \cdot OPT/(n\rho')$. By the Chernoff bounds, we have
\begin{eqnarray*}
\textrm{r.h.s.\ of Eqn.~\ref{eqn:appen-refine-proof}} & \le & \exp\left(-\frac{\delta^2}{2 + \delta} \cdot \rho'\theta'\right) \\
& = & \exp\left(-\frac{{\varepsilon'}^2\cdot OPT^2}{2 n^2 \rho' +  \varepsilon' n \cdot OPT}\cdot \theta'\right) \\
& \le & \exp\left(-\frac{{\varepsilon'}^2\cdot OPT^2}{2 n \cdot OPT +  \varepsilon' n \cdot OPT}\cdot \theta'\right) \\
%& = & \exp\left(-\frac{{\varepsilon'}^2\cdot OPT}{(2 +  \varepsilon')\cdot n}\cdot \theta'\right) \\
& = & \exp\left(-\frac{{\varepsilon'}^2\cdot OPT}{(2 +  \varepsilon')\cdot n}\cdot \frac{\lambda'}{KPT^*} \right) \\
& \le & \exp\left(-\frac{{\varepsilon'}^2\cdot \lambda'}{(2 +  \varepsilon')\cdot n} \right) \quad \le \:\: \frac{1}{n^{l}}.
\end{eqnarray*}
Therefore, $KPT'= f\cdot n/(1+\varepsilon') \le OPT$ holds with at least $1-n^{-l}$ probability. This completes the proof. \done

\pheader
{\bf Proof of Lemma~\ref{lmm:ext-rr}.}
Let $g$ be a graph constructed from $G$ by first sampling a node set $T$ for each node $v$ from its triggering distribution $\T(v)$, and then removing any outgoing edge of $v$ that does not point to a node in $T$. Then, $\rho_2$ equals the probability that $v$ is reachable from $S$ in $g$. Meanwhile, by the definition of RR sets under the triggering model, $\rho_1$ equals the probability that $g$ contains a directed path that ends at $v$ and starts at a node in $S$. It follows that $\rho_1 = \rho_2$. \done

\pheader
{\bf Proof of Lemma~\ref{lmm:compare-greedy}.}
Let $S$ be any node set that contains no more than $k$ nodes in $G$, and $\xi(S)$ be an estimation of $\E[I(S)]$ using $r$ Monte Carlo steps. We first prove that, if $r$ satisfies Equation~\ref{eqn:compare-r}, then $\xi(S)$ will be close to $\E[I(S)]$ with a high probability.

Let $\mu = \E[I(S)]/n$ and $\delta = \varepsilon OPT / (2kn\mu)$. By the Chernoff bounds, we have
\begin{eqnarray} \label{eqn:appen-greedy-proof-1}
\lefteqn{\Pr\left[\left|\xi(S)-\E[I(S)]\right|>\frac{\varepsilon}{2k}OPT\right]} \label{eqn:compare-greedy-proof-1} \nonumber \\
&=& \Pr\left[\left|r\cdot \frac{\xi(S)}{n}-r\cdot\frac{\E[I(S)]}{n}\right|>\frac{\varepsilon}{2kn}\cdot r\cdot OPT\right]  \nonumber \\
&=& \Pr\left[\left|r\cdot \frac{\xi(S)}{n}-r\cdot\frac{\E[I(S)]}{n}\right|>\delta \cdot r \cdot \mu\right]  \nonumber \\
&<& 2\exp\left(-\frac{\delta^2}{2+\delta}\cdot r\cdot \mu\right)\nonumber\\
&=& 2\exp\left(-\frac{\varepsilon^2}{(8k^2+2k\varepsilon)\cdot n}\cdot r\cdot \mu\right)\nonumber\\
&=& 2\exp\left((\ell+1)\log n+\log k\right)\nonumber\\
&=& \frac{1}{k \cdot n^{\ell+1}}
\end{eqnarray}

Observe that, given $G$ and $k$, {\em Greedy} runs in $k$ iterations, each of which estimates the expected spreads of at most $n$ node sets with sizes no more than $k$. Therefore, the total number of node sets inspected by {\em Greedy} is at most $kn$. By Equation~\ref{eqn:appen-greedy-proof-1} and the union bound, with at least $1 - n^{-\ell}$ probability, we have
\begin{equation} \label{eqn:appen-greedy-proof-2}
\left|\xi(S')-\E[I(S')]\right| \le \frac{\varepsilon}{2k}OPT,
\end{equation}
for all those $kn$ node sets $S'$ simultaneously. In what follows, we analyze the accuracy of {\em Greedy}'s output, under the assumption that for any node set $S'$ considered by {\em Greedy}, it obtain a sample of $\xi(S')$ that satisfies Equation~\ref{eqn:appen-greedy-proof-2}. For convenience, we abuse notation and use $\xi(S')$ to denote the aforementioned sample.

Let $S_0 = \emptyset$, and $S_i$ ($i \in [1, k]$) be the node set selected by {\em Greedy} in the $i$-th iteration. We define $x_i = OPT - I(S_i)$, and $y_{i}(v) = I\left(S_{i-1} \cup \{v\} \right) - I(S_{i-1})$ for any node $v$. Let $v_i$ be the node that maximizes $y_i(v_i)$. Then, $y_i(v_i) \ge x_{i-1}/k$ must hold; otherwise, for any size-$k$ node $S$, we have
\begin{eqnarray*}
I(S) & \le & I(S_{i-1}) + I(S \setminus S_{i-1}) \\
& \le & I(S_{i-1}) + k \cdot y_i(v_i) \\
& < & I(S_{i-1}) + x_{i-1}  \quad = \:\: OPT,
\end{eqnarray*}
which contradicts the definition of $OPT$.

Recall that, in each iteration of {\em Greedy}, it adds into $S_{i-1}$ the node $v$ that leads to the largest $\xi(S_{i-1} \cup \{v\})$. Therefore,
\begin{eqnarray} \label{eqn:appen-greedy-proof-3}
\xi(S_{i}) - \xi(S_{i-1}) &\ge& \xi(S_{i-1} \cup \{v_i\}) - \xi(S_{i-1}).
\end{eqnarray}
Combining Equations \ref{eqn:appen-greedy-proof-2} and \ref{eqn:appen-greedy-proof-3}, we have
\begin{eqnarray} \label{eqn:appen-greedy-proof-4}
\lefteqn{x_{i-1} - x_i} \nonumber \\
& = & I(S_{i}) - I(S_{i-1}) \nonumber \\
& \ge & \xi(S_{i}) - \frac{\varepsilon}{2k} OPT - \xi(S_{i-1}) + \Big(\xi(S_{i-1}) - I(S_{i-1})\Big)  \nonumber \\
&\ge& \xi(S_{i-1} \cup \{v_i\}) - \xi(S_{i-1}) - \frac{\varepsilon}{2k} OPT \nonumber \\
& & {} + \Big(\xi(S_{i-1}) - I(S_{i-1})\Big) \nonumber \\
&\ge& I\Big(S_{i-1} \cup \{v_i\}\Big) - I(S_{i-1}) - \frac{\varepsilon}{k} OPT \nonumber \\
&\ge& \frac{1}{k} x_{i-1} - \frac{\varepsilon}{k} OPT.
\end{eqnarray}

Equation~\ref{eqn:appen-greedy-proof-4} leads to
\begin{eqnarray*} %\label{eqn:appen-greedy-proof-4}
x_k & \le & \left(1 - \frac{1}{k}\right) \cdot x_{k-1} + \frac{\varepsilon}{k} OPT \\
& \le & \left(1 - \frac{1}{k}\right)^2 \cdot x_{k-2} + \left(1 + \left(1 - \frac{1}{k}\right)\right)\cdot \frac{\varepsilon}{k} OPT \\
& \le & \left(1 - \frac{1}{k}\right)^{k} \cdot x_{0} + \sum_{i = 0}^{k-1} \left( \left(1 - \frac{1}{k}\right)^{i} \cdot \frac{\varepsilon}{k} OPT\right) \\
& = & \left(1 - \frac{1}{k}\right)^{k} \cdot OPT + \left(1 - \left(1 - \frac{1}{k}\right)^k\right) \cdot \varepsilon \cdot OPT \\
& \le & \frac{1}{e} \cdot OPT - \left(1 - \frac{1}{e} \right) \cdot \varepsilon \cdot OPT.
\end{eqnarray*}
Therefore,
\begin{eqnarray*}
I(S_k) & = & OPT - x_k \\
& \le & (1 - 1/e) \cdot (1 - \varepsilon) \cdot OPT \\
& \le & (1 - 1/e - \varepsilon) \cdot OPT.
\end{eqnarray*}
Thus, the lemma is proved. \done

\end{sloppy}
\end{document}